\documentclass[apj]{emulateapj}

\newcommand{\msun}{$M_{\odot}$}

\newcommand{\Gam}{$\Gamma$}

\shorttitle{The IMF of NGC 3603}
\shortauthors{Harayama et al.}

\begin{document}

\title{The Initial Mass Function of the Massive Star-forming Region NGC 3603 \\
from Near-Infrared Adaptive Optics observations\altaffilmark{1}}

\author{Y. Harayama, F. Eisenhauer, and F. Martins}
\affil{Max-Planck-Institut f\"{u}r extraterrestrische Physik, Postfach 1312, 85741 Garching, Germany}
\email{yohei@mpe.mpg.de}
\altaffiltext{1}{Based on observations obtained at the Very Large Telescope (VLT) of the European Southern Observatory (ESO) on Paranal, Chile, under programs 70.C-0490 and 74.C-0764 (NACO), and on observations from the ESO archive (ISAAC).}

\begin{abstract}
We study the initial mass function (IMF) of one of the most massive Galactic star-forming regions NGC 3603 to answer a fundamental question in current astrophysics: is the IMF universal, or does it vary?
Using our very deep, high angular resolution $JHK_{S}L'$ images obtained with NAOS-CONICA at the VLT at ESO, we have successfully revealed the stellar population down to the subsolar mass range in the core of the starburst cluster.
The derived IMF of NGC 3603 is reasonably fitted by a single power law with index $\Gamma \sim -0.74$ within a mass range of $0.4 - 20$ \msun, substantially flatter than the Salpeter-like IMF. A strong radial steepening of the IMF is observed mainly in the inner $r \lesssim 30''$ field, indicating mass segregation in the cluster center.
We estimate the total mass of NGC 3603 to be about $1.0 - 1.6 \times 10^4$ \msun. The derived core density is $\geq 6 \times 10^4$ \msun pc$^{-3}$, an order of magnitude larger than e.g., the Orion Nebula Cluster.
The estimate of the half-mass relaxation time for solar-mass stars is about $10 - 40$ Myr, suggesting that the intermediate- and low-mass stars have not yet been affected significantly by the dynamical relaxation in the cluster. The relaxation time for the high-mass stars can be comparable to the age of the cluster.
We estimate that the stars residing outside the observed field cannot steepen the IMF significantly, indicating our IMF adequately describes the whole cluster.
Analyzing thoroughly the systematic uncertainties in our IMF determination, we conclude that the power law index of the IMF of NGC 3603 is $\Gamma = -0.74^{+0.62}_{-0.47}$. Our result thus supports the hypothesis of a potential top-heavy IMF in massive star-forming clusters and starbursts.
\end{abstract}


\section{Introduction}
\label{introduction}
One of the most interesting properties of massive star-forming regions is the stellar initial mass function (IMF).
Since the pioneering work by \citet{sal55}, which led to a standard picture of the IMF, the so-called Salpeter IMF ($dN/d\log \mathcal{M} \propto \mathcal{M}^\Gamma$ with $\Gamma = -1.35$, where $\mathcal{M}$ is stellar mass), numerous efforts have been made to understand the IMF of many types of objects such as field populations, stellar associations, and open and globular clusters, covering the whole stellar mass range from massive OB stars down to substellar brown dwarfs.
It is currently accepted that the IMF follows a single Salpeter-like power law in the high- to intermediate-mass range ($\mathcal{M} \gtrsim 1$ \msun). It becomes flatter towards subsolar masses and peaks at a characteristic mass of several tenths of a solar mass. It then declines towards the brown dwarf mass range.
Therefore, several analytical expressions have been proposed for the standard IMF, for example, a lognormal distribution \citep{mil79,sca86}, a segmented power law distribution \citep{sca98, kro01}, and a combination of both \citep{cha03}.
As these representations of the IMF fit reasonably well the various types of stellar populations, the idea of the \textit{universality} of the IMF was born.
A universal IMF basically suggests a universal star formation mechanism.
However, it is somewhat intuitive that different physical conditions such as the density, velocity fields, chemical composition, and tidal forces in the natal molecular clouds can lead to different star formation processes and, consequently, some variability in the IMF.
Indeed, there is growing evidence for the variable IMF.
There are significant variations in the power law index and the characteristic masses at which the distribution shows the peak or at which the power law breaks.
A typical example of an IMF that does not follow the Salpeter-like distribution is the so-called top-heavy IMF in starburst galaxies and young, massive star-forming regions.

Does the IMF really vary among stellar populations? To answer this question, we need to clarify, for at least some stellar populations, if the observed IMF variations are without any doubt true deviations from the Salpeter-like IMF or if they can simply be accommodated by the combined effects of observational, theoretical, and statistical uncertainties?
Many studies have so far addressed this question.
In order to address the question and to obtain new insights into the IMF of an intense starburst environment, we study the IMF of the massive star-forming HII region \objectname{NGC 3603} based on unprecedented spacial resolution observations of its central starburst cluster obtained by NACO at the Very Large Telescope (VLT), as well as a wider field with the Infrared Spectrometer and Array Camera (ISAAC) at VLT at the European Southern Observatory (ESO).
Compared to extragalactic starbursts (e.g. \objectname{M82}), young star-forming regions in the Milky Way and in the Magellanic Clouds -- so-called nearby starburst templates -- are close enough to resolve the individual stars with current powerful adaptive optics (AO)-assisted ground-based telescopes and space-based facilities.
These objects include, for example, the \objectname{Arches} cluster, the \objectname{Quintuplet} cluster, \objectname{R136}, and \objectname{NGC 3576}.
More nearby but less massive star-forming regions are the \objectname{Trapezium} cluster in the Orion Nebula Cluster (ONC), the \objectname{Pleiades} cluster, the \objectname{Taurus} cluster, and \objectname{IC 348}.

The IMF of NGC 3603 has been presented in several studies, and recent works have derived somewhat flat IMFs with a power law index of $\Gamma = -0.73$ for $1 - 30$ \msun\ in \citet{eis98}, $\Gamma = -0.9$ for $2.5 - 100$ \msun\ in \citet{sun04}, and $\Gamma = -0.91 \pm 0.15$ for $0.4 - 20$ \msun\ in \citet{sto06}.
As another example of the IMF of massive Galactic star clusters, the Arches cluster has shown a slightly flat IMF with $\Gamma = -0.6$ to $-1.1$ for intermediate- and high-mass stars \citep{fig99,sto05,kim06}.
Studies of the IMF of the Arches and other clusters are presented in \S~\ref{c-discussion}.

\subsection{NGC 3603}
NGC 3603 is one of the most luminous, optically visible HII regions in the Milky Way \citep{gos69} with its global properties such as $L$(H$_{\alpha}$) $\sim 1.5 \times 10^{39}$ ergs s$^{-1}$ \citep{ken84}, and the total mass of molecular clouds $\sim4 \times 10^{5}$ \msun\ \citep{gra88}.
Its total bolometric luminosity is as large as 10$^{7}$ $L_{\odot}$, 2 orders of magnitude larger than that of the ONC, and just an order of magnitude smaller than the other well known starburst template R136 in 30 Dor in the Large Magellanic Cloud (LMC).
In particular, the starburst cluster located in the northern part of the gigantic HII complex including \objectname{HD 97950} -- a very compact Trapezium-like system with plenty of high-mass components \citep{wal73} -- at its center, is one of the most massive and the densest star-forming regions known in the Galaxy.
The starburst cluster consists of three WNL stars, six O3 stars, and many other late O- and B-type stars in a volume of less than a cubic light year, providing most of the ionizing radiation in the giant HII region \citep{mof83, cla86, mof94, dri95,hof95, cro98}. It thus shows remarkable similarities with R136.
The distance of NGC 3603 from the Sun has been estimated in many earlier works by means of both photometric and kinematic analysis \citep[e.g.][]{vdb78,mel82}, and the currently accepted value is about 7 $\pm$ 1 kpc.
Owing to its intrinsic properties such as its proximity, relatively low visual extinction of only $A_{V} = 4 - 5$ mag, and extreme compactness and brightness, NGC 3603 is one of the most suitable Galactic templates of starburst phenomena in distant galaxies.

\section{Observations and data reductions}
\label{observation}
Observations were carried out in two periods, on 2003 March 18-21 and 2005 February 14-16, using the AO system NACO at the VLT at ESO on Paranal, Chile.
In the following analysis we mainly use the data obtained in the second observing run because of its better data quality. The first run data has been used for the estimate of statistical uncertainties in our point-spread function (PSF) photometry.

The central starburst cluster including the high-mass stellar complex HD 97950 in NGC 3603 was imaged in three bands with the NACO $J$ (1.27 $\mu$m), $H$ (1.66 $\mu$m), and $K_S$ (2.18 $\mu$m) filters in both runs, and an additional band with the $L'$ (3.80 $\mu$m) filter in the second run.
The bright cluster core consisting of several Wolf-Rayet (WR) stars and OB stars was used as the AO reference source. The pixel scale of $\sim27$ mas was used for all bands, corresponding to a field of view of about $28''\times 28''$.
In the $L'$-band imaging, a very short exposure time of 0.2 s was chosen so as to avoid saturation due to the substantially high sky background level.
In addition, we carried out short-exposure observations in $JHK_{S}$ bands to acquire photometry of the brightest members. We spent about one third of the whole integration time imaging a sky field in $JHK_{S}$ bands. In the $L'$ band we spent about one half of the integration time.

\subsection{Data reduction}
\subsubsection{NACO data}
The data reduction for the NACO data sets was performed in the standard manner of near-infrared (NIR) image reductions, which consists of a sky-subtraction, flat-fielding, bad-pixel correction, and a mosaic combination as major procedures.
We used the IRAF\footnote{\textit{IRAF} is distributed by the National Optical Astronomy Observatories, which are operated by the Association of Universities for Research in Astronomy, Inc., under cooperative agreement with the National Science Foundation (NSF).} package, IDL, and in part the ESO eclipse package.
Note that we have done several corrections, such as gain and pixel scale corrections, which were required because of the replacement of the CONICA detector between the first and the second data sets.
We selected only frames with reasonable resolutions to achieve the highest resolution for the mosaic frames.
Consequently, typical Strehl ratios measured on individual sources around the cluster center are $\sim$12\%, 18\%, and 32\% in the $JHK_{S}$-band mosaic frames, respectively. A typical Strehl ratio in the $L'$-band frame is more than 70\%.

\subsubsection{ISAAC data from the ESO archive}
\begin{figure*}
\plottwo{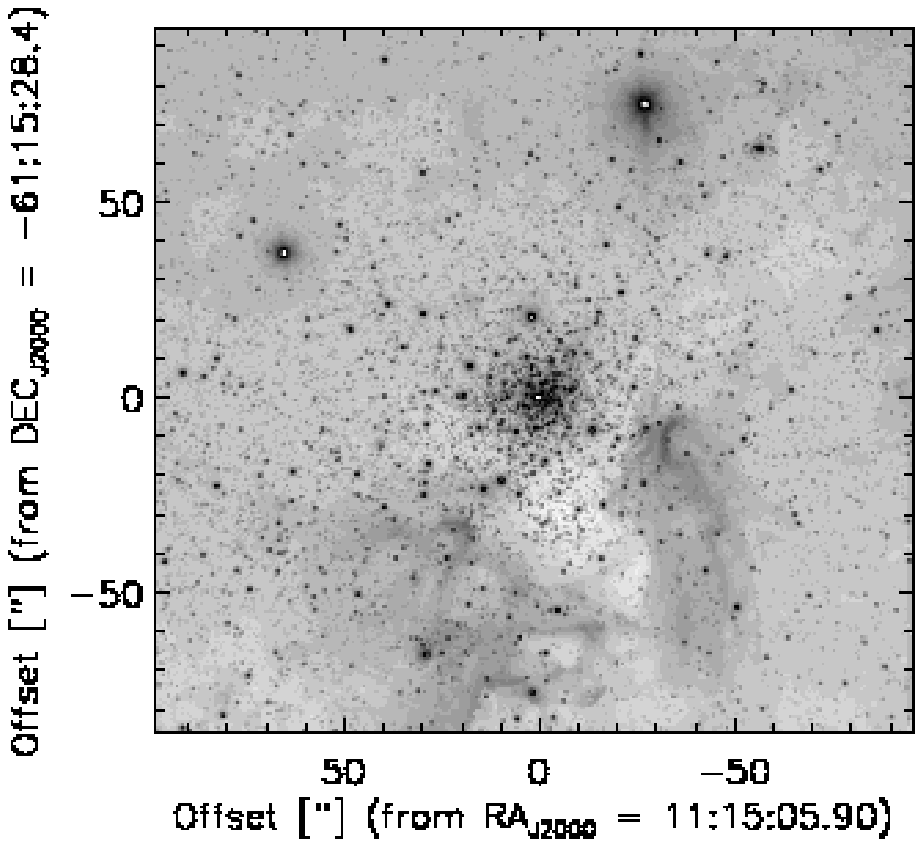}{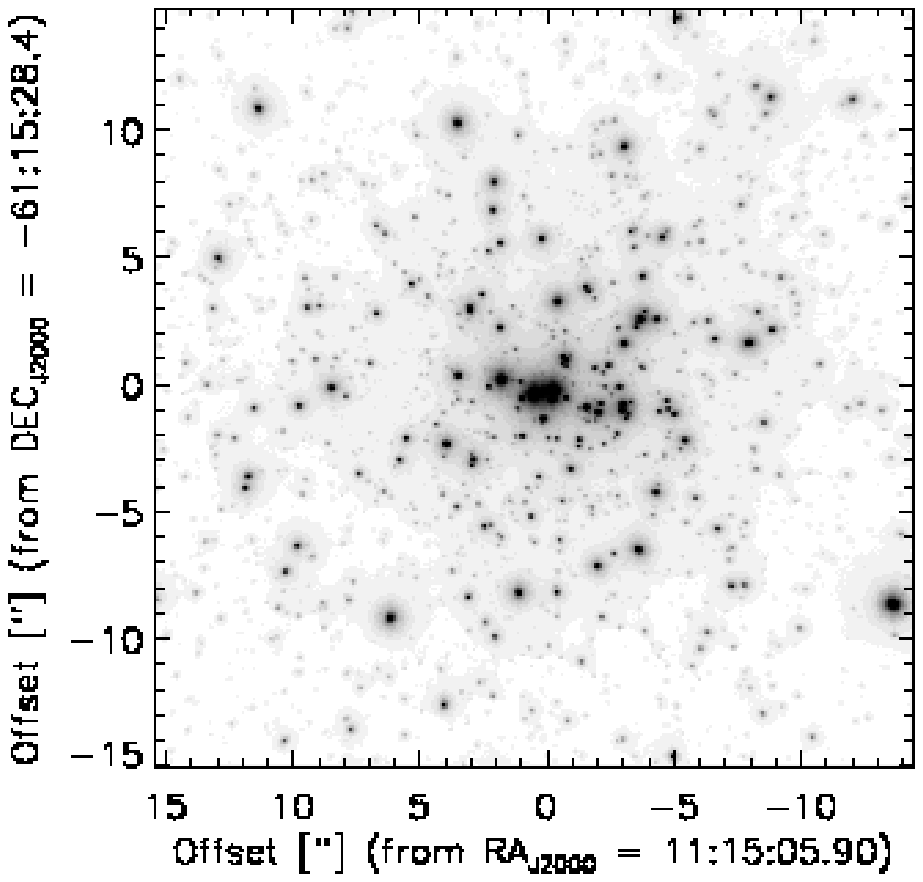}
\caption{The $K_{S}$-band mosaic frames. The wide field of $3'.2 \times 3'.2$ (5.2 $\times$ 5.2 pc) with the resolution of $\sim0.''46$ from the ISAAC data (\textit{left}) and the central field of $30'' \times 30''$ ($0.9 \times 0.9$ pc) with the resolution of $\sim0''.1$ from the NACO data (\textit{right}) are shown.
The long-exposure NACO data is scaled to enhance the faint stars; thus, the bright stars in the center are saturated. Stellar fluxes are shown in square-root and log scales for the NACO and the ISAAC frames, respectively. North is up and east is to the left.\label{fig-mos}}
\end{figure*}

In addition to the high-resolution NACO observations of the cluster center, we analyzed the VLT/ISAAC  $J_{S}HK_{S}$-band data sets of a wider field of view retrieved from the ESO archive facility.
These three-band data have the pixel scale of $0.147''$ pixel$^{-1}$, corresponding to a field of view of about $150'' \times 150''$. In addition to the basic reduction procedures, two particular procedures -- ghost removal and geometrical distortion correction -- were performed. Finally, a mosaic field of about $3'.2 \times 3'.2$ was used for the following PSF photometry.
The NACO and ISAAC $K_{S}$-band mosaic frames are shown in Figure~\ref{fig-mos}.

\section{Point Spread Function photometry}
\label{c-photometry}
We adopted the PSF photometry for the detection of stellar objects and the derivation of instrumental magnitudes using the STARFINDER package implemented in IDL \citep{dio00}. This package performs a PSF fitting using an empirical PSF extracted from sources in the image itself instead of using a PSF based on analytical models. This is particularly suitable for AO-assisted imaging data, which generally show a peculiar PSF, i.e., a sharply peaked core with widely extended wings.
We then performed photometric calibrations using frames of standard stars obtained during or close to the observing runs. Here we only corrected a non-negligible discrepancy between the ISAAC $J_{S}$-band filter and the $J$-band filter in the standard system. Hereafter we use $J$ band for the color-corrected ISAAC $J_{S}$-band photometry.
Our alternative calibration using $\sim60$ identical sources in the Two Micron All Sky Survey (2MASS)\footnote{The Two Micron All Sky Survey is a joint project of the University of Massachusetts and the Infrared Processing and Analysis Center/California Institute of Technology, funded by the National Aeronautics and Space Administration and the NSF.} catalogue yielded fairly close zeropoints ($\sigma \leq 0.04$ mag).
Moreover, for these calibrated NACO $JHK_{S}$-band lists, we performed an additional calibration with respect to the anisoplanatic effect by using the ISAAC photometry lists. In each annulus with a step of $2''$ at $8'' < r \leq 14''$, we computed a typical offset of $m_{NACO} - m_{ISAAC}$ magnitudes of a source that coincides in both NACO and ISAAC and corrected the offsets.

To derive detection limits in the ISAAC field of $r \gtrsim 30''$, where the source detection is not limited by the crowding but simply by the detector noise, we performed Monte Carlo tests of artificial source detections. Defining a 50\% recovery rate as a detection limit, the limits are about 20.5, 19.8, and 19.7 mag in $J$, $H$, and $K_{S}$ bands, respectively.
Photometry results are summarized in Tab.~\ref{tbl-mosaic} together with properties of each mosaic frame.
In both NACO and ISAAC data the source detection is primarily limited by the $J$-band data.

In addition, we construct a photometry list of $JHK_{S}(L')$-band detected sources of $m_{J} < 15.5$ mag in the innermost $r \leq 13''$ region ($\gtrsim95$\% completeness) from the NACO data. 

\begin{deluxetable*}{lccccc}
\tabletypesize{\footnotesize}
\tablecaption{Summary of mosaic frame properties \label{tbl-mosaic}}
\tablewidth{0pt}
\tablehead{
Band\tablenotemark{[a]} & Num. of & Total integration time\tablenotemark{[c]} & \textit{FWHM} of PSF\tablenotemark{[d]} & Detection & Num. of \\
& frames\tablenotemark{[b]} & at the center (s) & (arcsec) & thresholds\tablenotemark{[e]} ($\sigma$) & detected sources
}
\startdata
\multicolumn{3}{l}{NAOS-CONICA mosaics from the first run}\\
\tableline
$J$ (20 s $\times$ 2) & 11 & 440 & 0.15  & 7 & 1807 \\
$H$ (15 s $\times$ 3) & 17 & 765 & 0.13  & 7 & 3163 \\
$K_S$ (20 s $\times$ 2) & 16 & 640 & 0.096 & 7 & 3688 \\
\hline
\multicolumn{3}{l}{NAOS-CONICA mosaics from the second run}\\
\tableline
$J$ (20 s $\times$ 2)   & 127 & 5080  & 0.15  & 15 & 2589 \\
$H$ (15 s $\times$ 3)   &  36 & 1620  & 0.11  & 10 & 4228 \\
$K_S$ (20 s $\times$ 2) & 39  & 1560  & 0.098 & 7 & 4269  \\
$L'$ (0.2 s $\times$ 150)  & 73  & 2190  & 0.12  & 4 & 1216 \\
\multicolumn{3}{l}{Short exposure}\\
$J$ (0.35 s $\times$ 30) & 15  & 157.5 & 0.16 & 11 & 480  \\
$H$ (0.35 s $\times$ 30) & 15  & 157.5 & 0.12 & 10 & 1620 \\
$K_S$ (0.35 s $\times$ 30) & 16  & 168   & 0.11 & 7  & 1677 \\
\tableline
\multicolumn{3}{l}{ISAAC mosaics from the ESO archive}\\
\tableline
$J_S$ (1.772 s $\times$ 34) & 36 & 2160 & 0.47 & 17 & 10106 \\
$H$ (1.772 s $\times$ 34) & 59 & 3540 & 0.49 & 11 & 14089 \\
$K_S$ (1.772 s $\times$ 34) & 58 & 3480 & 0.46 & 8  & 16105
\enddata
\tablenotetext{[a]}{An integration time and a number of exposures (DIT $\times$ NDIT) for a single frame is given in the bracket.}
\tablenotetext{[b]}{Number of contributing frames to the final mosaic frames after frame selections with respect to the spatial resolution.}
\tablenotetext{[c]}{Total integration time of the most overlapping central region. This decreases towards outer regions from the center.}
\tablenotetext{[d]}{FWHM of the typical PSF extracted empirically from the mosaic frame and used in the flux measurement.}
\tablenotetext{[e]}{Detection threshold applied in the PSF photometry in unit of noise standard deviation derived from the mosaic frame. We chose an optimal threshold for each mosaic frame separately rather than using a uniform value so as to achieve an optimal detection.}
\end{deluxetable*}

\section{Completeness limits}
\label{ss-incomp-correction}
Even in our high spatial resolution NACO data, the central region of HD 97950 is extremely crowded.
To correct these incompleteness effects, we have employed the empirical method first introduced by \citet{eis98} for the ADONIS observation of NGC 3603. In this method we use the spatial distribution and measured photometry of detected sources. For this we assume (1) the probability of the detection of a source with a given magnitude is uniform across the whole field, (2) every object has a circular area around itself over which any sources fainter than $\Delta{m}$ mag relative to the object cannot be detected, and (3) at any point in the field, the blending effect is dominated only by the single brightest neighboring star.

\begin{figure*}
\plottwo{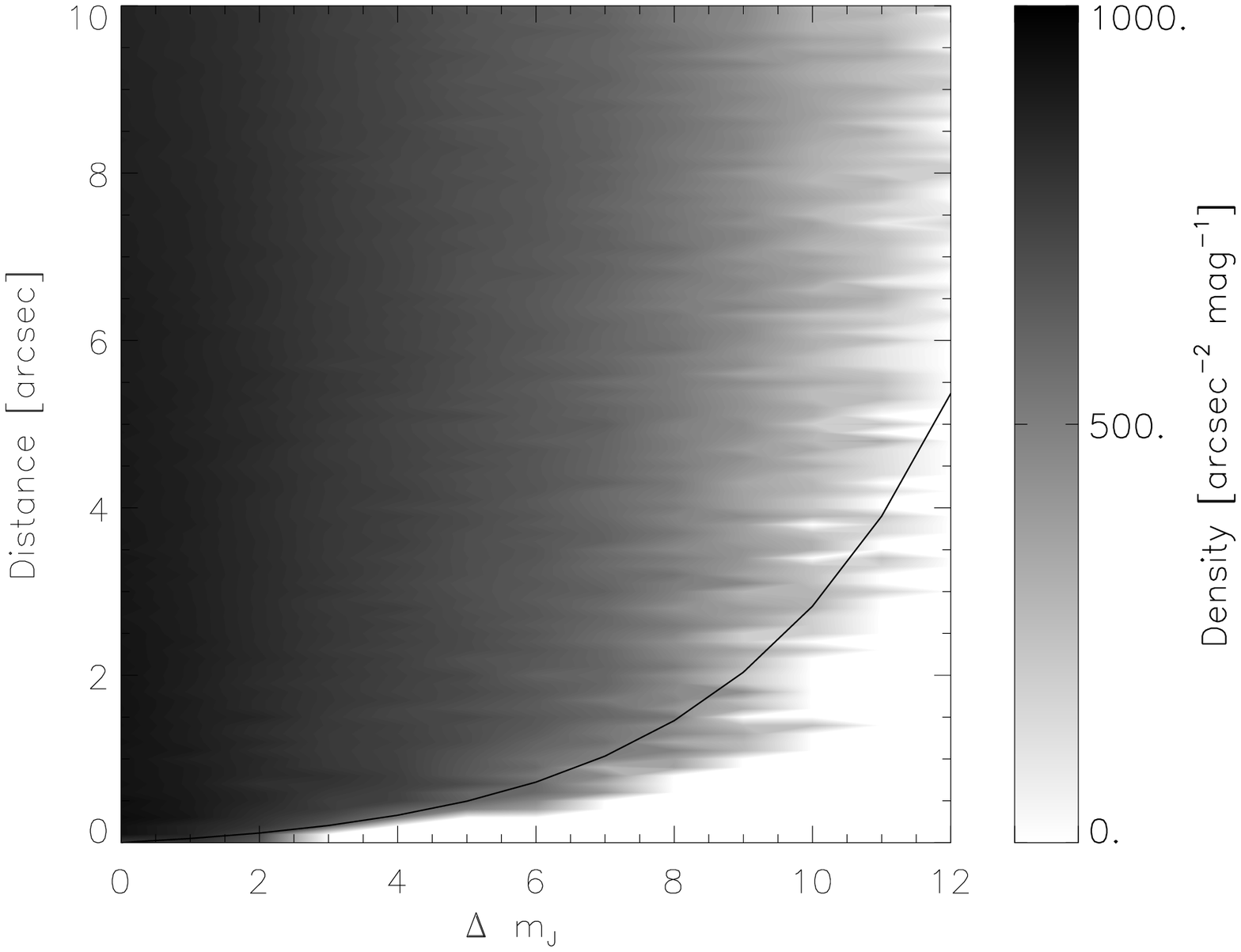}{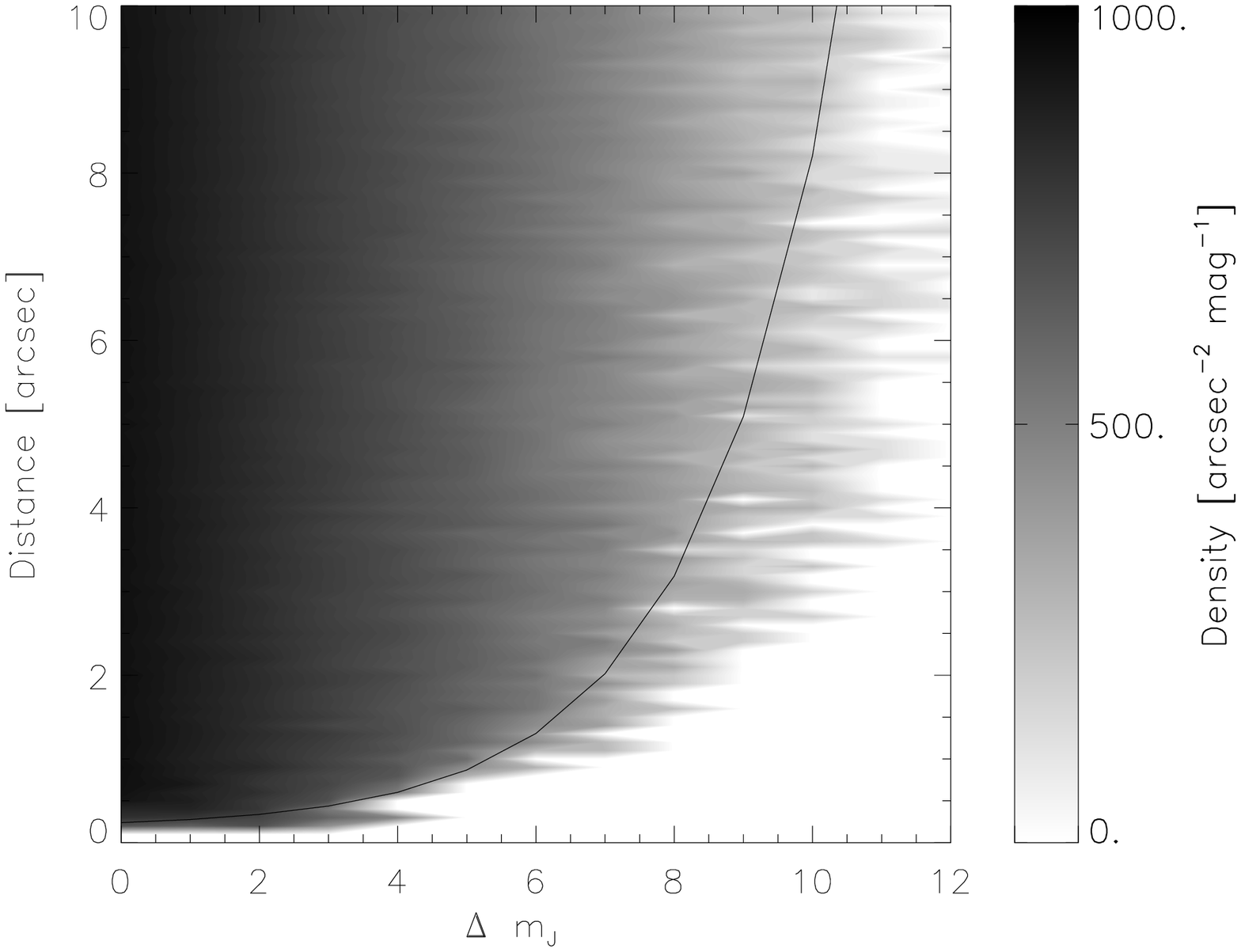}

\plottwo{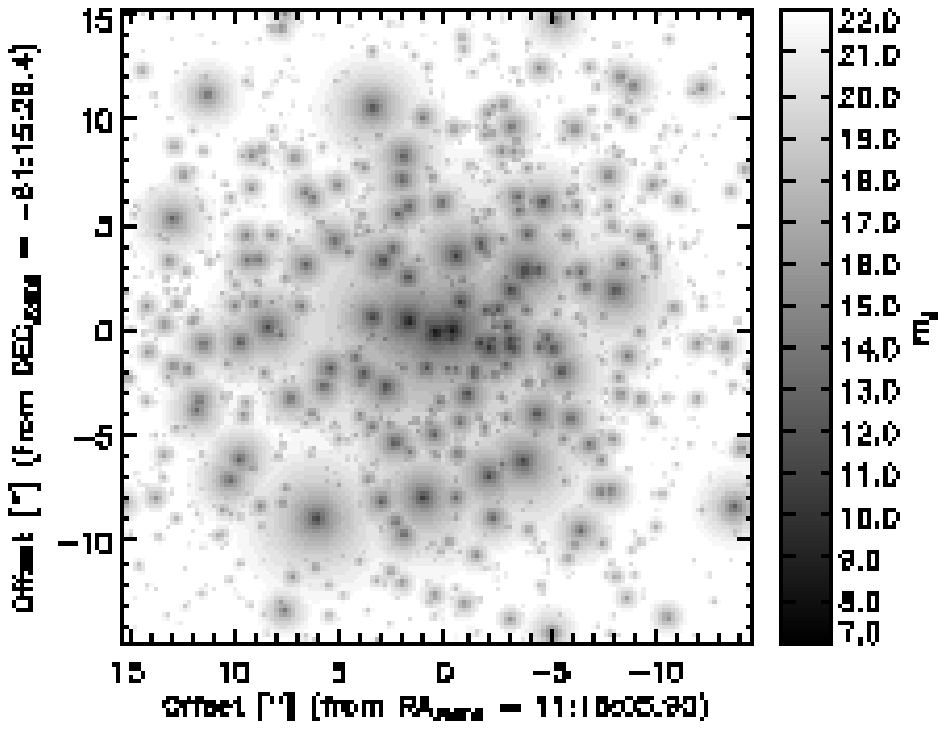}{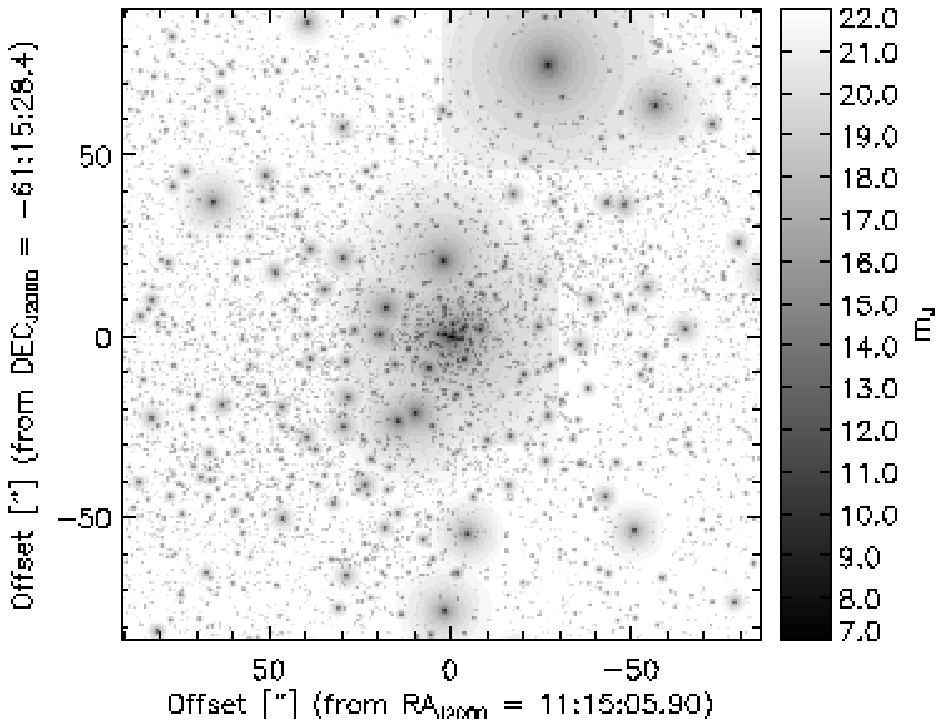}
\caption{Critical distance curve and the blending map derived from $J$-band photometry from the NACO (\textit{left}) and ISAAC (\textit{right}) data. Upper panels show the scaled distribution (i.e. density) $F(\Delta{m_J},d)$ in log scale and the critical distance determined based on a curve fit to points with half of the converging density in each $\Delta{m}$ bin. The color bar illustrates the density, and darker color corresponds to a higher probability to detect a star. Lower panels illustrate the blending map. The color bar indicates the limiting magnitude. \label{fig-comp}}
\end{figure*}

\begin{figure}
\epsscale{1.0}
\plotone{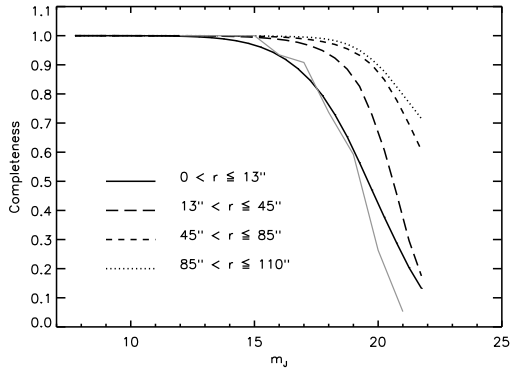}
\caption{Completeness as a function of the observed $J$-band magnitude calculated for four distinct radial regions. The gray curve shows the completeness curve from the Monte Carlo analysis of the innermost region.\label{fig-comp-frac}}
\end{figure}

The technique mainly consists of three steps: (1) the derivation of the critical distance using the obtained photometry, (2) the construction of the blending map using the critical distance, and (3) the computation of the incompleteness correction factor based on the map.
For this we use the $J$-band magnitude, which limits the source detections in both NACO and ISAAC data.
\begin{enumerate}
\item \textbf{Derivation of the critical distance}\\
We first computed a distribution function $f(\Delta{m_J},d)$. This $f$ illustrates how often two stars with a difference in magnitudes $\Delta{m_J}$ and a relative distance $d$ are found in the observation.
Scaling $f$ by the corresponding sampling annulus, the density distribution $F(\Delta{m_J},d)$ should converge to a certain value towards large radii because stars do not blend with each other any more. Then by fitting a curve to points at which the distribution has half the value of the plateau in each $\Delta{m_J}$ bin, a representative curve defining the critical distance can be determined. The distribution functions from both the NACO and ISAAC data are shown in Figure~\ref{fig-comp}. In this analysis we also derive another critical distance by taking 2/3 of the value of the plateau in order to provide \textit{larger} incompleteness corrections (see \S~\ref{s-robust-incomp}).
\item \textbf{Construction of the blending map}\\
Applying the derived critical distances back into the photometry, we then make a map. This blending map gives for each point the limiting magnitude. Thus in the map all detected sources have the concentric circular areas of the blending effect around themselves (Figure~\ref{fig-comp}). We adopted magnitudes of 2MASS photometry for very bright stars ($m_{J} \leq 10$ mag) in the ISAAC data.
\item \textbf{Computation of the incompleteness correction factors}\\
A completeness fraction of a magnitude bin is given by measuring the reciprocal proportion of the area over which a source with a given magnitude is detectable to the total area of the map. 
Then the inverse of the fraction is the incompleteness correction factor.
\end{enumerate}
Completeness is not uniform but is strongly dependent on the stellar density across the field, and it is predominantly associated with the radial distance from the cluster center. Therefore we applied variable correction factors in the analysis of the radial density profile, luminosity function (LF), and IMF. The radial variation of the completeness is shown in Figure~\ref{fig-comp-frac}.

Finally, in order to investigate how the resulting completeness in our method could differ from that based on the commonly applied Monte Carlo test of artificial sources, we perform such a test for the inner NACO region of $r \leq 13''$. In this test, we first generate artificial sources with a given magnitude using the original PSF, randomly distribute them onto the mosaic frame, perform the PSF photometry in the same manner as the original photometry list, and then derive the recovery rate at the magnitude. We perform this procedure in every magnitude within $m_J = 12 - 21$ mag. Here, we set the number of artificial sources at $\sim5$\% of the total number of stars in the field so as to keep low the degree of crowding.
The resulting completeness curve is shown in Figure~\ref{fig-comp-frac} (\textit{gray solid curve}). The curve shows fairly similar completeness to that of the inner field in our empirical technique (\textit{black solid curve}) up to $m_J \sim 19$ mag and a slight deviation at larger magnitude. One of the possible contributing factors to this deviation is the saturation of bright stars. In this recovery test, we use the NACO long-exposure observation, and the detection of faint artificial sources is to some extent restricted by the bright outskirts of saturated stars in the field, leading to somewhat lower completeness. In contrast, our completeness measurement in the empirical technique enables us to use both short- and long-exposure observations together, and hence we do not have the saturation problem in the measurement.
Considering that we have restricted ourselves within the 50\% completeness limit of $m_J \sim 19.4$ mag throughout our analysis of the LF and IMF in this study, we conclude that our incompleteness correction is in agreement with what is expected from the standard recovery test of artificial sources.


\section{Color-magnitude and color-color diagrams}
\label{c-cmd}

\subsection{Distance}
\label{sss-distance}
In the recent optical analysis \citet{sun04} derived a distance of 6.9 $\pm$ 0.6 kpc by fitting isochrones to the observed cluster sequence including the pre-main sequence (PMS) population in the color-magnitude diagram (CMD). Based on NIR photometry \citet{sto04} derived the distance as well as the foreground extinction focusing on the PMS-MS transition region. They derived a distance modulus of $(m - M)_{0}$ = 13.9 mag ($d$ = 6.0 $\pm$ 0.3 kpc). Their photometric distance is quite similar to the kinematic distance of 6.1 $\pm$ 0.6 kpc reported by \citet{dep99} based on radio continuum surveys and is slightly smaller than another kinematical distance estimate of 7.7 $\pm$ 0.2 kpc by \citet{nue02}.

Applying our best estimate of the isochrones (see \S~\ref{sss-age}) and the interstellar extinction (\S~\ref{sss-extinction}), the distance modulus of 13.9 mag provides a reasonable fit with the observed CMD. We therefore adopt $(m - M)_{0} = 13.9$ mag with an error of $\pm0.3$ mag from a visual inspection throughout the following analysis, corresponding to $d \sim6.0 \pm 0.8$ kpc.

\subsection{Age and mass-luminosity relation}
\label{sss-age}
In this section we describe how we estimate the age of the cluster and how we derive the mass-luminosity ($\mathcal{M}$-$L$) relation used later for the conversion of the observed luminosity to stellar mass.
First, an upper limit of the cluster age can be deduced from the evolutionary status of the massive stars. The presence of WR stars and numerous OB-type stars in the starburst cluster constrains the age to be not more than several million years \citep[e.g.][]{mey03}. Based on \textit{Hubble Space Telescope (\textit{HST})} observations, the three brightest members designated A1, B, and C in HD 97950 have been classified to be WN6ha stars, leading to the cluster age of $\lesssim$ 2.5 Myr \citep{mof94, dri95, cro98}.
Based on the isochrone fitting of the  $(J_{S} - K_{S}, J_{S})$ CMD, \citet{sto04} suggest 1 Myr as a common age for the MS and PMS populations. Also, \citet{sun04} derive an age of 1 ($\pm 1$) Myr by fitting isochrones to the optical CMD from the \textit{HST} observations.

To perform the isochrone fit, we create the ($J - K_{S}, J$) CMD of 1158 $JHK_{S}$-detected stars in the central region ($r \leq 13''$) of the NACO data as shown in Figure~\ref{fig-cmd-iso}.
In order to fully cover the observed mass range down to low-mass PMS stars with masses close to the hydrogen-burning limit, we have used three stellar evolutionary models.
The MS population is covered by the Geneva stellar evolution models \citep[][hereafter LS01]{lej01}. For the high-mass PMS population we use the isochrones from \citet[][hereafter PS99]{pal99}. For the low-mass PMS population we use the isochrones from \citet[][hereafter B98]{bar98}. We have assumed a solar metallicity for all isochrones.

\subsubsection{Pre-Main Sequence stars}
First we compare the PMS population with the PS99 isochrones. We restrict the comparison to the 0.3, 0.5, 1, and 3 Myr isochrones.
Focusing on the fit in the PMS-MS transition region (approx. $0.7 < J - K_{S} < 1.4$ and $14 < J < 16$ mag), we can readily choose 0.5 and 1 Myr isochrones. However, choosing between the 0.5 and 1 Myr is not straightforward. Therefore, we argue that the best-fit age is between 0.5 and 1 Myr, and in the following we use a 0.7 Myr isochrone constructed from a linear interpolation of these two isochrones.

Recently \citet{sto04} suggested a possible presence of a sequence of equal-mass binaries, which stretches along the MS and PMS-MS transition regions, suggesting a single star formation event $\sim$ 1 Myr ago in NGC 3603.
We have found no obvious secondary sequence but rather a scattered distribution in the region.
The scatter could be due to a combination of the photometric uncertainty and an intrinsic age spread of the PMS stars of several hundred thousand years.
Alternatively, this could reflect a possible uncertainty in the birth line. The zero point in the PS99 PMS models is defined to be the birth line. However, the collapse and accretion processes in the protostar formation are not fully understood, causing a significant uncertainty in the birth line. Thus, it is difficult to discuss an age spread in a young stellar population whose age is $\lesssim$1 Myr, for which the age discrepancy between the available PMS evolutionary models becomes larger partly due to dependence of the initial conditions \citep[e.g.][]{tou99,bar02}.

Considering these uncertainties, we summarize that the age of the PMS population is around $0.5 - 1.0$ Myr and assume an average age of 0.7 Myr. For the low-mass PMS stars with $\leq1.2$ \msun, we apply the youngest 1 Myr isochrone from the B98 model. Among currently available PMS evolutionary models, the B98 model has been reported to have the best consistency with dynamical mass determination for stellar masses below 1.2 \msun\ \citep{hil04}.

\subsubsection{Main sequence stars}
We compare the MS population to the 1, 1.5, 2, 2.5, and 3 Myr LS01 isochrones with a particular focus on the locations of the three bright members 1 (A1), 2 (B), and 3 (C) in the CMD.
Figure~\ref{fig-cmd-WR} shows the CMD of the three stars together with the isochrones.
To determine the age of high-mass MS stars, we need to assume initial masses of these three stars.
They are classified as WN6h+abs stars by \citet{cro98} and the authors derived the current masses of about $60 - 100$ \msun. We therefore assume initial masses of the WN stars to be about $80 - 120$ \msun\ (this is deduced from Figure~5 in the reference). Although the 2 Myr isochrone cannot be clearly excluded, the best match for all three stars is the 2.5 Myr. We thus select the 2.5 Myr isochrone.

It is worth mentioning that potential binary/multiple systems in the three WN stars would be insignificant in our age estimate.
Based on \textit{HST} NICMOS data, \citet{mof04} suggested that A1 is possibly an eclipsing binary system with a $30 - 90$ \msun\ H-rich WR component (WN6ha) and a $25 - 50$ \msun\ O star companion. The source B also has been suggested to be a binary system \citep[e.g.][]{mof83}. Even though the source B is assumed to be a binary system, \citet{cro98} derived the contribution from the likely O3-5 V star companion of $\sim0.1$ mag, being a negligible contribution in our determination. 
One potential uncertainty is the still unclear nature of the three WN stars. For example, strong hydrogen absorption features have been detected in their spectra, leading to the hypothesis that they are more likely hydrogen-rich, extremely luminous MS stars that have not evolved but have mimicked WR features by generating strong, hot stellar winds \citep{dri95, sch99, mof02}. \citet{wal02} reported that they identified two O2-3.5 III(f$^{\ast}$) stars in the core. Thus, if those sources are indeed such very luminous MS stars, the cluster's age would be to some extent smaller than our estimate, allowing an interpretation of a single star-forming event for both the PMS and MS stellar populations.

If our age estimate holds, we would be confronted with a puzzling inconsistency or a real age spread among the $0.5 - 1$ Myr PMS population and the $2 - 2.5$ Myr MS population.
One possible explanation for this is that several massive stars first formed in the center of the natal molecular cloud, and subsequently the strong winds and radiations from those massive stars have triggered the second generation of low-mass stars, leading to the star formation over several million years.
Indeed earlier studies have been suggesting sequential and ongoing star formation from north to south throughout the giant HII region \citep{mel89,hof95,dep99,tap01,nue02,sun04}.
Sher 25 \citep{she65,mof83}, for example, has turned out to be an evolved post-red supergiant with an hourglass-shaped nebula, which is excited by the HD 97950 stars \citep{bra97}.
If these stars indeed physically belong to NGC 3603, there should be at least two distinct star formation episodes separated by $\sim10$ Myr. Note that whether or not Sher 25 is a cluster member is not yet clear \citep{cro06}.

In summary, since the age estimate of the MS stars is based merely on the isochrone fitting to the three WN stars and hence is still uncertain, we conclude 2.5 Myr as an upper limit of the age of the cluster population.


\begin{figure}
\epsscale{1.}
\plotone{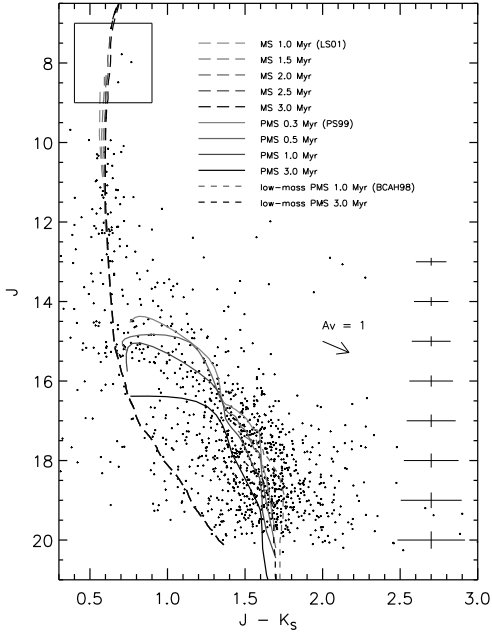}
\caption{The ($J - K_{S}, J$) CMD from 1158 $JHK_{S}$-detected sources in the NACO field ($r \leq 13''$). The long-dashed, solid, and dashed curves are isochrones from the LS01, PS99, and B98 models, respectively. The isochrones are plotted after corrections for a distance modulus of $(m - M)_{0} = 13.9$ mag and an average interstellar extinction of $A_{V} = 4.5$ mag. The arrow illustrates the reddening for $A_{V} = 1$ mag. The error bars on the right side give the photometric errors. The box indicates a zoom on the three brightest stars shown in Figure~\ref{fig-cmd-WR}.\label{fig-cmd-iso}}
\end{figure}


\begin{figure}
\epsscale{1.}
\plotone{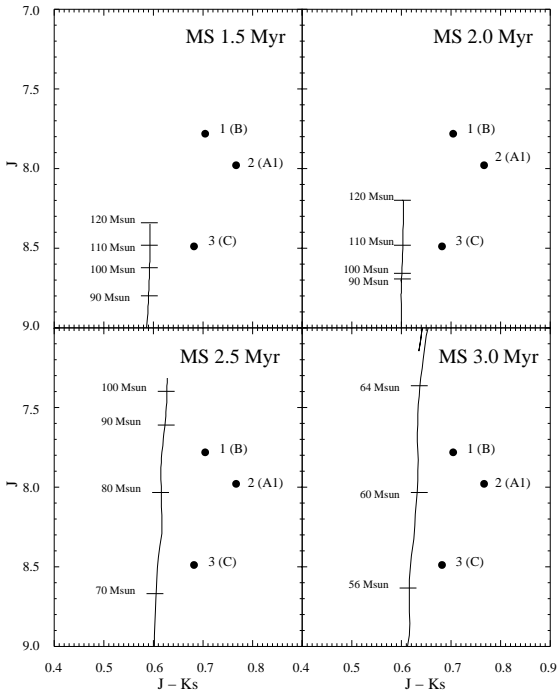}
\caption{CMD of the brightest stars. The panels show the 1.5, 2, 2.5, and 3 Myr LS01 isochrones together with the three brightest WR stars 1 (B), 2 (A1), and 3 (C). The designated source numbers correspond to the photometry list presented in the appendix. Stellar initial masses are indicated along the isochrones. \label{fig-cmd-WR}}
\end{figure}

\subsubsection{Mass-Luminosity relation}
\begin{figure}
\plotone{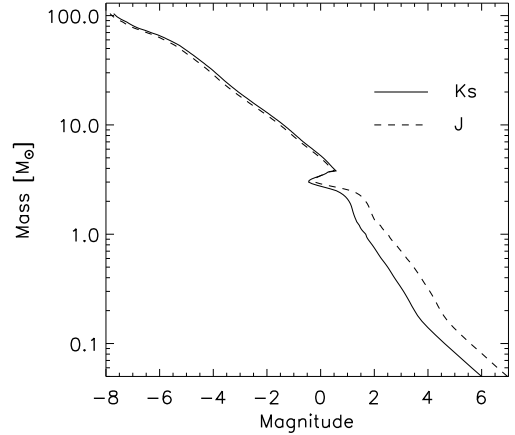}
\caption{Mass-luminosity conversion relation constructed from the best-fit isochrones. The single relation was created combining the 2.5 Myr LS01 isochrone for the MS population ($\gtrsim4$ \msun), the 0.7 Myr PS99 isochrone for the PMS population ($\sim4 - 1.2$ \msun), and the 1 Myr B98 isochrone for the PMS stars ($\lesssim1.2$ \msun). The $J$- and $K_S$-band plots are shown in the solid and dashed lines, respectively.\label{fig-l2m}}
\end{figure}

After the selections of the best-fit isochrones for three stellar mass ranges, we constructed a single $\mathcal{M}$-$L$ relation by connecting the three isochrones so as to convert the observed luminosities to stellar masses for the IMF determination. 
Figure~\ref{fig-l2m} shows the constructed single $\mathcal{M}$-$L$ relation in a stellar mass versus absolute magnitude plot.
In our $\mathcal{M}$-$L$ conversion we have used all $JHK_S$-band magnitudes defining the stellar mass of the closest point on the $\mathcal{M}$-$L$ relation, for a source, in the $JHK_S$ space. We adopted the distance modulus of $(m - M)_{0}$ = 13.9 mag and the uniform interstellar extinction of $A_V = 4.5$ mag.

It is worth mentioning that, although there are slight discrepancies between isochrones at the connecting points, these gaps are very small and thus are insignificant in our IMF determination.
At the connecting point of the PMS isochrone (0.7 Myr; PS99) and the low-mass PMS isochrone (1 Myr; B98) at $\sim1.2$ \msun, the gaps are ($\Delta{m_J}$, $\Delta{m_H}$, $\Delta{m_{Ks}}$) $\sim$ (0.19, 0.14, 0.14), fairly comparable to our photometric uncertainties at the corresponding magnitudes (e.g., error bars are shown in Figure~\ref{fig-cmd-iso}).
At the connecting point of the MS isochrone (2.5 Myr; LS01) and PMS isochrone at $\sim3.7$ \msun, the gaps are $\sim0.3$ mag in all bands but still much smaller than a magnitude range from a corresponding mass bin in our IMF. These gaps are thus insignificant in our IMF analysis.


\subsection{Interstellar extinction}
\label{sss-extinction}
The line-of-sight interstellar extinction towards NGC 3603 has been estimated in previous studies, and its average value is around $E(B-V) = 1.44$ mag with only little variation. \citet{sto04} derived $A_{V}$ = 4.0 and 4.5 mag for the MS and PMS populations, respectively. 
Performing an isochrone fitting in the $(J - K_{S}, J)$ CMD of the central cluster from our NACO field, we find that the isochrone with a shift of $A_{V}$ = 4.5 mag matches best the MS distribution for intermediate- to high-mass stars. The PMS population is also well fitted with this extinction. This can be seen in Figure~\ref{fig-cmd-iso} in which all isochrones are shifted by $A_{V}$ = 4.5 mag. We estimate an error of $\pm0.5$ mag.

\begin{figure}
\epsscale{1.}
\plotone{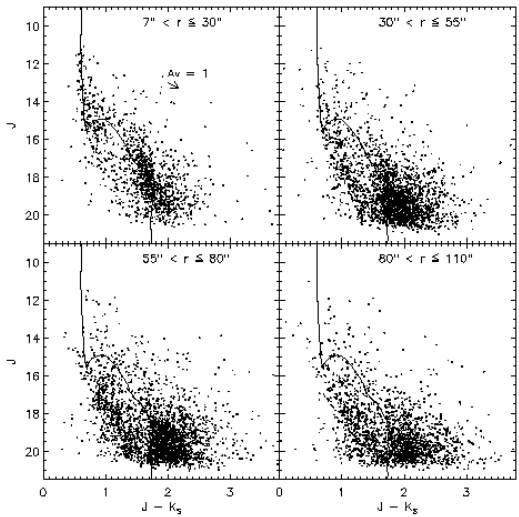}
\caption{Radial variations in the CMD from four radial regions with radii $7'' < r \leq 30''$, $30'' < r \leq 55''$, $55'' < r \leq 80''$, and $80'' < r \leq 110''$. The solid curve shows the combined isochrone.\label{fig-cmd-radial}}
\end{figure}

In order to explore a possible radial variation of the interstellar extinction, we also constructed CMDs for different radii from the wider ISAAC field (Figure~\ref{fig-cmd-radial}).
The stellar population in the innermost ISAAC annulus ($7'' < r \leq 30''$) reasonably resembles the central field from the NACO observations (Figure~\ref{fig-cmd-iso}), and the combined isochrone fits reasonably well the entire mass range from the bright MS stars to faint PMS stars.
Towards larger radii, however, we find an increasing redwards shift of the stellar distribution relative to the isochrone as well as an increasing scatter of the color distribution.
This trend is seen up to the $55'' < r \leq 80''$ annulus, and the distribution does not significantly change any more at $r > 80''$.
We find this trend in the $H - K_{S}$ versus $J - H$ color-color diagrams (CCDs) for the same four concentric annuli and conclude that the radial increase in extinction is $\Delta{A_{V}} \sim 2.0$ mag as the maximum at the outermost region.
Previous studies have also reported a radial trend of the interstellar extinction in NGC 3603 \citep{mel89,pan00}. \citet{sun04} derived $E(B-V) = 1.25$ mag at the core ($r \leq 0'.2$) and an increase of 1.8 mag or even higher at larger radii.
The radial trend can be explained by the strong stellar winds and radiation from the central high-mass stars which create a cavity in the interstellar matter, as has been suggested in earlier studies \citep{fro77,bal80,cla86}.


\subsection{Infrared excess emission and disks}
\label{sss-ccd-jhkl}
We construct a $K_{S} - L'$ versus $J - H$ CCD for the central region to examine the fraction of stars with infrared-excess emissions.
The $L$-band observation is a particularly effective tool to identify the excess emission from circumstellar disks \citep{lad92}.
The $L$-band observation has been used for tracing stars with circumstellar disks in star-forming regions such as the Taurus-Auriga molecular cloud \citep{ken95}, the Trapezium cluster \citep{lad00}, the Chamaeleon I dark cloud \citep{ken01}, 30 Dor \citep{mae05}, and NGC 3576 \citep{mae06}.

\begin{figure}
\epsscale{1.}
\plotone{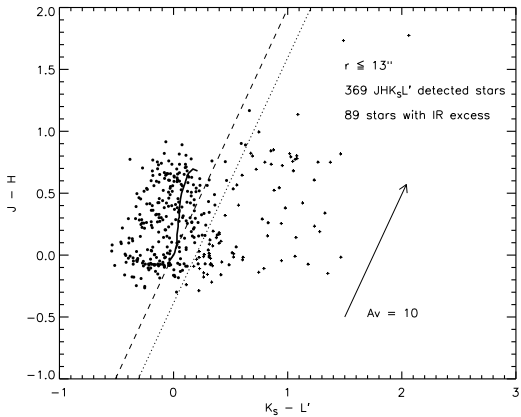}
\caption{The $K_{S} - L'$ vs. $J - H$ diagram from stars simultaneously detected in the $JHK_{S}L'$ bands in the cluster core $(r \leq 13'')$. The colors are dereddened using the extinction law by \citet{rie85}. The solid curve shows the empirical colors of MS stars from \citet{bes88} with spectral types from B8 to M0. The dashed line indicates the reddening vector passing through the B8 star. The dotted line indicates the selection criterion for the disk population, which is the reddening line of the B8 star shifted by the 1 $\sigma$ photometric error of $\Delta(K_{S} - L') = 0.20$ mag. Crosses are stars with disk emissions.\label{fig-ccd-naco-jhkl}}
\end{figure}

Figure~\ref{fig-ccd-naco-jhkl} shows the CCD for 369 stars detected simultaneously in the $JHK_{S}L'$ bands in the cluster center of $r \leq 13''$ after the rejection of field stars (see \S~\ref{c-field}). 
Because of our shallow $L'$-band detection, we set the $L'$-band limit of 15.2 mag below which the detection is reliable. Thus, this analysis is restricted to stars more massive than $\sim1.2$ \msun.
To classify stars with and without the $L'$-band excess emission, we define a threshold of the $K_{S} - L'$ color by adopting the reddening law and a typical error bar of $K_{S} - L' = 0.20$ mag, which is the 1 $\sigma$ scatter of $K_{S} - L'$ color for the bulk of the PMS population.

In total 89 out of 369 sources are classified as stars with disk emission, corresponding to a disk fraction of $\sim24$\%.
The resulting disk fraction, however, strongly depends on the adopted selection criterion. If we adopt a zero threshold instead, the disk fraction rises to $\sim38$\%. If we use a 2 $\sigma$ threshold, then the outcome is $\sim13$\%. Defining these values as a rough error bar, we conclude that the disk fraction for the given population and field is approximately $24 \pm 10$\%.

Parenthetically, we note that, since there are potential star-disk systems that are detectable only at longer wavelengths, e.g., at the $K_{S}L'$ bands or merely the $L'$ band, the use of stars detected in $JHK_{S}L'$ bands simultaneously could underestimate the actual disk fraction.
To roughly test this uncertainty, we compute a disk fraction using also stars that are detected only in $L'$ band simply assuming that such stars are star-disk systems.
Combining the $L'$-band-only detected stars and the $JHK_{S}L'$ stars with excess emissions increases the disk fraction by approximately 20\%.

In addition, we examine a mass dependency of the disk fraction.
The disk fraction in stellar clusters have shown some dependencies on the stellar mass or spectral types.
Based on N-body dynamical simulations and simulations of the mass loss from the star-disk encounters, \citet{pfa06} find that in the Trapezium cluster the circumstellar disks around high-mass stars dissipate much faster than those of intermediate-mass stars, largely due to gravitational interactions. This shorter disk dissipation timescale for higher-mass stars is generally consistent with most observations.
Based on the $JHKL$ observations of the Trapezium cluster, \citet{lad00} report that they find a considerably lower disk fraction for the high-mass O, B, and A stars than that for F-M type stars, and this could be due to either a lower probability for disk formation or more rapid disk dispersal times.
Recently, based on NIR and mid-IR data taken by the \textit{Spitzer} Space Telescope, \citet{her07} study the 3 Myr old $\sigma$ Orionis cluster, and derive a disk fraction of $\sim30$\% for T Tauri stars with $<1$ \msun\ and a much lower disk fraction for higher mass stars (e.g., $\sim 15 \pm 7$\% for Herbig AeBe stars with $\gtrsim2$ \msun), implying a more rapid disk evolution for stars with $>1$ \msun\ in the cluster.

Considering the low-mass limit of $\sim1.2$ \msun\ imposed by the $L'$-band detection limit, we here compute the disk fraction for high-mass MS stars and intermediate-mass PMS stars so as to test the mass dependency of disk fraction in this cluster. Setting the MS turnon mass of $\sim4$ \msun\ as the threshold, we derived the disk fraction of $\sim22$\% for stars with $\gtrsim 4$ \msun and $\sim27$\% for stars with $\sim1.2 - 4$ \msun. As is the case of the disk fraction of whole mass range, we derived an error bar of about $\pm10$\% for the both fractions.

Our result is consistent with earlier studies of NGC 3603, as well as with the general consensus on the circumstellar disk frequencies in intense star-forming regions. Based on the $L'$-band luminosity and the H$\alpha$ emission, \citet{sto04} derived a disk fraction of $20 \pm 3$\% for all stars in the inner region ($r < 20''$) and that of $12$\% for MS stars (O-A stars) only. They also reported a radial increase of disk fraction to $\sim40$\% in the outer regions ($r < 33''$).
As our data is restricted within the innermost $r \leq 13''$ region, our result is still compatible with the hypothesis that the disk fraction increases to some extent towards outer regions in the cluster.

Based on a study of evolutionary models for protoplanetary disks, \citet{arm03} reported that in young clusters up to 30\% of stars lose their disks within 1 Myr, the remainder, however, have disk lifetimes of typically in the range of $1 - 10$ Myr.
In their study of $\sigma$ Ori, \citet[][Fig.\~14]{her07} compile the disk fractions of T Tauri stars, as a function of age, for different stellar populations, showing a clear correlation between the disk fraction and the mean age.
Considering the age of NGC 3603 of $\lesssim 2.5$ Myr, our disk fraction of approx. 30\% for intermediate-mass PMS stars ($1.2 - 4$ \msun) appears be somewhat smaller than those of other clusters, $40 - 60$\% based on the correlation.
One reason could be that our analysis is limited for PMS stars with masses down to $\sim1.2$ \msun, and the disk fraction is expected to be much higher for lower mass T Tauri stars. Another reason could be that our analysis is restricted to the central $r \leq 13''$.
As mentioned above, \citet{sto04} report a radial increase of the disk fraction up to $\sim40$\%.


\section{Luminosity functions of NGC 3603}
\label{c-lf}
The LF described as $\psi(M) = dN/dM$ is one of the most important diagnostic tools in the study of star clusters, since it is not influenced by any theory-related uncertainties, but is purely an observational quantity. In contrast, the IMF relies on the transformation of the observed LF using a $\mathcal{M}$-$L$ relation and in that sense imposes additional assumptions.
The relation between the IMF $\xi (\mathcal{M})$ and the LF of a stellar population with a given age is 

\begin{equation}
\psi(M) \propto \xi (\log \mathcal{M}) \frac{1}{\mathcal{M}} \frac{d\mathcal{M}}{dM}.
\label{e-imf-lf}
\end{equation}

Since the stellar population in NGC 3603 is young enough not to be strongly affected by stellar evolution, it provides an empirical basis for the derivation of the \textit{initial} mass distribution.
Here we note that, since we construct the LFs of NGC 3603 by combining the NACO and ISAAC observations, we verify that this combination is feasible by comparing two LFs for an overlap region at $r = 10'' - 13''$ in both the NACO and ISAAC fields. We also checked that the two data sets are compatible for the following IMF construction.

\subsection{Cluster membership}
\label{c-field}
To derive the LF of the cluster stars, we need to correct the potential contribution from field stars.
For that we examine two different technical approaches.
The first approach is to directly measure the density and LF of the field stars. This is a commonly applied method in the analysis of stellar clusters and requires the observation of a so-called control field. Since we do not have a control field in our observations, we derive an upper limit for the field star density from the outermost region ($80'' < r \leq 120''$) of the ISAAC field. Another estimate is available from observations of NGC 3603 presented in \citet[][hereafter NPG02]{npg02}. Their field \textit{K}LF is derived from a field out of the cluster and is in agreement with estimates based on Galactic disk models.
The second approach is to directly differentiate the likely non-cluster members from the cluster members on the basis of their locations in the CMD. We hence call this method \textit{color cut} hereafter.

\begin{figure}
\epsscale{1.}
\plotone{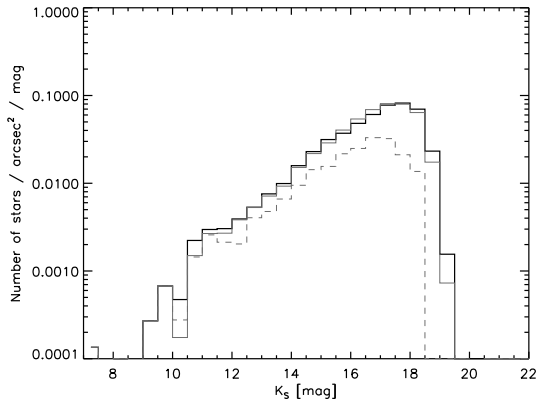}
\caption{Comparison of $K_{S}$-band LFs based on three different estimates for the field star subtraction. The observed $K_{S}$LF for 9158 stars detected simultaneously in the $JHK_{S}$ bands within $r \leq 110''$ is subtracted by the field LFs derived from the color cut (\textit{black line}), the outermost region ($80'' < r \leq 120''$; \textit{dashed line}), and the NPG02 \textit{K}LF (\textit{gray solid line}).\label{fig-lf-field}}
\end{figure}

Figure~\ref{fig-lf-field} compares the $K_{S}$-band LF after the correction based on the three estimates -- from the outermost region, the NPG02 field \textit{K}LF, and from the color cut.
We first construct an observed \textit{K}LF for stars simultaneously detected in the $JHK_{S}$ bands in the whole field and then subtract the \textit{K}LF by the three field \textit{K}LFs. We first find that the outermost field \textit{K}LF overestimates the field star density. This is most likely due to the fact that the outermost region in our ISAAC field is not far enough and is still within the cluster region.
While the \textit{K}LFs subtracted by the color cut \textit{K}LF and the NPG02 field \textit{K}LF show fairly similar distributions within the reliable magnitude range of $m_{K_S} \sim 12 - 18$ mag.
This verifies that, despite of the lack of control fields, the color cut method is a reasonable treatment for the field star correction. We thus use the color cut in the following LF and IMF analysis.


\begin{figure}
\epsscale{1.}
\plotone{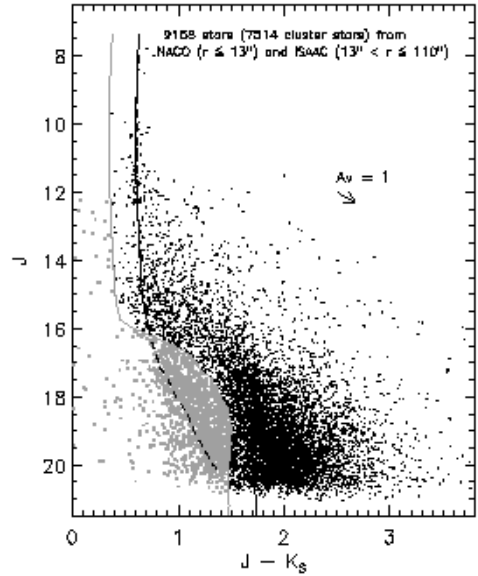}
\caption{The $(J - K_{S}, J)$ CMD for 9158 stars in the whole field $r \leq 110''$. The black solid curve is the combined isochrone. The gray curve illustrates the color cut created by smoothing out the turnover feature in the PMS-MS transition of the combined isochrone. The dashed curves represent the extended MS isochrones. The non-cluster members based on the color cut are shown in gray squares. \label{fig-cmd-cc}}
\end{figure}

The selection of the cluster stars by the color cut is illustrated in the CMD in Figure~\ref{fig-cmd-cc}.
The color cut (\textit{gray line}) is created by smoothing out the turnover feature in the PMS-MS transition of the combined isochrone and shifting this curve blueward by $\Delta(J - K_S) \sim 0.26$ mag. Note that the shift is applied only in the $J - K_S$ direction, i.e. no shift in the vertical $J$ direction.
In total 7514 out of 9158 stars are found to be cluster members.
Likely noncluster stars that are located around the MS (\textit{dashed curve}) at $m_{J} \geq 16$ mag and bluer than the color cut could be part of the following groups: (1) background MS stars with $A_{V}$ $\simeq$ 4.5 mag, (2) foreground low-mass MS stars, and (3) cluster stars formed in earlier generations. Thus, they are excluded from the LF and IMF derivations.
Sources in the first group would be MS stars with an interstellar extinction similar to those of the cluster members, but with a smaller brightness due to their larger distance. In this case the interstellar material such as gas and dust are assumed to exist locally along the line of sight, and hence the interstellar extinction behaves as a step function rather than a linear increase with distance.
Some supposedly noncluster stars could indeed be cluster members that have formed prior to the starburst $\lesssim 2.5$ Myr ago when the majority of the cluster members were born. Since the IMF study is basically the subject of a single star formation, such sources are correctly rejected from our analysis.
In fact, the number of stars from earlier generations is thought to be very small considering the lack of sources in the low-mass MS regime in optical observations. \citet{sun04} found no MS population for $M$ $\lesssim 4$ \msun\ in their optical observation. The absence of low-mass MS stars is also reported in \citet{gre04}.

One potential limitation of this color cut for the field star estimate is that the color cut curve is placed based on an eye inspection; therefore, the correction essentially relies on this judgment. However, as shown in Figure~\ref{fig-cmd-cc} (and also Figure~\ref{fig-cmd-radial}), there is a somewhat visible gap between the low-mass PMS population and the faint MS population, and the color cut curve is satisfactory in separating them.
Another limitation is the fact that the region redward of the color cut could still contain some intrinsically very red non-cluster objects such as the line-of-sight red giants.
However, the above analysis of the NPG02 field \textit{K}LF and our \textit{K}LF based on the color cut suggests that the contamination of red giants is statistically insignificant in our study. 
Nevertheless, we study the potential effect on the IMF applying a ``red cut'' in addition to our color cut in \S~\ref{s-robust-membership}.


\subsection{The cluster star LF}
\label{sss-cluster-lf}
\begin{figure}
\epsscale{1.}
\plotone{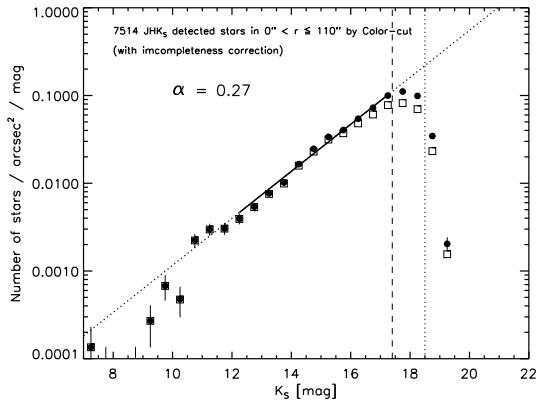}
\caption{The $K_{S}$-band LF of NGC 3603 derived from 7514 $JHK_{S}$-detected stars in the entire field ($r \leq 110''$) after the color cut. The circles and squares show the distribution with and without applying the incompleteness correction, respectively. The vertical dashed line represents the 50\% completeness limit of $m_{K} \sim 17.4$ mag (derived from $m_{J} \sim 19.4$ mag and typically $J - K_{S} = 2$ mag for low-mass PMS stars) within $r \lesssim 30''$. The vertical dotted line indicates the detection limit in the outer ISAAC field derived from the artificial source detection test. The best-fit power law slope (\textit{solid line}) is measured in a magnitude range from 12 mag up to the 50\% completeness limit.\label{fig-lf-best}}
\end{figure}

Figure~\ref{fig-lf-best} shows the \textit{K}LF of NGC 3603 applying the color cut and the incompleteness correction. The resulting LF shows a monotonic increase towards faint magnitude with a power law index of $\alpha = 0.27 \pm 0.01$. Because of the strong variation of the detection limit across the field, we cannot conclude if the observed sign of flattening of the LF at $m_{K_{S}} \sim 17.5 - 18$ mag is a real feature or a result of the observational limitation.
We find that the slope of our LF is consistent with those of the previous studies presented in NPG02 and \citet{bra99}, but our high angular resolution NACO data reveals $\sim50$\% more stars in the central region of $r \leq 33''$.


\section{Initial Mass Functions of NGC 3603}
\label{c-imf}

\subsection{Initial Mass Function determination}
\label{ss-imf-result}
\begin{figure}
\epsscale{1.}
\plotone{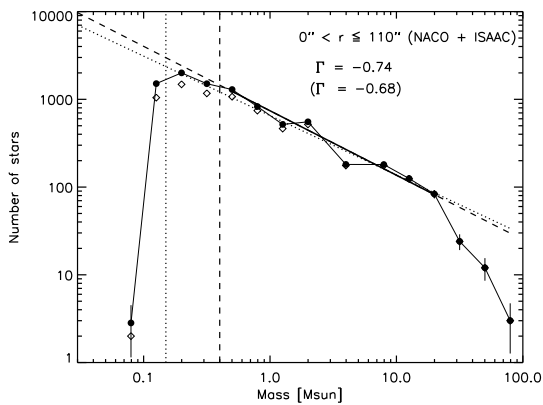}
\caption{IMF of NGC 3603 derived from 7514 cluster stars in the whole field extending out to $r \sim 110''$. A double size bin is used at around 4 \msun\ to smooth out a discontinuity from connecting the MS and PMS isochrones.
Diamonds and circles indicate the raw and the incompleteness corrected mass distributions, respectively.
The best-fit power law slopes, derived within $0.4 - 20$ \msun, are shown as a dotted and a dashed line. The Poisson counting error is used in a weighted linear least-square fit. The vertical dashed and dotted lines indicate the 50\% completeness limit within $r \sim 30''$ and the detection limit in the ISAAC outer fields, respectively.\label{fig-imf-best}}
\end{figure}

Applying the color cut and the incompleteness correction, we derive the IMF for the 7514 $JHK_{S}$-detected stars in the whole field. The resulting IMF is shown in Figure~\ref{fig-imf-best}.
To compute the stellar masses, we used the combined $\mathcal{M}$-$L$ relation created from the three best-fit isochrones.
As mentioned in \S~1, the IMF is generally known to follow a power law $\xi(\log \mathcal{M}) \propto \mathcal{M}^{\Gamma}$,
or alternatively expressed in the number of stars per unit mass interval,
\begin{equation}
\xi(\mathcal{M}) = \frac{dN}{d\mathcal{M}} = \frac{1}{\mathcal{M}(\ln 10)} \xi(\log \mathcal{M});
\label{e-imf2}
\end{equation}
thus, $dN/d\mathcal{M} \propto \mathcal{M}^{\gamma}$, with $\gamma = \Gamma - 1$.
The slope is derived for the mass range of $0.4 - 20$ \msun. This mass range is not affected by the saturation in the ISAAC data, and it guarantees a completeness of at least 50\% even in the most crowded central region.
The best-fit power law index is $\Gamma = -0.74$ ($\pm 0.02$). The error is merely a formal fit error. The real errors of the index involves many other systematic uncertainties, and we discuss them in detail in \S~\ref{c-systematic}.
Although we find a turnover at around 0.2 \msun, we can not answer if the turnover is intrinsic or it is simply caused by the detection limit.

In summary, our resulting IMF of NGC 3603 follows a power law with $\Gamma \sim -0.74$ within $0.4 - 20$ \msun. The detailed discussion in \S~\ref{c-systematic} will show if this finding is significant within the systematic uncertainties. In particular we will test if NGC 3603 as a whole or merely the observed region within $r \leq 110''$ has a shallow IMF.


\subsection{Radial variation of the IMF}
\label{ss-imf-radial}
\begin{figure}
\epsscale{1.}
\plotone{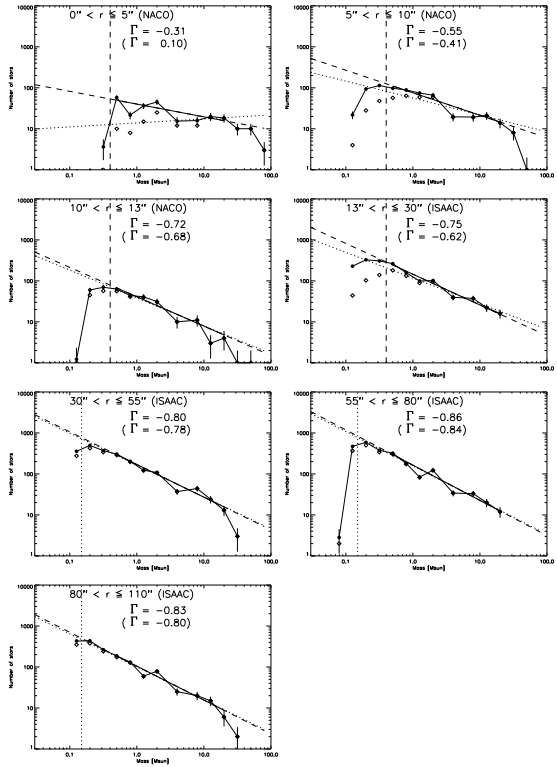}
\caption{IMFs for seven concentric annuli with increasing radius. The symbols and lines are identical to the global IMF in Figure~\ref{fig-imf-best}.\label{fig-imf-radial}}
\end{figure}

One of the characteristics of NGC 3603 that makes it a particularly interesting object is the high concentration of massive stars in the central starburst cluster. This brings up the question of a possibly mass-segregated stellar distribution in the cluster.
Utilizing the richness of our data, to answer the question, we have built IMFs of seven concentric annuli with radii of $5'',10''$, and $13''$ (from NACO) and $30'', 55'', 80''$, and $110''$ (from ISAAC) as shown in Figure~\ref{fig-imf-radial}.

Here we note that there is some indication of a broken power law at high masses for the inner three regions from the NACO data in which there is no saturation problem. We, however, adopt the single power law with the fixed mass range of $4 - 20$ \msun\ to keep consistency and to trace the radial variation.
The IMF power law indices are summarized in Tab.~\ref{tbl-radial-IMF}.
There is an obvious steepening of IMF with increasing radius up to $r \sim 30''$. For larger radii the slope stays more or less constant around $\Gamma \sim -0.8$. This is a clear indication of mass segregation in the cluster. 
The main characteristics of this mass segregation are as follows: (1) a strong concentration of high-mass stars in the very center at $r \lesssim 13''$, (2) a shallow IMF ($\Gamma \sim -0.8$) at $r \gtrsim 30''$, and (3) no evidence for further steepening at larger radii.
The third point allows us to conclude that the IMF of the whole cluster (including the regions not covered by our observations) cannot be steeper than an IMF with $\Gamma \sim -0.9$.


\begin{table}
\begin{center}
\caption{Radial variation of the IMF power law index\label{tbl-radial-IMF}}
\begin{tabular}{lc}
\tableline
\tableline
Regions & $\Gamma$ \\
\tableline
\multicolumn{2}{c}{NACO data} \\
\tableline
$r \leq 5''$         & -0.31 \\
$5'' < r \leq 10''$  & -0.55       \\
$10'' < r \leq 13''$ & -0.72       \\
\tableline
\multicolumn{2}{c}{ISAAC data} \\
\tableline
$13'' < r \leq 30''$ & -0.75 \\
$30'' < r \leq 55''$ & -0.80       \\
$55'' < r \leq 80''$ & -0.86       \\
$80'' < r \leq 110''$ & -0.83      \\
\tableline
\end{tabular}
\end{center}
\end{table}

\subsection{Comparison with previous studies}
\label{ss-imf-comp-ngc3603}
The IMF of NGC 3603 has been studied in earlier works based on both ground- and space-based observations.
Based on \textit{HST}/Planetary Camera 1 (PC1) observations \citet{mof94} derived the IMF of high-mass stars with a power law index of $\Gamma = -1.4 \pm 0.6$ within a mass range of $30 - 60$ \msun. \citet{hof95} derived the IMF with $\Gamma \sim -1.59$ for $15 - 50$ \msun\ based on their diffraction-limited speckle observations. These results are consistent with the standard Salpeter IMF; however, only the high-mass population is covered in these studies.

Here we compare our results with several recent works.
Based on high-resolution NIR observations using the AO system ADONIS/SHARPII+ at the ESO 3.6 m telescope, \citet{eis98} were the first to reveal the PMS population with masses as low as $\sim1$ \msun\ in the central starburst cluster. They derive a shallow IMF with $\Gamma \sim -0.73$ within $1 - 30$ \msun and find no obvious turnover in the IMF. Our resulting IMF with $\Gamma = -0.74$ over $0.4 - 20$ \msun\ is fairly similar to their result. Even though we can probe much fainter stars, we still see no sign of turnover.

Based on the seeing-limited \textit{UBVRI} CCD photometry, \citet{sag01} derived the MF for seven distant open star clusters, including NGC 3603. The slope of the MF is $\Gamma \sim -0.85$ for $7 - 75$ \msun. Although their IMF is limited for high-mass population, our result is in agreement with their result. The authors, however, regard it as being consistent with the Salpeter IMF within errors. They mentioned that, above 1 \msun, MF slopes of star clusters younger than 500 Myr (equivalent to the IMF) in the solar neighborhood had no dependence on Galactic longitude, Galactocentric distance, and cluster age, and they were in agreement with the Salpeter IMF within a 1 $\sigma$ error ($\sigma_{\Gamma} \sim 0.3$). As mentioned in \citet{sca05}, this implies that their actual uncertainties are very large. In contrast, our interpretation of the resulting slope is a case of a flat IMF.

Recently, combining multi-wavelength ground-based, \textit{HST}, and \textit{Chandra} X-ray observations, \citet{sun04} derive a moderately flat IMF with $\Gamma = -0.9$ for $2.5 - 100$ \msun. The IMF shows a gradual steepening towards the outer regions ($\Gamma = -0.5$ at $r \leq 0'.1$, $-0.8$ at $r = 0'.1 \sim 0'.2$, $-1.2$ at $r \geq 0'.2$).
Like us they find a radial variation in the IMF. Our IMFs, however, are somewhat flatter than their IMFs.
Their IMF of the outermost region ($r > 12''$) shows a slope of $\Gamma = -1.2$, which is $\sim0.3$ steeper than that of the whole field ($\Gamma = -0.9$), likely due to the mass segregation. In contrast, our IMF of the outer region ($\Gamma \sim -0.85$) is not substantially steeper than the IMF of the whole cluster ($\Gamma = -0.74$).
As for the innermost region,  our IMF with $\Gamma \sim -0.3$ for $r \leq 5''$ is slightly flatter than their IMF with $\Gamma \sim -0.5$ for $r \leq 6''$. This might be in part due to the fact that \citet{sun04} includes stars with masses up to 100 \msun, while our analysis is restricted to masses below 20 \msun. Moreover, our mass range covers down to 0.4 \msun, which is slightly lower than their low-mass coverage.

Most recently, based on the $JHK_{S}L'$ ISAAC data and H$_{\alpha}$ imaging from the \textit{HST}/Wide Field Planetary Camera 2 (WFPC2), \citet{sto06} derived the MF taking into account the field star population, individual reddening, and the potential binary contribution in the central region ($7''< r < 65''$). The derived IMF has a power law slope of $\Gamma \sim -0.91 \pm 0.15$ for $0.4 - 20$ \msun.
Compared to, for example, their result of $\Gamma = -0.87$ for $7'' < r \leq 33''$, our IMF shows a slightly shallower power law $\Gamma = -0.75$ for $13'' < r \leq 30''$. Since the same data set is used, this difference could arise from technical differences such as the field star subtraction and the incompleteness correction. Indeed, it appears that we applied a slightly smaller incompleteness correction than theirs. Although we present it in detail below (\S~\ref{s-robust-incomp}), we note that an application of a stronger incompleteness correction is found to steepen the IMF with $\Delta\Gamma \sim -0.13$. Thus, some part of the slight difference between both studies would be explained by the difference of the correction rates.
While \citet{sto06} claim no significant variation in power law index of the IMF but talk of a depletion of the high-mass tail of the stellar mass distribution with increasing radial distance, we see a strong variation with radial distance from $\Gamma \sim -0.3$ to $\sim-0.8$ over the mass range of $0.4 - 20$ \msun. A part of the reason for this discrepancy is that \citet{sto06} could not probe the central $r \leq 7''$ because of crowding, while our high resolution NACO observations resolve the central cluster even at $r \leq 5''$.

In summary, we find that our results are partly in agreement with previous studies, but the IMF shows a fairly shallower slope with $\Gamma = -0.74$ when compared to the Salpeter-like IMF.
However, to prove conclusively the shallowness of the IMF, we have to first analyze the observed mass segregation.

\section{Mass segregation in NGC 3603}
\label{c-massseg}
The mass segregation in a stellar cluster can potentially lead to a systematic error in the IMF determination \textit{if the observation does not cover the entire cluster}. This is because higher-mass stars are more concentrated towards the cluster center than lower mass stars.
Since we find the mass segregation in NGC 3603 based on the radial steepening of the IMF, we need to investigate whether the mass segregation is the cause of the shallow IMF. 
Therefore, in this section, we study the characteristics of the observed mass segregation in NGC 3603.
For that we investigate the dynamical evolutionary state of the cluster by estimating the relaxation time of the cluster (eq.~[\ref{e-relax}] in \S~\ref{s-evolutionary-status}).
We first derive the global properties of the cluster such as its size, total mass, core radius, and the half-mass radius. These quantities allow us to estimate the relaxation time and then to assess the potential impact of the mass segregation on the IMF determination.

\subsection{Radial mass density profile}
\label{s-radial-pro}
To determine the global parameters, we measure the projected radial mass density distribution and fit analytical models to the profile.
First we fit the profile using the standard empirical formula by \citet{kin62}.
This King model is commonly used to describe the radial stellar density and brightness density profiles of globular clusters and old Galactic clusters. It is described as
\begin{equation}
f(r) = k\left[ \frac{1}{\sqrt{1+(r/r_{c})^2}} - \frac{1}{\sqrt{1+(r_{t}/r_{c})^2}}\right]^2,
\label{e-king}
\end{equation}
where $k$ is a normalization factor, which is approximately the central mass density; $r_{c}$ is the core radius, and $r_{t}$ is the tidal radius.
Because our color cut may not exclude all field stars, we modify the King model by adding a constant term.

For comparison, we also fit a modified power law used by Elson, Fall, \& Freeman (1987; hereafter EFF87) in their study of the radial surface brightness profiles of young clusters in the LMC. They find that the young LMC clusters do not appear to be tidally truncated even at radii of several hundred arcsec.
The EFF87 model is described as
\begin{equation}
f(r) = f_{0}\left(1 + \frac{r^2}{a^2}\right)^{-\gamma/2},
\label{e-eff87}
\end{equation}
where $f_{0}$ is the central surface mass density, $a$ is a measure of the core radius, and $\gamma$ is the power law index at large radii. The parameter $a$ is related to the core radius $r_{c}$ of the equivalent King model by $r_{c} = a(2^{2/\gamma} - 1)^{1/2}$.

\begin{figure}
\epsscale{1.}
\plotone{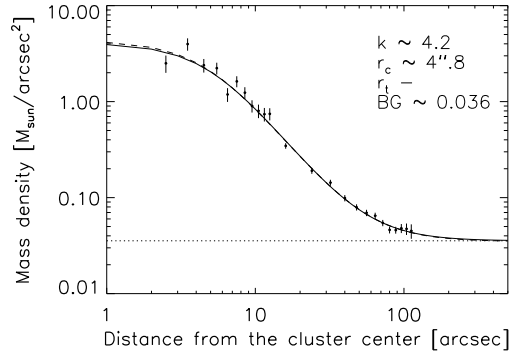}
\caption{Projected radial mass density profile of NGC 3603 for intermediate- and low-mass stars within the mass range of $0.5 - 2.5$ \msun. The mass density of the innermost ($r \leq 13''$) and outer field ($13'' < r \leq 110''$) is measured using $1''$ and $8''$ steps, respectively. The color cut and the incompleteness correction are applied. The solid and dashed curves show the best-fit King and EFF models, respectively. The horizontal dotted line shows the background level from the King model.\label{fig-rpro}}
\end{figure}

Figure~\ref{fig-rpro} shows the radial mass density profile of NGC 3603 for stars within the mass range of $0.5 - 2.5$ \msun, which corresponds to about $m_{K_S} = 15 - 17$ mag. Note that we use the barycenter, which we derive from the stellar density distribution, as the center of the cluster.
In the fit we omitted the innermost $r \lesssim 2''$ where the crowding effect cannot be corrected any more.
We select this mass range considering the following aspects. In a mass-segregated stellar cluster, the shape of the mass density profile and consequently its characteristic radii depend on the considered stellar mass range.
Since our eventual goal in this analysis is to determine how many intermediate- and low-mass stars potentially reside outside the observed field of view, we chose the upper mass limit of 2.5 \msun. The lower mass limit of 0.5 \msun\ is set to keep $>50$\% completeness.
From the fit of the King model we derive a core radius of $r_c \sim 4''.8$ ($\sim0.14$ pc at $d \sim 6$ kpc). Since the power law index in the best-fit EFF model is $\gamma = 1.97$, it is essentially identical to the King model with $r_{t} \to \infty$.
For comparison, \citet{gre04} report a core radius of $r_c \sim 0.25$ pc based on \textit{HST} observations, and NPG02 report $r_c \sim 23''$ as an upper limit. \citet{sun04} derive $r_c \sim 3''$ from their optical data.

We cannot directly deduce the tidal radius from our measurement.
As the stellar density is still decreasing at the edge of the observed field ($r \sim 110''$), the King model can satisfactory fit for any tidal radius of $r > 110''$. This difficulty is also expected from the fact that the EFF model, which does not take into account a tidal radius, fits the density profile equally well as the King model. We discuss other constrains for the tidal radius and the cluster size in the following section.


\subsection{Total mass of NGC 3603}
\label{s-total-mass}
\subsubsection{Size of NGC 3603}
In order to derive the total mass of the cluster by integrating the radial mass density profile, the size of the cluster has to be known.
Since fitting the King model to the density profile cannot realistically constrain the cluster's tidal radius, we adopt a cluster size from literature.
NPG02 give $r = 150'' \pm 15''$ as the radius where the stellar density falls below 3 $\sigma$ of the background variation. They also give a tidal radius $r_t \sim 1300''$ from the best-fit King model, but as in our study, this radius is outside their observed field, and accordingly very uncertain. Similarly, \citet{sun04} report a cluster radius of $2'$ based on the density profile of the bright stars identified in their optical and X-ray observations. They derive a tidal radius of $r_t \sim 900''$.
Following these two studies, we adopt a cluster size of $r \sim 150''$.

In addition, we give upper and lower limits of the cluster size to properly estimate the total mass and half-mass radius of the cluster with the uncertainties.
First, given the steady decrease of the density profile within our field, we can simply give a conservative lower limit of $r = 110''$.
The upper limit is derived from the cluster's tidal radius using the Galactic rotation curve and the measured density profile in a self-consistent way.
In the study of the Pleiades cluster \citet{pin98} give an approximate expression for the tidal radius of the cluster,
\begin{equation}
r_{t} = \left[\frac{G\mathcal{M}_{c}}{2(A-B)^2} \right]^{1/3} = 1.46\mathcal{M}_{c}^{1/3},
\label{e-rt2}
\end{equation}
where $\mathcal{M}_{c}$ is the mass of the cluster, and $(A - B)$ is the difference of the Oort constants, which describes the differential rotation of the Galaxy in the solar neighborhood, from \citet{ker86}.
Since the Galactocentric distance of NGC 3603 is similar to that of the Sun, we adopt this relation for our estimate.
Using equation~(\ref{e-rt2}) and the total mass calculated from the mass density profile (which will be explained below), we can iteratively determine a self-consistent tidal radius.
As a result we derive $r_{t} \sim 1260''$ for the upper limit of the cluster size.

\subsubsection{Deprojection of the surface mass density distribution}
Integrating the best-fit King model, in which our estimate of the tidal radius $r_t = 1260''$ is set, up to the three radii $r = 110'', 150''$, and $1260''$, we derive the best estimate of the total mass and half-mass radius with error bars.
Here we use a volume density distribution derived from the deprojection of the best-fit King model.
The calculation is done in the following way: (1) deprojection of the King model into a volume density, (2) integration of the volume mass density up to the cluster radius, and (3) scaling the stellar mass from the mass range used for the mass density ($0.5 - 2.5$ \msun) to the full stellar mass range (assuming $0.1 - 100$ \msun) using the derived IMF.
In a spherical system, the surface mass density $f(R)$ is related to the volume mass density $m(r)$ by the Abel integral
\begin{equation}
f(R) = 2 \int^{\infty}_{R}\frac{rm(r)dr}{\sqrt{r^2-R^2}}.
\label{e-abel}
\end{equation}
Inverting this equation gives the volume mass density
\begin{equation}
m(r) = -\frac{1}{\pi} \int^{\infty}_{r}\frac{df(R)}{dR}\frac{dR}{\sqrt{R^2-r^2}},
\label{e-abel-inv}
\end{equation}
where $r$ and $R$ are the spatial and the projected radii, respectively. The enclosed mass in a sphere of radius $r$ is then obtained by integrating equation~(\ref{e-abel-inv}) as $\mathcal{M}(r) = 4 \pi \int^{r}_{0} m(x) x^2 dx$.

First we derive the upper limits.
Integration of the de-projected King model up to the tidal radius of $1260''$ gives a total mass of $\sim2520$ \msun\ for the stars within the mass range of $0.5 - 2.5$ \msun. Assuming a power law IMF with $\Gamma = -0.74$ over the mass range $0.1 - 100$ \msun, we then obtain a total stellar mass of $\mathcal{M}_{tot} \sim 16,000$ \msun. The half-mass radius of NGC 3603 is $r_{hm} \sim 52''$ ($\sim0.7$ pc at 6 kpc). As we have measured the radial mass density profile for stars within $0.5 - 2.5$ \msun, this half-mass radius applies only to stars in this mass range and does not take into account any mass segregation.
For the best-guess radius of $r = 150''$, we derive $\mathcal{M}_{tot} \sim 11800$ \msun\ and $r_{hm} \sim 30''$. For the lower limit of $r = 110''$, we derived $\mathcal{M}_{tot} \sim 10700$ \msun\ and $r_{hm} \sim 25''$.

To get a handle on the uncertainty of the derived total mass, we analyze the potential effect due to systematic errors in the assumed IMF power law index and the stellar mass range.
As our goal is to estimate the total stellar mass outside the observed field of view $r \geq 110''$, it may be more suitable to use the IMF of the outermost ISAAC fields ($\Gamma = -0.85$) instead of the average IMF.
In this case we derive $\mathcal{M}_{tot} \sim 9700$ \msun, i.e., an $\sim18$\% decrease.
Next, if we assume a slightly wider mass range of $0.01 - 120$ \msun\ instead of $0.1 - 100$ \msun, we derive $\mathcal{M}_{tot} \sim 13500$ \msun, an $\sim15$\% increase.
Taking into account the lower and upper limits for the cluster radius and the errors from the IMF slope and the stellar mass range, we estimate the total mass of NGC 3603 to be $1.0 - 1.6 \times 10^4$ \msun, and the half-mass radius to be about $25'' - 50''$ ($0.7 - 1.5$ pc at $d = 6$ kpc).

Using the core radius of $\sim5''$ we also derive the central mass density of $\sim6\times10^4$ \msun\ pc$^{-3}$. It is worth mentioning that the core density measured directly from the surface density profile (without deprojection) would give a factor of $\sim2$ larger value.

\subsection{Dynamical evolutionary status}
\label{s-evolutionary-status}
Here we estimate the relaxation time of the cluster.
As the relaxation time depends on the stellar density, it is a function of radial distance from the cluster center, and it varies by several orders of magnitude across the cluster. While the inner parts relax most quickly, the outer parts take much longer to relax. The relaxation time also depends on the stellar mass. Higher-mass stars relax more quickly than lower-mass stars.
In our study we restrict ourselves to the \textit{average} relaxation time of the cluster.
Following \citet{bin87}, the median relaxation time $t_{rel}$ of a stellar population is given by 
\begin{equation}
t_{rel} = \frac{6.5 \times 10^8}{\ln (0.4N)} \left(\frac{\mathcal{M}}{10^5 M_{\odot}} \right)^{1/2} \left( \frac{1 M_{\odot}}{m_{\star}} \right) \left( \frac{r_{\mathrm{ch}}}{1 \mathrm{pc}} \right)^{3/2} \mathrm{yr}
\label{e-relax}
\end{equation}
where $\mathcal{M}$ is the total mass within the characteristic radius $r_{ch}$, $m_{\star}$ is the characteristic stellar mass, and $N$ is a total number of stars in the cluster. 
Usually the half-mass radius $r_{hm}$ is taken as the characteristic radius \citep[e.g.,][]{por02}.
The according half-mass relaxation time is then a good estimate for the dynamical timescale of the whole cluster.

Here we estimate the relaxation time for the two extreme cases with upper and lower limits of the total mass and half-mass radius.
For the upper limit half-mass radius $r_{hm} = 50''$, a total mass of $\sim8000$ \msun\ within the radius, a characteristic stellar mass of 1 \msun, and a total of $\sim16,000$ stars, we get a half-mass relaxation time of $\sim39$ Myr. As a lower limit, we derive $\sim11$ Myr. The uncertainty in the number of stars could change the relaxation time by $\pm5 - 10$\%.
In any case it is safe to say that the dynamical timescale for stars with masses around 1 \msun\ is of order 10 million years. This is about an order of magnitude larger than the age of the intermediate- and low-mass stars ($\lesssim2$ \msun) in the cluster, which is less than $\sim1$ Myr. Thus, solar-mass stars are still young enough such that the dynamical evolution has not yet changed substantially their initial kinetic energy.

As the dynamical time scale is inversely proportional to the stellar mass, the high-mass population is expected to have a much shorter relaxation time. The smaller half-mass radius of the high-mass stars further decreases the relaxation time proportionally to the power of 1.5 (see eq.~[\ref{e-relax}]).
For example, if we simply apply the lower limit of $r_{hm} = 25''$, which is derived based on the $0.5 - 2.5$ \msun\ stars, we obtain a rough estimate of the relaxation time of about 1 Myr for stars typically of 10 \msun.
We therefore conclude that very high-mass stars in the central cluster in NGC 3603 have a dynamical timescale comparable to the cluster age. Thus, we cannot conclude if the observed mass segregation in the high-mass population is dynamical or primordial.
In fact it might be due to the mixture of both effects, since the position of massive stars in young clusters generally reflects the cluster's initial conditions \citep{bon98}.
A similar dynamical state is seen in other young stellar populations. For example, the young ONC ($\lesssim1$ Myr) has been reported to show a mass segregation, in which higher mass stars are preferentially located in the cluster center due to primordial effects. But as in NGC 3603, there is no evidence of mass segregation for stars below $1 - 2$ \msun\ \citep{hil98}.

\section{Systematic uncertainties of the IMF determination}
\label{c-systematic}
There are some systematic uncertainties that could potentially affect the characteristics of the derived IMF.
A first class of such uncertainties could arise from the age, distance, and foreground extinction of the cluster.
The selection of stellar evolutionary model among various available choices is also subject to an uncertainty, and once a set of evolutionary model is chosen, the metallicity selection is also a potential uncertainty.
From a technical point of view, our incompleteness correction and the selection of cluster stars are potential uncertainties.
Moreover, the stellar mass outside the observed field of view is a source of uncertainty.
As for the intrinsic properties of the cluster, the presence of variable extinction and unresolved binary/multiple systems could also affect the derived IMF characteristics.
We thus investigate how these uncertainties could vary the index of the power law slope by scrutinizing them one by one.

For the sake of simplicity, we use several techniques depending on the parameters.
To examine the uncertainties due to age, distance, foreground extinction, metallicity, evolutionary model, and the stellar mass outside the observed field, we construct a simplified two-mass bin IMF in each case.
Here we count only the stars in two mass ranges, $1.7 - 2.6$ \msun\ and $8 - 19$ \msun\ as low- and high-mass bins, respectively. Another simplification is that we use only the $J$-band magnitude for the luminosity-mass conversion, in contrast of using all three $JHK_{S}$ bands for the best estimate of the IMF in \S~\ref{c-imf}.
For investigating the rest of the uncertainties, we follow the technique used in the determination of the best IMF.
As some of the errors have asymmetric distributions, we give the median and the upper and lower limits in each measurement.
In the end we combine all resulting errors to obtain the uncertainty in the best estimate of the power law of the IMF.

\subsection{Age}
\label{s-robust-age}
As presented in \S~\ref{sss-age}, we have used the 2.5 Myr LS01 isochrone for the MS population and the 0.7 Myr PS99 isochrone for the higher mass PMS population ($ \geq 1.2$ \msun).
To safely cover the uncertainties in the current age estimate of $\lesssim3$ Myr, we study the effect of a selection among the 0.2, 1, 2.5 and 5 Myr LS01 isochrones for the high-mass bin, and the 0.3, 0.7, 1, and 1.5 Myr PS99 isochrones for the low-mass bin.
By counting the number of stars in the two mass bins, we compute the IMF power law indices for all possible combinations of the MS and PMS isochrones.
We summarize the results in Tab.~\ref{tbl-iso-age-gamma}.
There are several points to highlight.
First, there is a correlation between the age and the resulting power law index. If the age of the MS isochrone is fixed to 2.5 Myr, a younger age of the PMS isochrone results in a shallower IMF slope.
In contrast, for a fixed PMS isochrone age, a younger age of the MS isochrone results in a steeper slope of the IMF.
This outcome can be explained by re-writing the definition for the IMF $\xi (\mathcal{M})$ using a $\mathcal{M}$-$L$ relation
\begin{equation}
\xi (\log \mathcal{M}) \propto \psi(M) \mathcal{M} \left(\frac{d\mathcal{M}}{dM}\right)^{-1},
\label{e-imf-lf2}
\end{equation}
where the last term is the derivative of the $\mathcal{M}$-$L$ relation.
In the $\mathcal{M}$-$L$ diagram $(\log \mathcal{M} - M_{J})$ of the various isochrones, a younger age shows a steeper profile. Hence, a given mass range covers a narrower magnitude range for younger isochrones, resulting in smaller number counts. This leads to the observed correlation.
Another point is that the biggest change of the index originates from the selection of the PMS isochrone. This is because the $\mathcal{M}$-$L$ relation of PMS stars is more sensitive to the age variations than that of MS stars. We derive the most extreme change for the youngest 0.3 Myr PMS isochrone.

We note that the IMF power law index derived from the two mass bins is only little different from fitting the complete IMF. For our best age estimate with 2.5 Myr MS and 0.7 Myr PMS isochrones, the index from the simple two-mass bin analysis gives $\Gamma = -0.82$ in comparison to $\Gamma = -0.74$ from the complete fit.

As a result, we find that the age uncertainty leads to a systematic uncertainty in the IMF power law index of $\Delta\Gamma = +0.60$ and $-0.41$.
We note that the systematic error in the above analysis provides the upper and lower limits, but a realistic uncertainty is expected to be much smaller. We are aware that the 1.5 Myr PMS isochrone clearly fails to fit the MS turnon point and that the 0.3 Myr PMS isochrone fails to reproduce the bulk of the population in the PMS-MS transition region.


\begin{table}
\begin{center}
\caption{Variation of the power law index due to age\label{tbl-iso-age-gamma}}
\begin{tabular}{ccc}
\tableline
\tableline
\multicolumn{2}{c}{Age of isochrone (Myr)} & \Gam  \\
\tableline
High-mass bin (MS)   & Low-mass bin (PMS) &         \\
\tableline
0.2             & 0.3               & -0.35   \\
1.0             & 0.3               & -0.34   \\
2.5             & 0.3               & -0.30   \\
5.0             & 0.3               & -0.28   \\
0.2             & 0.7               & -0.87   \\
1.0             & 0.7               & -0.86   \\
2.5             & 0.7               & -0.82   \\
5.0             & 0.7               & -0.80   \\
0.2             & 1.0               & -1.10   \\
1.0             & 1.0               & -1.09   \\
2.5             & 1.0               & -1.05   \\
5.0             & 1.0               & -1.03   \\
0.2             & 1.5               & -1.28   \\
1.0             & 1.5               & -1.27   \\
2.5             & 1.5               & -1.23   \\
5.0             & 1.5               & -1.21   \\
\tableline
\normalsize
\end{tabular}
\end{center}
\end{table}

\subsection{Distance}
\label{s-robust-distance}
Earlier studies have measured the distance to be within $6 - 8$ kpc, and $d \sim 6$ kpc yields a reasonable fit in our isochrone fitting in the CMD. We therefore study distances of 5, 6, 7, and 8 kpc to see the resulting change of the IMF slope. Note that we do not adjust the color cut for the rejection of field stars but only change the distance in the apparent magnitude-mass conversion.
The resulting power law indices are $\Gamma = -0.85, -0.84, -0.86$, and $-0.85$ for $d = 5 - 8$ kpc, respectively, with a negligible variation of $\Delta\Gamma = \pm0.01$. The small variation of the index is expected from the fact that a change in distance simultaneously shifts both low- and high-mass magnitude ranges, and thus does not yield any significant differences in the ratio of the high- to low-mass counts.

\subsection{Foreground extinction}
\label{s-robust-fore-ex}
As described in \S~\ref{sss-extinction}, our best estimate of the interstellar extinction towards NGC 3603 is $A_{V} = 4.5 \pm 0.5$ mag.
Considering the observed radial increase of $\Delta{A_{V}} \sim 2.0$ mag from the cluster center to the outermost ISAAC fields, we apply three different values $A_{V} = 4.0, 4.5$, and $6.5$ mag to derive the uncertainty in the IMF slope.
Again we do not adjust the color cut for the rejection of field stars.
The resulting power law indices of the IMF are $\Gamma \sim -0.86, -0.83, -0.86$ for $A_{V} = 4.0, 4.5, 6.5$ mag, respectively. The slope does not show any trend and is fairly insensitive to the adopted foreground extinction with an error of only $\Delta\Gamma ={}^{+0.02}_{-0.01}$.

\subsection{Theoretical model dependence}
\label{s-robust-model}
Here we adopt several currently available MS and PMS evolutionary models to derive the number counts for the low- and high-mass bins and calculate the change in the power law index of the resulting IMF.

For the $8 - 19$ \msun\ high-mass bin, in addition to the 2.5 Myr MS isochrone in the Geneva model (LS01), we apply an isochrone of the Padova evolutionary model computed with the 2MASS filter system \citep{ber94,gir00,bon04}. We use the youngest available 4 Myr isochrone here.
For the $1.7 - 2.6$ \msun\ low-mass bin, in addition to the PS99 set, we apply a PMS model published by \citet[][hereafter SDF00]{sie00}, and the $Y^2$ isochrones based on the revised Yale models \citep{yi01,dem04}.
As discussed in \S~\ref{sss-age}, it is well known that the various available PMS evolutionary models result in considerable discrepancies in the age estimate of a star cluster, in particular, for a very young age (less than a few megayears).
Although we have applied the 0.7 Myr PS99 isochrone for the PMS stars, for this analysis, it is necessary to individually select the best-fit age for each PMS model instead of simply applying 0.7 Myr.
We select the 1.5 Myr SDF00 isochrone and the 1.5 Myr $Y^2$ isochrone for the PMS.
The resulting IMF power law indices are summarized in Tab.~\ref{tbl-model-gamma}, showing a slightly asymmetric distribution with $\Delta\Gamma ={}^{+0.07}_{-0.13}$.


\begin{table}
\begin{center}
\caption{Variation of the power law index due to the selection of stellar evolutionary models\label{tbl-model-gamma}}
\begin{tabular}{llc}
\hline
\hline
\multicolumn{2}{c}{Isochrone model} & \Gam  \\
\hline
High-mass bin (MS)   & Low-mass bin (PMS) &    \\
\hline
LS01 (Geneva) 2.5 Myr & PS99 0.7 Myr      & -0.82   \\
                      & SDF00 1.5 Myr     & -0.75   \\
                      & $Y^2$ 1.5 Myr     & -0.95   \\
\hline
Padova 4 Myr          & PS99 0.7 Myr      & -0.83   \\
                      & SDF00 1.5 Myr     & -0.76   \\
                      & $Y^2$ 1.5 Myr     & -0.95   \\
\hline
\normalsize
\end{tabular}
\end{center}
\end{table}

\subsection{Metallicity}
\label{s-robust-metallicity}
We have so far adopted a solar metallicity $Z = 0.02$ in the isochrone selections. This is based on the fact that the Galactocentric distance of NGC 3603 is similar to that of the Sun, so that the radial gradient of the metallicity in the Milky Way does not need to be taken into account \citep[e.g.][]{rud06}.
To estimate the potential effect of the metallicity selection, we test a half-solar, solar, and twice-solar metallicity.
Since no isochrones with different metallicities are available in the PS99 set, we employ the 1.5 Myr SDF00 PMS models with $z$ = 0.01, 0.02, and 0.04 for the low-mass stars. For the high-mass stars we use 2.5 Myr LS01 MS isochrones with $z$ = 0.008, 0.02, and 0.04.
The resulting power law indices are $\Gamma = -0.88$ (half-solar), $-0.75$ (solar) and $-0.76$ (twice-solar).
The application of the solar and twice-solar metallicity yield almost the same value, and a major difference is seen for the half-solar metallicity, which steepens the IMF by $\Delta\Gamma\sim -0.13$. 
Note that, although we confirmed that there were clear correlations between the metallicity and the number counts in the both low- and high-mass bins, no trend eventually was seen in the power law index.
Although we do not find any trend in our analysis, a possible dependence of the IMF of stellar systems on the metallicity has been discussed in several recent studies \citep[e.g.,][]{sol07}.

\subsection{Individual extinction}
\label{s-robust-variableEX}
We have so far applied a uniform foreground extinction of $A_V = 4.5$ mag in the conversion from the observed luminosities to stellar masses. However, potential local variations from circumstellar gas and dust, and larger scale variations on the size scale of the cluster could affect the derived stellar masses. In fact we find a radial variation of $\Delta{A_V} \sim 2.0$ mag from the cluster center towards the outer regions.
To derive the impact of this uncertainty, we construct the IMF of NGC 3603 with and without the treatment of a variable extinction. For this we shift the stars along the reddening direction in the $JHK_{S}$ magnitudes space as close as possible to the isochrone, rather than picking the closest points on the isochrone for a fixed average extinction.

As a result, we find that the treatment of the variable extinction flattens the IMF slope by $\Delta\Gamma \sim +0.15$. We also find that the change of the IMF slope is mostly caused by the almost parallel alignment of the reddening direction and the isochrone of the PMS-MS transition region. Because of this alignment, sources in the scattered PMS-MS region are dereddened to the intermediate-mass MS, which causes an artificial bump at around 5 \msun\ and results in a slightly flatter IMF slope within $0.4 - 20$ \msun.
As the change of the power law index mostly arises from this local feature, the real impact of the presence of the individual extinction on the IMF slope is expected to be even smaller.

\subsection{Incompleteness correction}
\label{s-robust-incomp}
Here we evaluate the potential influence of our empirical-based incompleteness correction on the IMF by considering two extreme cases: one is without any correction, and the other is with an over-correction.
The first case has already been analyzed in Figure~\ref{fig-imf-best}, the power law index being $\Gamma = -0.68$. This means that without the incompleteness correction the IMF would come out shallower by $\Delta\Gamma \sim +0.06$.
For the second case, we prepared a larger correction option in the computation of the correction factor (see detailed explanation in \S~\ref{ss-incomp-correction}).
Using the same technique as for the best IMF ($\Gamma = -0.74$), but applying a larger correction factor, we find the IMF showing a power law index of $\Gamma \sim -0.87$. Thus, the larger correction steepens the IMF by $\Delta\Gamma \sim -0.13$. 
Here we note that a correction by simply a factor of 2 of the normal correction in each mass bin steepens the IMF power law index by $\Delta\Gamma \sim -0.04$.

In summary we derive a potential error of $\Delta\Gamma = ^{+0.06}_{-0.13}$ in our incompleteness correction.
As mentioned in \S~\ref{ss-imf-comp-ngc3603}, the slight difference of the resulting IMF between our study and that by \citet{sto06} would be partly explained by the difference in the correction rates between two studies.

\subsection{Cluster membership}
\label{s-robust-membership}
In order to deal with line-of-sight field stars in the observed fields, we have so far applied the color cut method instead of a statistical approach based on control field measurements.
In the color cut we exclude blue stars in the $(J - K_S, J)$ CMD because those stars can most likely be regarded as non-cluster members.
As mentioned in \S~\ref{c-field}, a potential source of uncertainty in this application is the fact that the region redward of the color cut could still contain some field stars whose $J - K_S$ colors are either similar to those of cluster stars, or are redder than cluster stars, e.g., red giants\footnote{We note that extragalactic contaminations are expected to be negligible because of the relatively small field of view. Quantitatively, integrated galaxy number densities are expected to be approx. $10^4$ galaxies deg$^{-2}$ up to $\sim20$ mag in the $H$ and $K$ bands \citep[e.g.][]{djo95,yan98}, corresponding to only a few galaxy counts in our field of $r \leq 110''$}.

In fact, we verify that the ``blueward only'' color cut is a reasonable treatment based on the analysis of LF in \S~\ref{c-field} adopting the field star \textit{K}LF in NPG02 which is derived from a control field and is in agreement with what is expected from Galactic models.
Although we are not able to make any statistical estimate of field stars from our own data because of the lack of a control field in our observations, we are still able to test the potential effect of red stars by simply cutting them out ("red cut").

We determine the red cut criteria as follows.
For MS stars down to $m_J \leq 14$ mag, stars located within $\Delta(J - K_S) = +0.35$ mag redward from the combined isochrone are regarded as cluster stars, and stars redder than the boundary are excluded from the IMF determination.
For PMS stars with $m_J > 15$ mag, we apply $\Delta(J - K_S) = +0.50$ mag considering larger photometry errors for fainter stars.
As for the region of $m_J = 14 - 15$ mag, the line is drawn by directly connecting the curves in the upper and lower magnitude ranges so as to deal with the turnover feature of the isochrone in the PMS-MS transition region.
In addition, since we measure the radial increase of the interstellar extinction across the cluster (as shown in Figure~\ref{fig-cmd-radial}), we separate inner and outer regions with $r \leq 30''$ and $r = 30'' - 110''$ and apply an additional shift of +0.25 mag for the outer region. Thus, we use $\Delta(J - K_S) = +0.60$ and $+0.75$ mag for MS and PMS stars, respectively.
Here we note that, as the blueward color cut, the shift is applied only in the $J - K_S$ direction.

\begin{figure*}
\plottwo{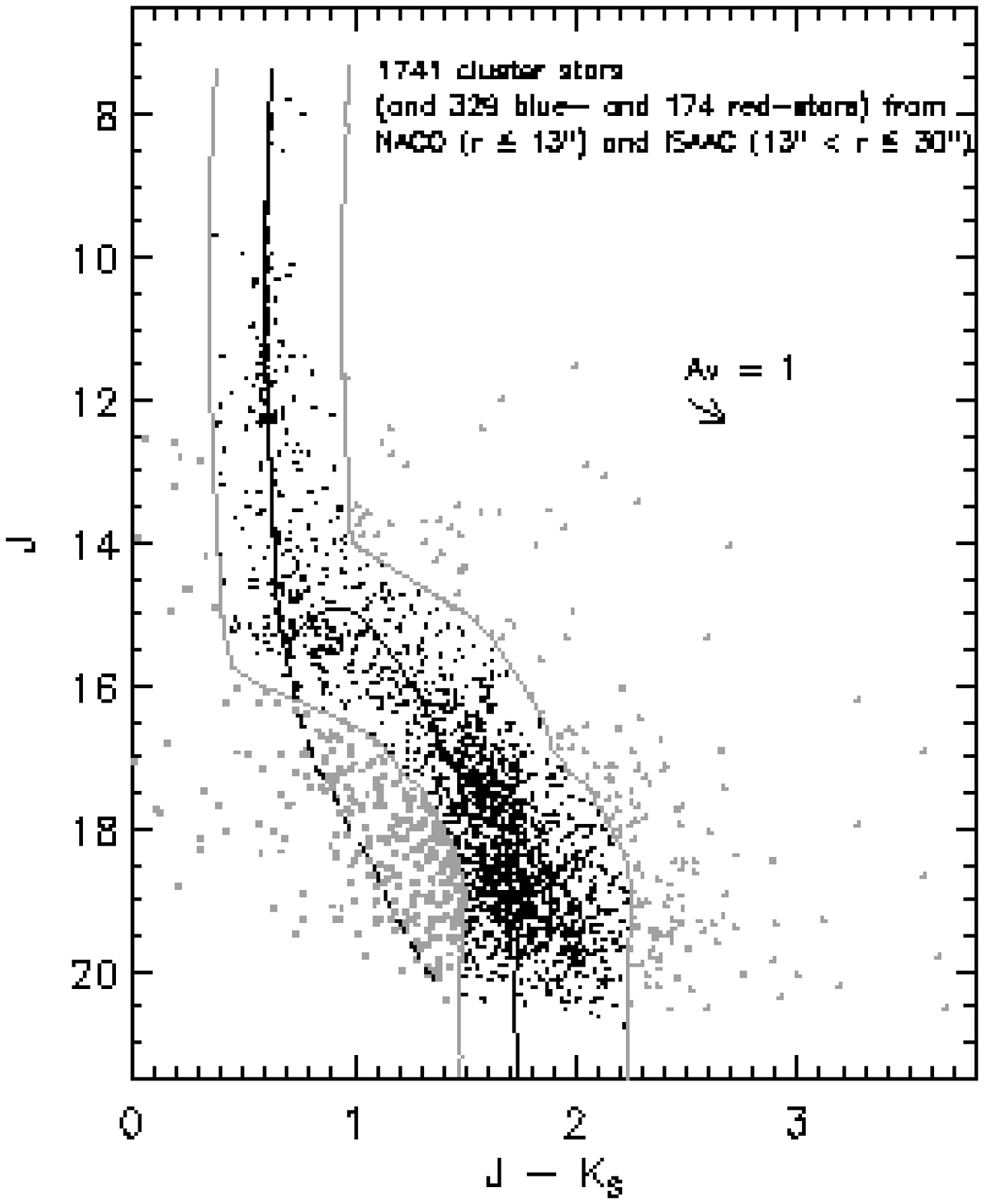}{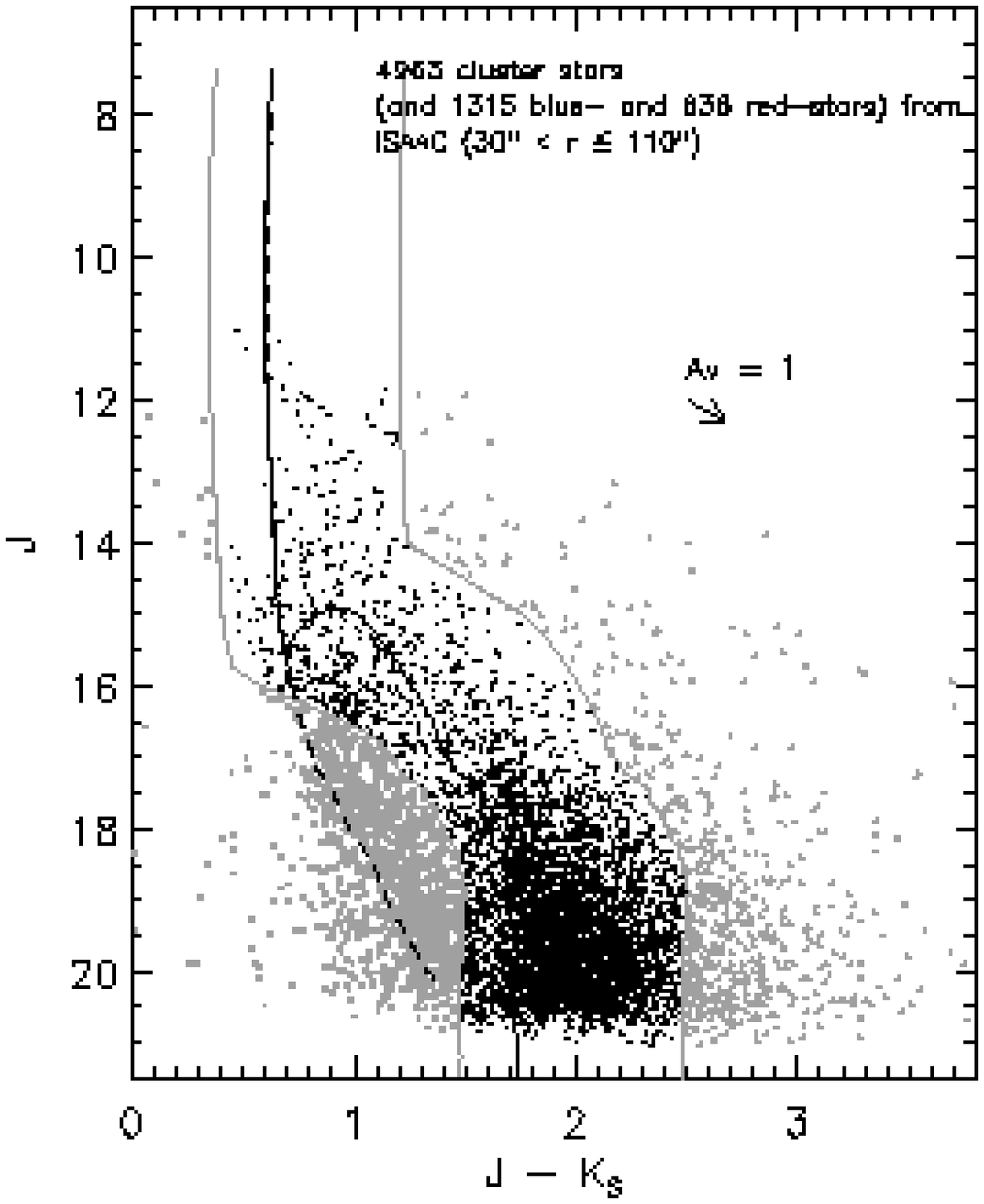}
\caption{Selection of cluster stars by the red cut in addition to the blue cut for the inner field of $r \leq 30''$ (\textit{left}) and outer field of $r = 30'' - 110''$ (\textit{right}). The gray solid curve redwards of the isochrone is the boundary of the red cut.
Red stars regarded as non-cluster members and thus removed for the IMF determination are shown in gray triangles. As Figure~\ref{fig-cmd-cc}, blue stars based on the original color cut are shown in gray squares.\label{fig-robust-redcut}}
\end{figure*}

The red cut is illustrated in the $(J - K_S, J)$ CMDs in Figure~\ref{fig-robust-redcut}.
As a result, 810 stars out of 7514 stars, which are used for the best estimate of the IMF, are classified as red stars and excluded. Thus, 6704 stars are used for the IMF calculation, and the resulting IMF shows the power law index of $\Gamma = -0.76$, being fairly similar to that of the best IMF. Thus we conclude that the potential error of the IMF power law index due to the presence of red stars is $\Delta\Gamma \sim -0.02$.

We note that our criteria of the red cut is similar to those in the previous study by \citet{sto06}, and the only difference is that we apply two different thresholds depending on the radial distance.


\subsection{Unresolved binary systems}
\label{s-robust-binary}
Unresolved binaries (or higher-order multiple systems) have been identified as an important uncertainty in the IMF determination \citep{kro01,kro02,mal01}.
Any unresolved source composed of two or more stars mimics a slightly brighter single star in the photometry, resulting in an underestimate (overestimate) of the low-mass (high-mass) population. 
A mass function derived from a population with unresolved systems, the so-called system mass function, thus shows a slightly larger power law index than the actual stellar mass function, and an appropriate treatment of the unresolved sources will steepen the derived IMF. 

In this section we first present our estimate of the binary fraction based on the analysis of the $(J - K_{S}, J)$ CMD.
In a second step we then model the impact of the unresolved binaries on the IMF by applying the well known binary frequency from the ONC.

\subsubsection{Unresolved binary fraction from the observed CMD}
\label{ss-bf-cmd}
Several studies have already reported the potential bias from unresolved binaries on the IMF of Galactic young star clusters. For example, \citet{fig99} derived a significantly flat IMF with $\Gamma \sim -0.65$ for the Arches cluster, for which a binary fraction of unity would steepen the IMF by $\Delta\Gamma \sim -0.3$.

\citet{sto04,sto06} report the presence of a secondary sequence in NGC 3603, which shows up in their $(J - K_{S}, J)$ CMD towards the bright red side of the MS above the PMS-MS transition region.
They interpret this sequence as equal-mass binary stars, formed together with the other PMS stars in a single star formation epoch $\sim1$ Myr ago. They find no IR excess emission for the stars of this secondary sequence in their $L$-band analysis but rather find an anti-correlation between those sources and H-$\alpha$ emission stars.
This suggests that the secondary sequence is not caused by the contribution from circumstellar disks, but by unresolved binaries.
Counting stars on the secondary sequence, they derive a fraction of unresolved binaries of 30\% in the central cluster. From the typical offset of the secondary sequence of $\Delta{J} \sim -0.75$ mag, they argue that most of the candidates are likely to be close to equal-mass binary systems.
Using the unresolved binary fraction of 30\% and assuming equal-mass components, the authors apply a binary correction in the IMF determination, steepening the IMF slope by $\Delta\Gamma \sim -0.06$.

In contrast, we do not find a distinct secondary sequence, but merely a scattered distribution towards the red side of the MS and to the top of the PMS-MS transition region. Therefore we think it is not feasible to confidently distinguish potential unresolved binaries (which can have non-uniform mass-ratio) from reddened MS stars or younger PMS stars. The scatter can be partly due to an intrinsic age spread in NGC 3603 as discussed in our age estimate in \S~\ref{sss-age}.

\begin{figure*}
\begin{center}
\plottwo{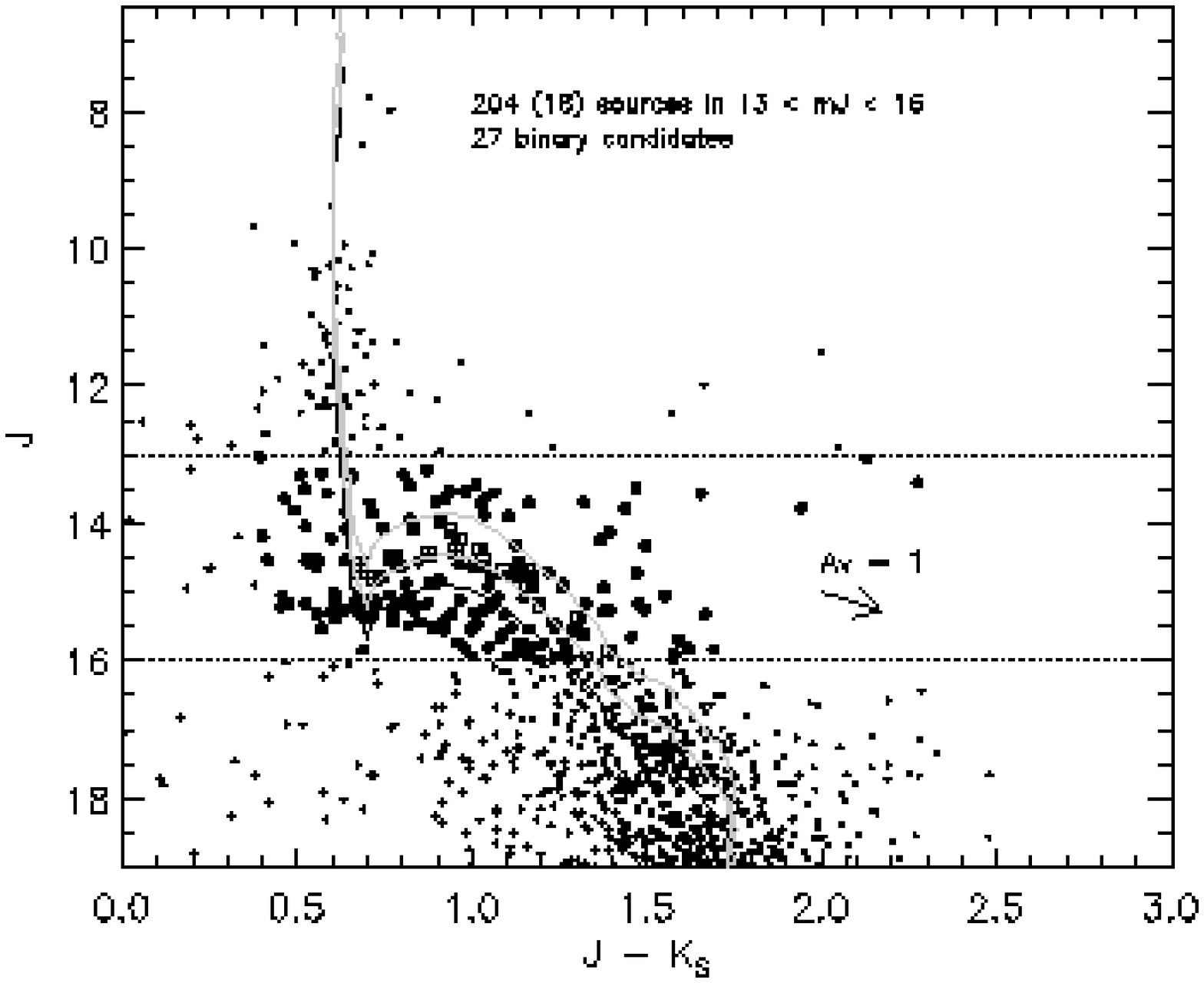}{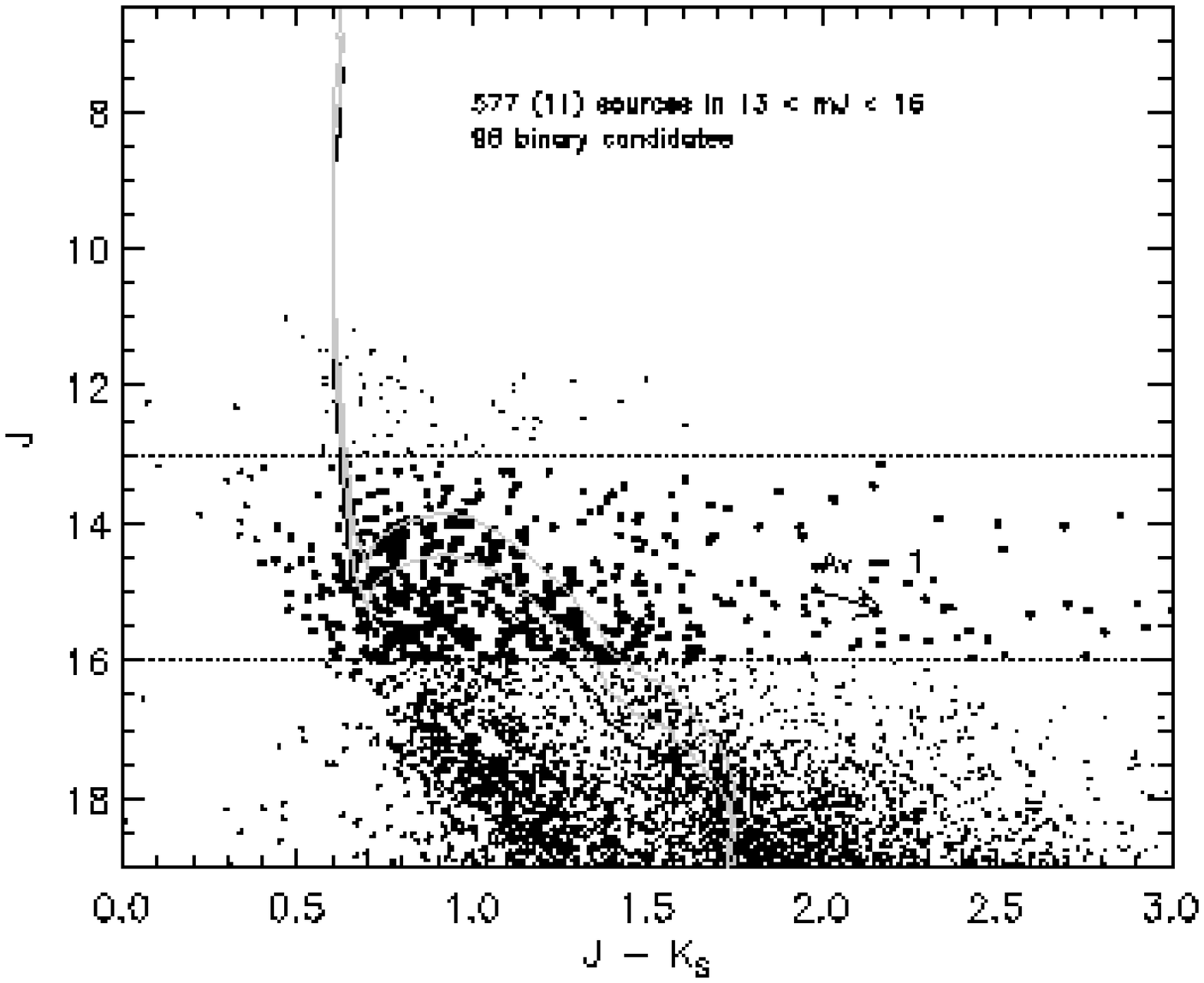}
\end{center}
\caption{Unresolved binary candidates in the $(J - K_{S}, J)$ CMDs for stars from the NACO field of $r \leq 13''$ (\textit{left}) and the ISAAC field of $13'' < r \leq 110''$ (\textit{right}). The analysis is restricted to a range of $m_{J} = 13 - 16$ mag. The binary band is indicated in gray solid curves. In total 186 cluster stars (\textit{large dots}) plus 18 non-cluster stars (\textit{plus signs}) are identified in the magnitude range, and 27 sources are located within the binary band (\textit{points and squares}), i.e., a binary fraction of $\sim13$\%. In the outer field, 96 binary candidates among 577 stars are detected, resulting in $\sim17$\%.\label{fig-imf-binary}}
\end{figure*}

While it is difficult to reliably estimate the fraction of unresolved binaries from our data, we can give a rough estimate by counting sources located in the upper part of the PMS-MS transition in the $(J - K_{S}, J)$ CMD.
The selection criterion for potential binaries is illustrated in the CMDs in Figure~\ref{fig-imf-binary}.
The location of the binary band is defined based on the following idea: in case of an equal-mass system, the real magnitude of the two stars is 0.75 mag larger (i.e., fainter) than the system magnitude.
Assuming a $\pm0.3$ mag error in the $J$ magnitude to define the width of the band, the limits of the binary band are set by shifting the isochrone by $-1.05$ and $-0.45$ mag.
Among the total of 204 sources at $r \leq 13''$ within $m_{J} = 13 - 16$ mag, we find 27 stars within the binary band, corresponding to a binary fraction of $\sim13$\% (Figure~\ref{fig-imf-binary} left).
The same experiment for the outer ISAAC field of $13'' < r \leq 110''$ yields a similar value of $\sim17$\% as 96 binary candidates are found among 577 sources in the magnitude range.
However, since this measurement is based on the assumption of equal-mass components and the typical uncertainty of $\pm0.3$ mag, the true number is subject to large uncertainties. For example, if we adopt a magnitude range of $\Delta{m_{J}} = 0.75\pm0.4$ mag to define the binary band, the binary fraction increases to $\sim21$\% for the NACO field and $\sim22$\% for the ISAAC field.

Here we note that the potential unresolved binaries in our data also include multiple sources from projection effects, i.e., optical binaries. Using the spatial resolution of $\sim490$ AU in the NACO field and $\sim2200$ AU in the ISAAC field, we derive a likelihood for optical binaries of about 6 and 9\% for the NACO and ISAAC data, respectively.


\subsubsection{Systematic error of the IMF from unresolved binaries}
\label{ss-binary-simulation}

As outlined above, we cannot reliably identify unresolved binaries in our data in the CMD analysis.
To quantify the systematic error of the IMF from such unresolved binaries, we thus choose to simulate the effect adopting the measured binary distribution in the ONC considering the similar age and high-mass stellar content of the ONC and NGC 3603.

\paragraph{Binary fraction of high- and low-mass stars in the ONC.}
The binary fraction in the ONC including the Trapezium cluster is among the best studied.
Note that, although some binary/multiple surveys mentioned below differentiate between binary and higher-order systems, we do not follow this distinction but simply use the term binary fraction.
\citet{pro94}, for example, derived a binary fraction of $\sim12$\% for a projected linear separation of $26 - 440$ AU in the Trapezium cluster, and \citet{pet98} derived $\sim9.5 \pm 4.5$\% for $63 - 225$ AU for stars with masses down to $\sim0.04$ \msun.
Combining their new observations with the results from \citet{pet98}, \citet{koe06} derive a binary fraction of $\sim14 \pm 7$\% and $\sim23 \pm 10$\% for stars with $\mathcal{M} > 2$ \msun\ in the periphery and the central core, respectively, and $\sim4.8 \pm 1.8$\% and $\sim3.6 \pm 3.2$\% for stars with $\mathcal{M} = 0.1 - 2$ \msun\ for a projected separation of $60 - 500$ AU. The higher binary fraction for high-mass stars in the ONC is consistent with earlier results from \citet{pre99}.

Here we adopt the binary fractions from \citet{koe06}. To simplify our analysis, we use their typical values of $\sim5$\% for low-mass stars ($0.1 - 2$ \msun) and $\sim15$\% for intermediate- to high-mass stars ($>2$ \msun).
To also cover the various other estimates of the binary fraction in young stellar clusters and to estimate the \textit{upper limit} of the potential impact of unresolved binaries on the IMF determination, we also analyze the case for a 2 times larger binary fraction, i.e., $\sim10$\% for $0.1 - 2$ \msun\ and $\sim30$\% for $>2$ \msun.

\paragraph{Binary fraction as a function of separation.}
The angular resolution of our NACO and ISAAC data correspond to minimum detectable binary separations of $\sim490$ and $\sim2200$ AU, respectively.
In their binary analysis of nearby solar-type stars, \citet{duq91} show that the orbital period follows a lognormal distribution. Using this distribution \citet{koe01} derives the distribution for the projected separation, which follows the form $ f(x) \propto \exp[-\frac{1}{2}(x-\bar{x})^2/\sigma^2]$, where $x$ is the logarithm of the projected separation in AU with $\bar{x} = 1.44$ and $\sigma = 1.55$ (see Figure~6 in the reference).
Integrating this distribution up to the minimum separations, we can estimate that $\sim79$\% of all binaries are unresolved in the NACO data, and $\sim89$\% are unresolved in the ISAAC data.
Considering the rather larger uncertainty in the adopted binary fractions, we ignore this difference in the following experiment.

\paragraph{Binary fraction as a function of mass ratio.}
The mass ratio ($q = m_2/m_1$) distribution has shown various forms in various types of stellar populations.
\citet{rei97} find evidence for preferential formation of nearly equal-mass binary systems, i.e., a peak at $q \sim 1$ among nearby M dwarfs.
In contrast, \citet{duq91} derive a distribution that peaks near $q = 0.2 - 0.3$ for solar-type field stars. \citet{fis92} derive a peak at $q = 0.4 - 0.6$ and $0.6 - 0.8$, respectively, for the field G  and M dwarf binaries. \citet{duc99} find a distribution peaking at $q = 0.5 - 0.75$ for binaries with a primary mass of $\leq 1$ \msun\ in the young star cluster IC 348.
For some clusters the mass ratio distribution can be fitted by an inverse power law $f(q) \propto q^{-\Gamma}$, for example, with $\Gamma = 0.33 - 0.5$ in Sco OB2 \citep{sha02,kou05} and $\Gamma = 1.3$ in the Hyades cluster \citep{pat98}.

Thus, it is not straightforward to assume a unique mass ratio distribution; we hence examine two cases.
In the first model, which favors equal-mass binaries, 75\% of the unresolved binary systems have a mass ratio of $q = 1$ and 25\% have $q = 0.4$. In the second model, which favors non-equal-mass binaries, 25\% of the population have $q = 1$ and 75\% have $q = 0.4$.

\paragraph{Simulation of the unresolved binary correction.}
Using the adopted binary fractions and the two cases of the mass ratio distribution mentioned above, we study the potential influence of the unresolved binaries on our IMF determination by means of a simple simulation. We perform the simulation as follows. First we randomly generate a set of binaries for each mass bin according to the observed number counts and the assumed binary fractions. We then split those binary systems according to the mass ratio distribution, subtract the system from the original mass bin, and then add the split masses into the corresponding mass bins. Finally we compute the power law index of this simulated IMF within the mass range of $0.4 - 20$ \msun.
As a result, the power law index of the IMF becomes slightly steeper by $\Delta\Gamma \sim -0.02$.
Even assuming twice the binary fraction of the ONC, the IMF steepens by only $\Delta\Gamma \sim -0.04$. The two cases of the mass ratios yield almost identical power law indices. We conclude that the systematic error in the IMF power law index from the presence of unresolved binaries is $\Delta\Gamma \lesssim -0.04$.

\subsection{The stellar mass outside the observed field of view}
\label{s-mass-coverage}
In the analysis of the dynamical state of the cluster in \S~\ref{c-massseg} we have outlined that the high-mass stars in NGC 3603 appear to be strongly concentrated towards the cluster center. In contract, the intermediate- and low-mass stars in the cluster appear neither dynamically nor primordially mass-segregated at this stage and extend beyond the observed field of $r \leq 110''$. Therefore, we need to quantify how much these intermediate- and low-mass stars outside the field of view can change the IMF.

For that we use the de-projected mass density distribution from \S~\ref{s-total-mass}.
Integrating the volume mass density of the stars within the mass range of $0.5 - 2.5$ \msun\ up to the observation limit $r = 110''$, we find that our observations cover more than 90\% of these stars if the cluster radius is $r = 150''$, and still $\sim67$\% when using the upper limit of the cluster size, i.e., the tidal radius of $r_t = 1260''$.
Qualitatively we conclude that our data indeed covers the majority of the cluster mass, and that the observed IMF is thus representative for the whole cluster.
To quantify how much the intermediate- and low-mass stars outside our field of view can change the IMF power law index, we simply add the maximum missing mass of 33\% to the number counts of the low-mass bin ($1.7 - 2.6$ \msun) in the simplified two mass bins IMF. As a result, the power law index decreases from $\Gamma = -0.82$ to $-0.98$, that is, a steepening of the IMF with $\Delta\Gamma \sim -0.16$.

\subsection{Error combination}
\label{s-robust-error-combine}
Tab.~\ref{tbl-error-combine} summarizes the individual systematic errors from \S~\ref{s-robust-age} - \ref{s-mass-coverage}. We combine them into a single systematic error to obtain the final power law index in our IMF determination.
To combine those asymmetric errors, we apply the method by \citet{bar03}\footnote{This method is implemented in the program from http://www.slac.stanford.edu/~barlow/java/statistics1.html}. The program constructs, for each set of positive and negative errors, an asymmetric distribution consisting of two half-Gaussian distributions and then convolves these distributions to derive a combined asymmetric error.
We note that in this combination we also add the fitting error $\pm0.02$ in the power law index calculation.
The resulting combined error is $\Delta\Gamma ={}^{+0.62}_{-0.47}$. For comparison, the simple quadratic sum of the positive and negative errors yields fairly similar values of $+0.63$ and $-0.50$, respectively.

In summary, we conclude that the IMF power law index of NGC 3603 for the mass range $0.4 - 20$ \msun\ including all systematic errors is $\Gamma = -0.74^{+0.62}_{-0.47}$. 
Considering a Gaussian probability distribution with the above 1 $\sigma$ errors, we calculate the probability that the IMF of NGC 3603 is as steep as the Salpeter IMF, resulting in Pr$(\Gamma \leq -1.35) \sim 0.1$, i.e., about 10\%. This leads us to interpret that the IMF of NGC 3603 is distinguishably flatter than the Salpeter-like IMF.


\begin{table}
\begin{center}
\caption{Summary of the error estimates and the combination\label{tbl-error-combine}}
\begin{tabular}{ll}
\tableline
\tableline
 Parameters & Errors  \\
\tableline
Age                    & +0.60, -0.41 \\
Distance               & $\pm0.01$ \\ 
Foreground extinction  & +0.02, -0.01 \\
Evolutionary model     & +0.07, -0.13 \\ 
Metallicity            & -0.13   \\
Individual extinction  & +0.15 \\
Incompleteness correction & +0.06, -0.13 \\
Cluster membership     & -0.02 \\
Unresolved binary      & -0.04 \\
Stellar mass outside the FOV & -0.16 \\
\tableline
Combined               & +0.62, -0.47 \\
\tableline
\normalsize
\end{tabular}
\end{center}
\end{table}

\section{Discussion}
\label{c-discussion}
We find that the IMF of NGC 3603 is well described by a single power law with $\Gamma ={-0.74}^{+0.62}_{-0.47}$ within $0.4 - 20$ \msun. We also find that the IMF steepens towards outer region with the power law index from $\Gamma \sim -0.3$ to $-0.8$. The steepening mainly occurs in the inner $r \lesssim 30''$.
The average IMF is slightly flatter but still consistent with previous studies (see discussions in \S~\ref{ss-imf-comp-ngc3603}).
Here we discuss the resulting IMF of NGC 3603 with a particular focus on the question: is the IMF of NGC 3603 an example for top-heavy IMF in starbursts?

\subsection{IMF in various stellar populations}
There is growing evidence that the IMF varies between different stellar populations.
Here we summarize some recent IMF studies of young stellar populations.

Several stellar clusters have been suggested to have a somewhat flat IMF.
For example, the Arches cluster near the Galactic Center (GC) shows a comparatively shallow IMF with $\Gamma = -0.6$ to $-1.1$ for the high-mass stars \citep{fig99,sto05,kim06}.
The Quintuplet cluster -- another massive young cluster near the GC -- has also been suggested to have a mass distribution potentially flatter than the field IMF \citep{fig99}.

Other stellar clusters show standard Salpeter-like IMFs, but there are substantial differences between various studies.
For example, the IMF of the massive star cluster R136 in 30 Dor has been reported to have a power law index of $\Gamma \sim -1.0$ to $-1.6$ \citep{bra96, mas98}, and even substantially flatter for solar-mass stars \citep{sir00}.
There are many other examples such as the double cluster h and $\chi$ Persi with power law indices of $\Gamma \sim -1.3$ \citep{sle02}, and NGC 1960 (M36) and NGC 2194 with $\Gamma \sim -1.2$ to $-1.3$ \citep{san00}. Recently, the IMF of the Galactic massive cluster Westerlund 2 was measured to fit with $\Gamma \sim -1.2$ \citep{asc07}. Yet other clusters have IMFs slightly steeper, for example, NGC 2422 with $\Gamma \sim -2.0$ ($0.9 - 2.5$ \msun) \citep{pri03}, NGC 3576 with $\Gamma = -1.62$ ($> 3$ \msun) \citep{fig02}, and the Sco OB2 association with $\Gamma = -1.9$ \citep{bro98}.

From studies of nearby star-forming regions, it has been found that the variations in the IMF power law index and in the characteristic masses become even larger towards very low stellar masses. Examples of objects which show such variations are the Trapezium cluster \citep{mue02}, the Pleiades cluster \citep{bou98}, the Taurus star-forming region \citep{bri02}, the embedded cluster IC 348 \citep{pre03,luh03}, and the NGC 1333 molecular cloud \citep{wil04}.

Thus, we have so far known that there are moderate variations in the IMF characteristics in intermediate- to high-mass stars among young stellar clusters and substantial variations at very low- and substellar masses.

\subsection{Universal or variable IMF?}
The question arises, is the observed IMF variation a true deviation from the Salpeter IMF? Or can the variation be explained by systematic uncertainties in the measurements?

\citet{kro01}, for example, argues that the so-far observed IMF variations are caused by combined effects of observational, theoretical, and statistical uncertainties.
Prior to \citet{kro01}, \citet{sca98} summarized the state of the IMF research by compiling the IMFs of many clusters and associations. Although no systematic trend is seen, there is a substantial scatter of at least unity ($\pm0.5$) in the $\Gamma$ versus stellar mass plot, indicative of IMF variations. The large uncertainties prevent the author from giving any conclusive statement about a varying IMF. As an average IMF, the author proposes a three-segment power law IMF with the index $\Gamma = -0.2\pm0.3$ for the mass range of $0.1 - 1$ \msun, $\Gamma = -1.7\pm0.5$ for $1 - 10$ \msun, and $\Gamma = -1.3\pm0.5$ for $10 - 100$ \msun. Here the IMF is steepest in the intermediate-mass range.

\citet{kro01} defines the Galactic field IMF adding the results from local star counts to the compilation of the MF power law indices in \citet{sca98}. Assuming the defined average IMF, the author investigates the uncertainty inherent in any observational IMF estimate through $N$-body model calculations. The author concludes that no true variations can be detected within the fundamental limit of uncertainties of a universal IMF described by a segmented power law with index of $\Gamma = 0.7\pm0.7$ for $0.01 - 0.08$ \msun, $\Gamma = -0.8\pm0.5$ for $0.08 - 0.5$ \msun, $\Gamma = -1.7\pm0.3$ for $0.5 - 1$ \msun, and $\Gamma = -1.3\pm0.7$ for $>1$ \msun.

Recently \citet{sca05} again summarized the current IMF estimates from field star counts and from observations of open clusters. The author found that the variation of the IMF power law index among the field star studies within $1 - 15$ \msun\ is not negligible and that the variation among the cluster IMFs is considerable.
These variations might be accommodated by the combined effects of various study-specific uncertainties among studies.
However, there are also IMF variations when observing, analyzing, and interpreting the data of several clusters in the same manner. For example, \citet{phe93} derive the IMFs of eight young open clusters in a consistent way. Although the average value of the power law index $\Gamma = -1.4 \pm 0.13$ is in agreement with the Salpeter IMF, two of the clusters show strong deviations with $\Gamma \sim -1.8$ (NGC 581) and $\Gamma \sim -1.1$ (NGC 663). \citet{sca05} mentioned that these observed variations in cluster IMFs might still be explained by the uncertainties and, if that is the case, then the cluster data are consistent with a universal IMF but with \textit{sizeable} variations, preventing us from determining an average IMF or $\Gamma$.

How does our resulting IMF of NGC 3603 with $\Gamma = -0.74$ fit with the fundamental limit in \citet{kro01}?
\citet{kro01} explains the observed scatter is mainly by the following three uncertainties: (1) Poisson noise due to the finite number of stars, (2) dynamical evolution of the star cluster, and (3) unresolved binaries.
In our determination the Poisson noise is taken into account when fitting the power law. It is only $\Delta\Gamma \sim \pm0.02$ for our $\sim10,000$ stars.
As discussed in \S~\ref{c-massseg}, we estimate that the dynamical evolution in NGC 3603 is expected to be insignificant for the intermediate- and low-mass stars and thus does not affect significantly the determination of the IMF from the field covering up to $r \leq 110''$. This is supported by the fact that the IMF does not steepen significantly beyond $r \gtrsim 30''$.
As for unresolved binaries, \citet{kro01} derives that the correction for unresolved binaries will typically steepen the IMF by $0.5 \lesssim \Delta\Gamma \lesssim 0.8$ for stars within $0.08 - 1$ \msun, but it does not change significantly for $\gtrsim1$ \msun. 
Our simulation of the potential impact of unresolved binaries on the IMF within $0.4 - 20$ \msun\ yields $\Delta\Gamma \sim -0.04$ (see \S~\ref{s-robust-binary}). Since the mass range is different in our analysis and the simulation of \citet{kro01}, it should not directly be compared. However, it is worth mentioning that there is a difference in the adopted binary frequencies. From the studies of the ONC we adopt a binary frequency of 15 and 30\% for $0.1 - 2$ and $\gtrsim2$ \msun, respectively, while \citet{kro01} adopts 100\% for $\gtrsim3$ \msun. Our result is similar to that of \citet{sto06}, in which their correction of unresolved binaries steepens, by $\Delta\Gamma \sim -0.06$, the IMF for stars within $0.4 - 20$ \msun\ in the field $7''< r < 33''$.

We conclude that our analysis of the three main uncertainties find the IMF of NGC 3603 to be distinguishably flatter than the Salpeter IMF. But it is still within the fundamental limit in \citet{kro01}.
Therefore, as long as the intrinsic uncertainty by \citet{kro01} accommodates an IMF with a power law index of $\Gamma = -0.74$, we are not yet able to give a conclusive answer to the question of the IMF variation.

\subsection{Is the IMF top heavy in young and massive starburst clusters?}
The above IMF studies suggest that the IMF is almost universal in many different star-forming environments but still with nonnegligible variations in the power law index. Although there is no clear systematic trends in these variations, there is some evidence that these variations are related to the stellar \textit{density}.
Compared to the field IMF, the IMF for intermediate- and high-mass stars tends to be slightly steeper in sparsely populated star-forming regions, while it is sometimes slightly flatter in very dense star clusters.

Among young massive star-forming clusters, we find NGC 3603, the Arches cluster, and the GC cluster with a slightly flatter IMF.
For the Arches cluster \citet{fig99} derive an IMF power law index of $\Gamma \sim -0.65$ for stellar masses down to 10 \msun. \citet{sto05} find the present-day mass function (PDMF) with $\Gamma = -0.86$ for $6 - 60$ \msun. \citet{kim06} derive the PDMF with $\Gamma \sim -0.91$ for $1.3 - 50$ \msun. Using numerical simulations, they also correct for the dynamical evolution of the cluster and trace back the IMF to have a power law index $\Gamma \sim -1.0$ to $-1.1$. Interestingly, their PDMF has a flatter slope ($\Gamma = -0.71$) if only high-mass stars are considered ($5 - 50$ \msun). \citet{dib07} recently reproduce this shallow IMF in the high-mass range in the Arches cluster in their models investigating the coalescence and subsequent collapse of pre-stellar cores in molecular clouds. This suggests a primordial origin of the IMF in the Arches cluster.
Also the young stellar population in the GC cluster exhibits a top heavy IMF. From the observed $K$-band LF of the high-mass stars orbiting the central massive black hole in two counter-rotating disks, \citet{pau06} report that the IMF is likely substantially flatter (by $1 - 1.5$ dex) than the Salpeter IMF. However, these stars probably have formed in very dense gas disks rather than in a self-contracting cloud, and thus a direct comparison with NGC 3603 and other clusters is somewhat problematic.

A common characteristic of these clusters, which always show a slightly flatter IMF, is the high central mass density.
The core mass density of NGC 3603 is at least $\sim6\times10^4$ \msun\ pc$^{-3}$. Note that this value is derived from the volume mass density profile, and the central density directly measured from the surface density (as sometimes given in literature for other clusters) is a factor of $\sim2$ larger (see \S~\ref{s-total-mass}).
The Arches cluster has a total mass of $\sim1\times10^4$ \msun\, similar to that of NGC 3603, and a core mass density $\sim3\times10^5$ \msun\ pc$^{-3}$ \citep{fig99}, several times larger than NGC 3603.
The stellar density in the core of the GC cluster is expected to be much larger due to the presence of the central massive black hole. For example the stellar density in the central arcsecond of the GC is estimated to be $> 3\times10^7$ \msun\ pc$^{-3}$ \citep{gen03}.

In contrast, the central mass densities in the other Galactic or Local Group young stellar clusters are about an order of magnitude smaller than those in NGC 3603 and the Arches cluster.
The ONC, for example, has a core radius of $\sim0.2$ pc, which is similar to NGC 3603, but has an order of magnitude smaller total mass of $\sim10^3$ \msun. It has thus an order of magnitude smaller central mass density of $2\times10^4$ \msun\ pc$^{-3}$ \citep{hil98b}. The IMF of the Trapezium cluster in the ONC follows the Salpeter IMF above a solar mass \citep{mue02}.

An example of a cluster that does not exactly follow this line of argument is the R136 cluster in 30 Dor.
Having a central mass density comparable to those in NGC 3603 and the Arches cluster, the IMF of R136 is only marginally different from a Salpeter-like IMF. But there is still a considerable variation among different studies.
Hosting more than 50 O3-type, WN6, and O/WN6 transition stars \citep{mas98} and, having a total mass of $2 - 3\times10^{4}$ \msun, R136 has been estimated to have a central mass density of $\sim5.5\times10^4$ \msun\ pc$^{-3}$ from stars $\geq2.8$ \msun\ within $r \leq 0.11$ pc, which could be 3 times larger for $\geq0.1$ \msun\ \citep{hun95}. \citet{mac03} derive $3\times10^4$ \msun\ pc$^{-3}$ within $r \leq 0.32$ pc.
As for its IMF, \citet{bra96} derive power law indices of $\Gamma = -1.3$ at $r \leq 0.4$ pc and $\Gamma = -2.2$ at $r > 0.8$ pc, and find an average IMF with $\Gamma = -1.6$ for the high-mass stellar population. \citet{mas98} measure the IMF with $\Gamma = -1.3$ to $-1.4$ for $2.8 - 120$ \msun.
In contrast, \citet{sir00} derive a broken power law IMF with $\Gamma = -1.28$ for $2.1 - 6.5$ \msun\ and $\Gamma = -0.27$ for $1.35 - 2.1$ \msun, indicating a substantially flat IMF over the entire mass range.
Although there is ample evidence for a flattening of the IMF at subsolar masses, the observed flattening at such high masses of $\sim2$ \msun\ in this study is somewhat peculiar.
Thus it might be the case that R136 is in conflict with our hypothesis of the flat trend of the IMF in the dense star clusters. It is not yet possible for us to argue in more detail about the IMF of R136, as a reliable measurement for low- and intermediate-mass stars is more difficult because of its much larger distance than those of the Galactic clusters.

As a summary, although with low statistical significance, the variations of the IMF could be linked to the stellar density of the population.
In particular the most massive starburst clusters like NGC 3603 and the Arches cluster with their dense central regions tend to have a distinguishably flatter IMF compared to the standard Salpeter-like IMF.

\section{Summary and concluding remarks}
\label{c-summary}
Our study is aiming at a fundamental question of current star formation research. Is the IMF universal, or does it vary with environment? To answer this question, we measured the IMF of NGC 3603 -- one of the most massive Galactic star-forming regions -- from our NIR observations with the AO system NACO at the VLT/ESO. In the following we summarize the main results from our study.

\begin{enumerate}
\item \textbf{NIR Photometry}\\
Our very deep, high angular resolution $JHK_{S}L'$-band images obtained by NACO show unprecedented details of the core of the starburst cluster in NGC 3603. Together with the wider field ISAAC $JHK_{S}$ images, we could successfully derive magnitudes and positions of almost 10,000 stars in the dense cluster up to $r \sim 110''$ covering the mass range from the most massive stars down to $\sim0.4$ \msun.
The brightest 256 stars in the NACO images of the inner $r \leq 13''$ are listed and cross-identified with potential counterparts in previous studies.

\item \textbf{Age, extinction, and disk fraction}\\
Based on the fitting of the stellar evolution models in the CMDs and CCDs, we derived the age of the PMS stars of $0.5 - 1.0$ Myr and the upper limit for the age of the MS stars of $\sim2.5$ Myr, suggesting a slight age spread in the cluster. The derived average foreground extinction is $A_{V} = 4.5 \pm 0.5$ mag, and the foreground extinction increases by $\Delta{A_{V}} \sim 2.0$ mag towards larger radii ($r \gtrsim 55''$). Using the $K_{S} - L'$ versus $J - H$ CCD, we derived a circumstellar disk fraction of $\sim25 \pm 10$\% for stars with a mass of $\geq 0.9$ \msun\ in the central cluster ($r \leq 10''$).

\item \textbf{IMF and its radial variation}\\
Applying the field star rejection and the incompleteness correction, the \textit{K}LF for 7514 stars simultaneously detected in the $JHK_{S}$ bands follows a power law with no obvious turnover or truncation within the detection limit of $m_{K_{S}} \sim 17.4$ mag (based on the $J$-band 50\% completeness of $\sim19.4$ and the typical color of $J - K_{S} \sim 2$). Within the mass range of $0.4 - 20$ \msun\, the IMF is well described by a single power law with a power law index $\Gamma \sim -0.74$.
We found the power law index decreasing from $\Gamma \sim -0.3$ at $r \leq 5''$ to $\Gamma \sim -0.8$ at $r \sim 30''$. The strong steepening occurs in the inner $r \lesssim 13''$, pointing towards mass segregation in the very center of the cluster. No significant variation of the IMF is found for larger radii ($r \gtrsim 30''$).

\item \textbf{Size, mass, and dynamical status}\\
Fitting a King model to the radial density profile of stars with a mass of $0.5 - 2.5$ \msun, we derived a core radius of $\sim4''.8$ ($\sim0.14$ pc at $d \sim 6$ kpc).
As the radial density decreases even at the limits of our field of view, we can give a firm lower limit of $r = 110''$ ($\sim3.2$ pc) for the cluster size. We also derive an upper limit of $r = 1260''$ ($\sim37$ pc) for the tidal radius of the cluster.
The de-projected King model allowed us to extrapolate to the total mass of NGC 3603. Assuming a single power law IMF with index $\Gamma = -0.74$ within the mass range of $0.1 - 100$ \msun, we found a total mass of about $1.0 - 1.6 \times 10^4$ \msun. The half-mass radius is found to be within $25'' - 50''$ ($0.7 - 1.5$ pc). The derived core mass density of the cluster is $\geq 6 \times 10^4$ \msun\ pc$^{-3}$.

We estimate a half-mass relaxation time of approximately $10 - 40$ Myr for stars with a typical mass of 1 \msun, an order of magnitude larger than the age of the PMS population in the cluster ($\lesssim1$ Myr). This implies that the intermediate- and low-mass stars have not yet experienced significant dynamical relaxation. However, the relaxation time of the high-mass stars is expected to be an order of magnitude shorter and is comparable to the cluster age. We could thus not conclude if the observed mass segregation of the high-mass stars is caused by dynamical evolution or if it is primordial. Indeed it can be due to the combination of them.
We compute that our images with a maximum radius of $r \sim 110''$ cover at least $\sim67$\% of intermediate- and low-mass stars of NGC 3603. The stars outside the observed field cannot steepen the IMF by more than $\Delta\Gamma \lesssim 0.16$.
Considering also the fact that the IMF does not significantly change beyond $r \gtrsim 30''$, we conclude that the observed IMF is representative for the whole NGC 3603 stellar cluster, irrespective of the mass segregation in the very center.

\item \textbf{Systematic uncertainties of the IMF}\\
We thoroughly analyzed the systematic errors in the IMF determination. In particular we derived the errors from uncertainties in the age, distance, foreground extinction, stellar evolution models, metallicity, individual extinction, incompleteness correction, cluster membership, unresolved binaries, and the stellar mass outside the observed field.
Combining all errors we derived the power law index of $\Gamma = -0.74^{+0.62}_{-0.47}$.
Assuming a Gaussian probability distribution, we conclude that the probability that the IMF is as steep as the Salpeter IMF ($\Gamma = -1.35$) is less than $\sim10$\%.
\end{enumerate}

Our result thus supports the hypothesis of a top-heavy IMF in massive star-forming clusters and starburst galaxies.
One potentially common property among such clusters showing flat IMFs would be the high stellar density in the core of the cluster.

\acknowledgments
We are grateful to the staff of ESO Paranal Observatory for their support in observations.
We thank F. Palla and S. Stahler for providing us with the PMS evolutionary models and isochrones.
We also thank D. N\"{u}rnberger for helpful discussions about NGC 3603.
\facility{VLT:Yepun (NACO)}, \facility{VLT:Antu (ISAAC)}



\appendix

\section{Photometry list}
\label{appendix-photolist}

Thanks to our AO-assisted NACO data with the unprecedented high resolution, we successfully identify objects in the highly crowded cluster center and perform photometry from the brightest WN stars to the faintest stars in subsolar masses.
Here we present a photometry list (Table~6) of $JHK_{S}(L')$-band detected stars for magnitude $m_{J} < 15.5$ mag ($\geq95$\% completeness) in the inner $r \leq 13''$ region (diameter of $\sim0.75$ pc at $d \sim 6$ kpc). This list encompasses all sources within the magnitude range on the projected plane, and any rejections of field stars have not been performed.

The list contains 256 detected sources. Each source has an ID number, a coordinate, $JHK_{S}$ magnitudes, and a $L'$ magnitude in the case of a source detected in the all four bands. The sources are aligned according to increasing $m_{J}$. The coordinates are indicated in their offsets from the cluster center [RA = $11^{h}15^{m}05^{s}.90$, Dec = $-61^{\circ}15'28''.4$ (J2000)]. We note that, since the photometric calibration of the $L'$-band photometry was done by merely one data set of a standard star, the presented $L'$-band magnitude has to be considered as a roughly calibrated photometry. The spectral types are adopted from \citet[][Table~8 therein]{cro98}, which consists of their stellar spectral types combined with those from \citet{mof83} and \citet{dri95}.

In case of sources which are identical or likely to be counterparts to objects in other studies, the designations or IDs are presented. We perform cross-identifications of sources with the \textit{HST} optical detections presented in \citet{mof94} (MDS94), with the ground-based mid-IR detections in \citet{ns03} (NS03), and with the optical detections ($V < 14$ mag) from the \textit{HST} data set in \citet{sun04} (SB04). The identifications of sources are done by means of a visual inspection after making maps of the examining lists. We obtain the impression that there seems to be some degree of position mismatches across fields among the maps due perhaps to differences of instrumental properties such as resolutions, sizes of field of view, and geometric distortions. Thus, we chose to use the eye inspection rather than applying any solid criteria, e.g., a tolerant distance of a coincidence.
As for designations of the counterparts, the source IDs in \citet{mel89} (MTT), as well as the letters A1-3 to F \citep{vdb28,hof86} are simply adopted from \citet{mof94}.
In addition, the X-ray counterparts obtained by the \textit{Chandra} satellite in \citet{mof02} are indicated. We simply adopt the MDS94 sources with X-ray counterparts presented in \citet[][Figure~2 therein]{mof02}. Only bright WR and early O-type stars with X-ray counterparts are presented, and since the authors suggested that many detected X-ray sources might be low-mass PMS members, many more sources in our NACO list are expected to have X-ray counterparts.
As the spatial scale of our NACO NIR data is much smaller than that of the mid-IR data in \citet{ns03}, a single mid-IR source could be superposed with several NIR sources. In such a case, the source that would have a dominant flux contribution in the NIR is presented.



\begin{longtable}{rrrrrrrllll}
\caption{Photometry list}\\
\hline
\hline
Num. & $\Delta$R.A.$^{\mathrm{[a]}}$ & $\Delta$Decl.$^{\mathrm{[a]}}$ &
$J~~~$ &$H~~~$& $K_{S}~~$ & $L'~~$ & Type$^{\mathrm{[b]}}$ & MDS94$^{\mathrm{[c]}}$ & NS03$^{\mathrm{[d]}}$ & SB04$^{\mathrm{[e]}}$ \\
\hline
\endhead
\hline 
\endfoot
       1 &   0.53& -0.39 &   7.78 &   7.70 &   7.08 &   6.45 &WN6h+abs &H23(B) & &10582 \\
       2 &  -0.23& -0.24 &   7.98 &   7.79 &   7.21 &   6.75 &WN6h+abs &H$^{\mathrm{[c9]}}$30(A1)  & &10558 \\
       3 &   1.83&  0.18 &   8.49 &   8.13 &   7.81 &   7.49 &WN6h+abs &H18(C)  & &10635 \\
       4 &  -0.17& -0.60 &   9.38 &   8.95 &   8.78 &   8.80 &O3 V &H$^{\mathrm{[c9]}}$31(A2)  & &10559 \\
       5 &   0.11& -0.26 &   9.68 &   9.45 &   9.31 &   9.32 &O3 III &H$^{\mathrm{[c9]}}$26(A3)  & &10570 \\
       6 &   6.20& -9.19 &   9.94 &   9.64 &   9.45 &      - &O5 III(f) &H22(17) & & \\
       7 &  -2.00& -1.10 &   9.95 &   9.51 &   9.31 &   9.19 &O3 III &H42 & &10478 \\
       8 &  -3.67&  2.57 &  10.09 &   9.64 &   9.37 &   8.93 &O5.5 III(f) &H39(6/F)  & &10392 \\
       9 &  -0.40&  3.26 &  10.16 &   9.76 &   9.54 &   9.34 &O5 V &H19(E)  & &10543 \\
      10 &  -2.98& -0.78 &  10.24 &   9.78 &   9.54 &   9.45 &O4 V &H$^{\mathrm{[c10]}}$49(D$^{\mathrm{[c5]}}$) & &10422 \\
      11 &  -1.53& -0.90 &  10.25 &   9.84 &   9.65 &   9.60 &O3 V &H40  & &10498 \\
      12 &   1.15& -8.19 &  10.29 &   9.67 &   9.64 &   9.78 &O4 V(f) &H51(23)  & &10616 \\
      13 &  -2.95& -1.01 &  10.30 &   9.97 &   9.75 &   9.65 &O5 V &H$^{\mathrm{[c10]}}$50(D$^{\mathrm{[c5]}}$)  & &10423 \\
      14 &   3.52&  0.35 &  10.33 &   9.92 &   9.78 &   9.74 &O3 V &H16  & &10706 \\
      15 &  -0.63&  1.06 &  10.40 &  10.02 &   9.85 &   9.83 &O4 V &25  & &10534 \\
      16 &  -3.02&  1.60 &  10.54 &  10.14 &   9.93 &   9.73 &O3 V &38  & &10419 \\
      17 &  -3.58& -6.51 &  10.59 &  10.20 &   9.94 &   9.78 &O4 V &H62(15$^{\mathrm{[c3]}}$) &6N$^{\mathrm{[d3]}}$ &10396 \\
      18 &   8.50& -0.13 &  10.74 &  10.11 &  10.13 &  10.18 &O4 V &H9(21$^{\mathrm{[c4]}}$)  &6K$^{\mathrm{[d2]}}$ &10895 \\
      19 &  -7.98&  1.62 &  10.90 &  10.26 &  10.25 &   9.76 &O5 V &H61$^{\mathrm{[c1]}}$(10/G)  & &10207$^{\mathrm{[e1]}}$ \\
      20 &   2.91& -2.95 &  10.95 &  10.52 &  10.41 &  10.49 &O4 V &20  &6G$^{\mathrm{[d1]}}$ & \\
      21 &   3.05&  3.01 &  10.98 &  10.58 &  10.44 &  10.33 &O4 V &H14  & &10683 \\
      22 &  -1.96& -7.15 &  11.08 &  10.70 &  10.47 &  10.43 &O5.5 V &58(15)$^{\mathrm{[c3]}}$ &6N$^{\mathrm{[d3]}}$ & \\
      23 &   1.86&  2.24 &  11.13 &  10.73 &  10.57 &  10.50 &O4 V &17  & & \\
      24 &  -4.26& -4.24 &  11.19 &  10.78 &  10.52 &  10.39 &O4 V &H60  & &10363 \\
      25 &   3.53& 10.28 &  11.20 &  10.71 &  10.52 &   9.97 &O4 V &H7(26)  & &10705 \\
      26 &  -2.10& -0.75 &  11.22 &  10.81 &  10.64 &  10.62 &O4 V &41  & & \\
      27 &  -0.92& -3.32 &  11.27 &  10.83 &  10.67 &  10.75 & &46  & & \\
      28 &  -2.84& -0.10 &  11.33 &  10.92 &  10.75 &  10.64 &O8 V-III &45  & & \\
      29 &  -0.76&  0.97 &  11.37 &  10.78 &  10.59 &  10.71 &O4 V &27  & & \\
      30 &  -5.41& -2.20 &  11.37 &  10.88 &  10.66 &  10.45 &O4 V &59  & &10319 \\
      31 &  -2.40&  0.75 &  11.39 &  10.98 &  10.80 &  10.70 &O6.5 V+? &37  & & \\
      32 &  -4.29&  2.57 &  11.39 &  10.94 &  10.73 &  10.42 &O4 V &43  & &10361 \\
      33 &  -3.17& -1.15 &  11.39 &  11.10 &  10.99 &  10.87 &O4 V &52(D$^{\mathrm{[c5]}}$) & & \\
      34 &   3.97& -2.35 &  11.52 &  10.39 &   9.52 &   8.31 & &  &6G$^{\mathrm{[d1]}}$ & \\
      35 &   0.98& -2.03 &  11.55 &  11.09 &  10.95 &  11.10 & &28  & & \\
      36 &  -5.00& -1.16 &  11.56 &  11.11 &  10.86 &  10.65 &O4 V &57  & & \\
      37 &  -1.86&  0.65 &  11.66 &  11.26 &  11.09 &  11.02 & &36  & & \\
      38 &   2.10&  7.97 &  11.67 &  10.82 &  10.71 &  10.45 &O4 V &(H7b)10(22$^{\mathrm{[c7]}}$)  & & \\
      39 &   5.58& -2.11 &  11.68 &  11.28 &  11.16 &  11.19 & &68(44$^{\mathrm{[c6]}}$)  & & \\
      40 &  -1.56&  3.85 &  11.76 &  11.36 &  11.13 &  10.91 & &  & & \\
      41 &   9.79& -0.83 &  11.79 &  11.27 &  11.25 &  11.20 & &8(21$^{\mathrm{[c4]}}$)  &6K$^{\mathrm{[d2]}}$ & \\
      42 &  -1.25& -2.19 &  11.81 &  11.36 &  11.19 &  11.26 & &77  & & \\
      43 &   9.82& -6.37 &  11.83 &  12.90 &  13.04 &  11.43 & &12(30$^{\mathrm{[c8]}}$)  & & \\
      44 &   1.06&  0.14 &  11.88 &  11.59 &  11.44 &  11.48 & &  & & \\
      45 &  -8.05&  1.65 &  11.97 &  12.05 &  12.62 &      - & &H61$^{\mathrm{[c1]}}$(10/G)  & &10207$^{\mathrm{[e1]}}$ \\
      46 &  -4.50&  5.76 &  11.98 &  11.57 &  11.26 &  10.80 & &34  & & \\
      47 &   1.02& -0.55 &  11.99 &  11.25 &  10.32 &   9.14 & &  & & \\
      48 &   2.48& -5.59 &  12.02 &  11.59 &  11.43 &  11.54 & &70  &6M & \\
      49 &  10.30& -7.35 &  12.07 &  11.77 &  11.67 &  11.80 & &13(30$^{\mathrm{[c8]}}$)  & & \\
      50 &   2.15&  6.88 &  12.10 &  11.56 &  11.28 &  10.99 & &11(22$^{\mathrm{[c7]}}$)  & & \\
      51 &  -3.50&  2.27 &  12.10 &  11.78 &  11.57 &  11.33 & &  & & \\
      52 &  -4.71& -0.64 &  12.12 &  11.65 &  11.44 &  11.27 & &56  & & \\
      53 &  -3.79&  4.28 &  12.12 &  11.71 &  11.45 &  11.11 & &35  & & \\
      54 &   2.97& -3.15 &  12.14 &  11.66 &  11.50 &  11.55 & &21  &6G$^{\mathrm{[d1]}}$ & \\
      55 &  11.87& -4.06 &  12.16 &  11.58 &  11.63 &  11.51 & &67$^{\mathrm{[c2]}}$(36)  & & \\
      56 &   1.86&  5.56 &  12.19 &  11.82 &  11.57 &  11.32 & &  & & \\
      57 &  11.76& -3.63 &  12.19 &  11.85 &  11.29 &  10.30 & &67$^{\mathrm{[c2]}}$(36) & & \\
      58 &   2.58&  3.55 &  12.20 &  11.78 &  11.56 &  11.42 & &15  & & \\
      59 &   3.12& -8.35 &  12.21 &  11.63 &  11.63 &  11.85 & &71  & & \\
      60 &  -3.10& -1.31 &  12.24 &  11.74 &  11.61 &  11.42 & &53(D$^{\mathrm{[c5]}}$)  & & \\
      61 &  -1.63&  3.68 &  12.28 &  11.87 &  11.63 &  11.42 & &  & & \\
      62 &  -2.75& -1.93 &  12.28 &  11.88 &  11.70 &  11.62 & &54  & & \\
      63 &   5.85& -2.93 &  12.28 &  11.86 &  11.69 &  11.71 & &69(44$^{\mathrm{[c6]}}$)  & & \\
      64 &   0.65& -5.18 &  12.30 &  11.89 &  11.74 &  11.85 & &  & & \\
      65 &  -4.41& -1.06 &  12.32 &  11.94 &  11.70 &  11.54 & &55  & & \\
      66 &   2.07& -9.93 &  12.33 &  12.00 &  11.94 &  12.13 & &72  & & \\
      67 &   6.72&  2.80 &  12.38 &  13.05 &  11.87 &  11.62 & &  & & \\
      68 &  -0.71& -0.48 &  12.38 &  11.67 &  10.81 &   9.33 & &  & & \\
      69 &   2.30& -0.08 &  12.40 &  11.96 &  11.23 &  10.04 & &  & & \\
      70 &  -2.19&  0.47 &  12.42 &  12.66 &  12.48 &  12.38 & &  & & \\
      71 &  -1.24& -2.42 &  12.43 &  11.92 &  11.74 &  11.82 & &44  & & \\
      72 &   5.33&  3.95 &  12.54 &  12.12 &  11.92 &  11.72 & &73  & & \\
      73 &   2.31&  5.25 &  12.54 &  12.75 &  12.48 &  12.22 & &  & & \\
      74 &  -5.83& -4.45 &  12.55 &  12.95 &  12.70 &  12.45 & &  & & \\
      75 &  11.55& -0.93 &  12.58 &  11.92 &  12.39 &  11.00 & &  & & \\
      76 &  -6.58&  1.79 &  12.60 &  12.15 &  11.86 &  11.30 & &  & & \\
      77 &  -2.14& -9.15 &  12.61 &  12.02 &  11.92 &  12.94 & &  & & \\
      78 &  -6.29& -9.75 &  12.65 &  12.13 &  12.71 &  12.76 & &  &6O & \\
      79 &  -0.75&  0.76 &  12.69 &  12.56 &  12.28 &  11.51 &O4 V &29  & & \\
      80 &   7.42& -3.51 &  12.74 &  12.13 &  12.09 &  12.20 & &  & & \\
      81 &  -3.28& -0.69 &  12.75 &  12.37 &  11.95 &  10.92 & &  & & \\
      82 &   4.37& -1.57 &  12.78 &  12.68 &  12.56 &  12.50 & &  & & \\
      83 &  -0.38& -2.13 &  12.82 &  12.36 &  12.21 &  12.32 & &  & & \\
      84 &  -8.29&  2.86 &  12.86 &  12.26 &  12.54 &  12.08 & &  & & \\
      85 &  -3.85&  2.88 &  12.89 &  12.55 &  12.16 &  11.65 & &  & & \\
      86 &  -0.39& -8.19 &  12.89 &  12.42 &  11.66 &  10.35 & &  & & \\
      87 &   0.27&  5.74 &  12.90 &  11.88 &  10.85 &   9.19 & &  & & \\
      88 &   9.44&  3.04 &  12.95 &  12.49 &  12.05 &  10.34 & &  & & \\
      89 &  -3.30&  6.03 &  12.96 &  12.81 &  12.38 &  11.91 & &  & & \\
      90 &  -0.04& -2.11 &  12.98 &  12.51 &  12.35 &  12.49 & &  & & \\
      91 &  -0.59& -4.58 &  12.98 &  12.53 &  12.36 &  12.43 & &  & & \\
      92 &  -2.27& -0.08 &  12.99 &  12.51 &  12.35 &  12.27 & &  & & \\
      93 &  -6.33&  2.51 &  13.03 &  12.91 &  12.64 &  12.25 & &  & & \\
      94 &  -3.04&  9.37 &  13.07 &  13.09 &  10.94 &   9.18 & &  &6J & \\
      95 &   2.45& -9.33 &  13.21 &  13.23 &  13.02 &  12.79 & &  & & \\
      96 &  -7.76& -7.89 &  13.23 &  12.64 &  12.36 &  12.13 & &  & & \\
      97 &  -3.41&  5.40 &  13.28 &  12.82 &  12.71 &  12.25 & &  & & \\
      98 &  -0.49& -6.12 &  13.29 &  12.81 &  12.49 &  12.42 & &  & & \\
      99 &   8.98&  3.09 &  13.30 &  12.50 &  12.79 &  12.61 & &  & &  \\
     100 &  -3.69&  0.73 &  13.30 &  12.90 &  12.65 &  12.48 & &  & &  \\
     101 &   7.92& -0.45 &  13.33 &  12.98 &  12.68 &  12.59 & &  & &  \\
     102 &  -6.74& -5.69 &  13.35 &  12.56 &  14.21 &  11.29 & &  & &  \\
     103 &   3.05&  2.88 &  13.40 &  12.05 &  11.13 &   9.50 & &  & &  \\
     104 &   6.73&  6.26 &  13.45 &  12.51 &  12.43 &  12.48 & &  & &  \\
     105 &   9.12&  1.00 &  13.45 &  12.87 &  12.62 &  12.77 & &  & &  \\
     106 &  -7.25& -7.92 &  13.48 &  12.58 &  12.02 &  10.64 & &  & &  \\
     107 &   6.40&  5.96 &  13.50 &  13.20 &  12.98 &  14.29 & &  & &  \\
     108 &  -2.24& -2.34 &  13.53 &  12.78 &  12.59 &  12.59 & &  & &  \\
     109 &  -8.53& -1.49 &  13.53 &  13.13 &  12.56 &  11.13 & &  & &  \\
     110 &  -2.61& -6.65 &  13.55 &  12.77 &  12.49 &  12.38 & &  & &  \\
     111 &   2.98& -2.72 &  13.56 &  13.16 &  12.98 &  12.95 & &  & &  \\
     112 &   0.33& -3.62 &  13.57 &  12.53 &  11.91 &  11.59 & &  & &  \\
     113 &  -3.77&  0.65 &  13.62 &  12.99 &  12.58 &  11.96 & &  & &  \\
     114 &   2.09& -2.07 &  13.64 &  12.87 &  12.72 &  12.87 & &  & &  \\
     115 &   1.23& -2.04 &  13.64 &  13.28 &  13.18 &  13.15 & &  & &  \\
     116 &  -7.62&  7.08 &  13.68 &  13.23 &  12.78 &  12.74 & &  & &  \\
     117 &   3.56& -4.81 &  13.70 &  13.07 &  12.38 &  11.06 & &  & &  \\
     118 &   1.13& -0.94 &  13.70 &  12.79 &  12.54 &  12.25 & &  & &  \\
     119 &  -0.71&  2.10 &  13.70 &  13.19 &  13.00 &  12.95 & &  & &  \\
     120 &   3.76&  0.56 &  13.71 &  14.18 &  14.08 &  13.70 & &  & &  \\
     121 &   2.96& -0.67 &  13.71 &  12.99 &  12.69 &  12.64 & &  & &  \\
     122 &  -4.76& -0.96 &  13.77 &  12.74 &  11.83 &  10.35 & &  & &  \\
     123 &  -2.79&  2.21 &  13.78 &  12.96 &  12.35 &  10.78 & &  & &  \\
     124 &  -9.19&  4.22 &  13.81 &  13.55 &  13.32 &  12.74 & &  & &  \\
     125 &  -1.86&  3.39 &  13.82 &  13.40 &  13.18 &  12.90 & &  & &  \\
     126 &   9.54&  4.20 &  13.86 &  13.34 &  13.14 &  13.44 & &  & &  \\
     127 &   6.97&  0.85 &  13.88 &  12.96 &  12.84 &  12.70 & &  & &  \\
     128 &   1.15&  9.82 &  13.89 &  13.27 &  12.79 &  12.63 & &  & &  \\
     129 &  -6.03&  9.27 &  13.93 &  15.01 &  13.92 &  13.09 & &  & &  \\
     130 &   5.22&  6.60 &  13.94 &  13.17 &  13.12 &  13.21 & &  & &  \\
     131 &  12.32& -2.12 &  13.98 &  13.39 &  13.08 &  13.29 & &  & &  \\
     132 &  10.05&  0.91 &  14.05 &  13.65 &  13.52 &  13.80 & &  & &  \\
     133 &   1.81& -6.02 &  14.06 &  13.76 &  13.42 &  13.14 & &  & &  \\
     134 &  -5.62&  4.22 &  14.06 &  13.66 &  13.32 &  13.22 & &  & &  \\
     135 &  -5.76&  2.45 &  14.08 &  13.56 &  13.25 &  12.90 & &  & &  \\
     136 &  -5.24& -2.54 &  14.12 &  13.50 &  13.18 &  12.84 & &  & &  \\
     137 &   2.23& -5.52 &  14.13 &  12.88 &  12.74 &  12.88 & &  & &  \\
     138 &  -6.69&  1.80 &  14.16 &  14.30 &  14.54 & - & &  & & \\
     139 &   3.51& -0.56 &  14.18 &  14.38 &  13.86 &  13.32 & &  & & \\
     140 &  -4.37& 12.12 &  14.19 &  15.22 &  13.79 &  13.36 & &  & &  \\
     141 &  -3.84& -2.76 &  14.22 &  13.90 &  13.62 &  12.97 & &  & &  \\
     142 &   1.25&  1.39 &  14.22 &  13.49 &  13.26 &  13.03 & &  & &  \\
     143 &  -1.61& -1.42 &  14.24 &  14.81 &  12.87 &  12.80 & &  & &  \\
     144 &  -6.03&-10.39 &  14.31 &  13.50 &  13.19 &  13.24 & &  & &  \\
     145 &  -1.34&-10.91 &  14.32 &  13.37 &  12.82 &  12.88 & &  & &  \\
     146 &   4.06& -2.83 &  14.38 &  13.64 &  13.36 &  12.99 & &  & &  \\
     147 &  -6.83& -5.71 &  14.39 &  14.62 &  14.79 & - & &  & &  \\
     148 &   4.69&  3.48 &  14.40 &  13.76 &  13.45 &  12.93 & &  & &  \\
     149 &  -8.39&  2.87 &  14.42 &  14.23 &  14.84 & - & &  & &  \\
     150 &   3.74& -7.34 &  14.45 &  13.77 &  13.58 &  13.40 & &  & &  \\
     151 &  -2.78& -4.39 &  14.46 &  13.96 &  13.70 &  13.40 & &  & &  \\
     152 &  -4.72&  6.05 &  14.49 &  14.10 &  13.72 &  13.13 & &  & &  \\
     153 &   0.50& -1.39 &  14.49 &  13.87 &  13.72 &  13.30 & &  & &  \\
     154 &   1.23& -0.88 &  14.50 &  13.95 &  13.73 &  13.37 & &  & &  \\
     155 &   3.49&  2.79 &  14.50 &  14.10 &  13.96 &  13.66 & &  & &  \\
     156 &   8.84& -8.60 &  14.53 &  14.30 &  14.01 &  14.09 & &  & &  \\
     157 &   7.24&  7.91 &  14.53 &  13.69 &  13.97 &  13.92 & &  & &  \\
     158 &  -2.67& 10.39 &  14.53 &  14.49 &  13.90 &  13.42 & &  & &  \\
     159 &  -2.59&  4.01 &  14.54 &  14.69 &  14.12 &  13.30 & &  & &  \\
     160 &  -4.36& -3.96 &  14.56 &  14.03 &  13.71 &  13.70 & &  & &  \\
     161 &  -8.21& -3.34 &  14.56 &  13.75 &  13.53 &  13.04 & &  & &  \\
     162 &  -1.02&  0.02 &  14.56 &  13.70 &  13.43 &  13.27 & &  & &  \\
     163 &  -5.92& -4.47 &  14.57 &  15.44 &  15.39 & - & &  & & \\
     164 &   2.72& -4.73 &  14.59 &  14.25 &  13.92 &  13.61 & &  & &  \\
     165 &   9.24&  8.05 &  14.60 &  14.28 &  14.07 &  13.53 & &  & &  \\
     166 &  -6.06&  5.64 &  14.61 &  13.91 &  13.64 &  13.23 & &  & &  \\
     167 &  -2.55&  8.23 &  14.62 &  13.79 &  13.59 &  13.04 & &  & &  \\
     168 &  -3.46&  6.11 &  14.62 &  14.31 &  14.06 &  13.56 & &  & &  \\
     169 &   7.81& -8.48 &  14.63 &  14.21 &  13.83 &  13.62 & &  & &  \\
     170 &  -9.68&  3.60 &  14.63 &  13.93 &  13.63 &  13.64 & &  & &  \\
     171 &  -0.06&-10.27 &  14.65 &  14.80 &  14.40 &  14.23 & &  & &  \\
     172 &   8.34&  4.23 &  14.65 &  13.95 &  13.86 &  14.07 & &  & &  \\
     173 &  10.40& -1.83 &  14.67 &  14.68 &  14.43 &  14.04 & &  & &  \\
     174 &   1.93& -5.90 &  14.70 &  14.31 &  14.02 &  13.66 & &  & &  \\
     175 &  -0.24&  1.86 &  14.70 &  13.86 &  13.54 &  13.12 & &  & &  \\
     176 &   6.43& -4.20 &  14.71 &  13.96 &  13.49 &  13.03 & &  & &  \\
     177 &  -7.26& -6.42 &  14.71 &  13.90 &  13.62 &  13.53 & &  & &  \\
     178 & -10.72&  0.81 &  14.72 &  13.54 &  13.25 &  13.14 & &  & &  \\
     179 &   5.03&  4.13 &  14.76 &  13.96 &  13.71 &  13.38 & &  & &  \\
     180 &  -2.64& -2.11 &  14.77 &  14.03 &  13.64 &  13.24 & &  & &  \\
     181 &   4.12& -4.33 &  14.78 &  13.94 &  13.61 &  13.32 & &  & &  \\
     182 &   0.20&  0.44 &  14.78 &  14.27 &  14.04 &  13.57 & &  & &  \\
     183 &  -5.47& -5.42 &  14.80 &  14.40 &  14.08 &  13.73 & &  & &  \\
     184 &   0.87& -4.62 &  14.84 &  14.35 &  13.94 &  13.43 & &  & &  \\
     185 &   0.78& -4.06 &  14.84 &  14.37 &  13.93 &  13.37 & &  & &  \\
     186 &  -5.82&  6.25 &  14.84 &  14.21 &  13.96 &  13.56 & &  & &  \\
     187 &  -4.17& -5.15 &  14.88 &  14.05 &  13.62 &  13.33 & &  & &  \\
     188 &  -2.38& -1.66 &  14.88 &  14.34 &  14.07 &  14.01 & &  & &  \\
     189 &  -1.93& 10.03 &  14.91 &  14.73 &  14.54 &  14.24 & &  & &  \\
     190 &  -7.51&  7.05 &  14.91 &  15.04 &  15.72 & - & &  & & \\
     191 & -11.80& -0.93 &  14.91 &  14.15 &  13.87 &  13.92 & &  & &  \\
     192 &   1.36&  3.06 &  14.91 &  13.96 &  13.66 &  13.26 & &  & &  \\
     193 &  -7.80& -5.42 &  14.92 &  14.04 &  13.76 &  13.71 & &  & &  \\
     194 &   3.90&  7.44 &  14.93 &  13.86 &  13.76 &  13.39 & &  & &  \\
     195 &   2.16&-12.24 &  14.94 &  16.54 &  13.56 &  13.67 & &  & &  \\
     196 &  -1.75& -4.84 &  14.95 &  14.91 &  14.77 &  13.42 & &  & &  \\
     197 &  -9.76&  6.64 &  14.98 &  16.41 &  16.92 & - & &  & &  \\
     198 &  12.03&  0.17 &  15.02 &  14.60 &  14.42 &  14.59 & &  & & \\
     199 &   0.61&-12.82 &  15.04 &  14.58 &  14.28 & - & &  & &  \\
     200 &  -9.15& -3.56 &  15.05 &  14.03 &  13.50 &  13.50 & &  & &  \\
     201 & -10.89&  5.90 &  15.06 &  14.93 &  14.60 &  14.05 & &  & &  \\
     202 &   9.64& -4.31 &  15.06 &  14.44 &  14.24 &  14.03 & &  & &  \\
     203 &  -6.44&  2.52 &  15.06 &  15.15 &  15.64 & - & &  & &  \\
     204 &  -2.08&  9.08 &  15.07 &  14.40 &  14.00 &  12.48 & &  & &  \\
     205 &  -9.83&  2.76 &  15.08 &  14.86 &  14.25 &  14.26 & &  & &  \\
     206 &   0.07&  2.31 &  15.08 &  14.60 &  14.41 &  14.26 & &  & &  \\
     207 &   9.29&  6.48 &  15.09 &  14.53 &  14.29 &  14.58 & &  & &  \\
     208 &   2.36& -2.98 &  15.10 &  14.32 &  13.95 &  13.52 & &  & &  \\
     209 &   9.23& -6.76 &  15.10 &  14.78 &  14.39 &  14.14 & &  & &  \\
     210 &  -0.86& -0.69 &  15.13 &  14.56 &  14.52 & - & &  & &  \\
     211 &  -1.70& -2.53 &  15.13 &  14.63 &  14.17 &  13.41 & &  & &  \\
     212 &  -1.21&  1.78 &  15.13 &  14.58 &  14.41 &  14.22 & &  & &  \\
     213 &   1.46& -4.57 &  15.14 &  15.25 &  14.48 &  13.88 & &  & &  \\
     214 &   1.65&  1.24 &  15.15 &  14.69 &  14.42 &  14.26 & &  & &  \\
     215 &  -3.13&  8.19 &  15.15 &  16.16 &  14.28 &  13.62 & &  & &  \\
     216 &  -3.96&  5.85 &  15.16 &  14.16 &  13.83 &  13.39 & &  & &  \\
     217 &  -0.30&  9.25 &  15.16 &  14.98 &  14.64 &  14.37 & &  & &  \\
     218 &  -9.11& -0.36 &  15.16 &  14.86 &  14.56 &  14.30 & &  & &  \\
     219 &  -9.69&  6.57 &  15.17 &  15.08 &  14.69 &  14.03 & &  & &  \\
     220 &  -3.83& -9.70 &  15.17 &  14.17 &  13.69 &  13.71 & &  & &  \\
     221 &  -4.38& -0.30 &  15.18 &  14.50 &  14.38 &  13.79 & &  & &  \\
     222 &   0.51& -6.21 &  15.18 &  14.40 &  13.99 &  13.60 & &  & &  \\
     223 &   7.64& -7.76 &  15.20 &  14.61 &  14.36 &  14.07 & &  & &  \\
     224 &  -8.11& -2.49 &  15.20 &  14.62 &  14.48 &  13.84 & &  & &  \\
     225 &  -2.53& -3.43 &  15.21 &  14.68 &  14.39 &  13.84 & &  & &  \\
     226 &  -0.16& -2.65 &  15.21 &  14.57 &  14.15 &  13.69 & &  & &  \\
     227 &  -1.50& -4.39 &  15.22 &  15.24 &  14.77 &  14.47 & &  & &  \\
     228 &  -0.68&  1.64 &  15.24 &  14.81 &  14.57 &  13.90 & &  & &  \\
     229 &   2.25&  2.85 &  15.25 &  14.15 &  13.86 &  13.59 & &  & &  \\
     230 &   8.33&  1.04 &  15.25 &  13.96 &  13.76 &  13.47 & &  & &  \\
     231 &  -7.38&  0.41 &  15.26 &  14.72 &  14.23 &  13.44 & &  & &  \\
     232 &   2.46&  0.58 &  15.27 &  15.02 &  14.63 &  14.28 & &  & &  \\
     233 &   8.80&  8.29 &  15.27 &  14.57 &  13.90 &  13.84 & &  & &  \\
     234 &  -0.20&  2.11 &  15.28 &  14.91 &  14.51 &  14.04 & &  & &  \\
     235 &   8.65& -2.30 &  15.29 &  14.71 &  14.46 &  14.19 & &  & &  \\
     236 &  -1.33&  4.92 &  15.29 &  14.69 &  14.29 &  13.73 & &  & &  \\
     237 &   3.77&  4.45 &  15.30 &  15.16 &  14.70 &  14.27 & &  & &  \\
     238 &   0.70&  6.96 &  15.31 &  15.29 &  14.75 &  14.37 & &  & &  \\
     239 &  -7.64&  5.60 &  15.31 &  14.83 &  14.64 &  14.08 & &  & &  \\
     240 &  -7.80&  4.68 &  15.31 &  14.90 &  14.37 &  14.38 & &  & &  \\
     241 &   3.72& -0.75 &  15.32 &  14.51 &  14.23 &  14.11 & &  & &  \\
     242 &  -5.70&-11.08 &  15.32 &  15.02 &  13.66 &  13.76 & &  & &  \\
     243 &   2.67&-11.71 &  15.35 &  14.83 &  14.34 &  14.12 & &  & &  \\
     244 &  12.79&  0.97 &  15.36 &  14.88 &  14.67 &  14.84 & &  & &  \\
     245 &  -3.25&  8.73 &  15.36 &  14.95 &  14.65 &  14.13 & &  & &  \\
     246 &  -7.87& -5.45 &  15.37 &  16.42 &  15.82 & - & &  & & \\
     247 &  -9.85&  3.17 &  15.37 &  14.38 &  14.07 &  13.87 & &  & &  \\
     248 &  -5.18&  0.53 &  15.39 &  14.99 &  14.78 &  14.06 & &  & &  \\
     249 &  -1.78&  9.00 &  15.39 &  14.93 &  14.69 &  13.84 & &  & &  \\
     250 &  -4.63& -2.72 &  15.40 &  14.95 &  14.47 &  13.63 & &  & &  \\
     251 &  -8.13& 10.08 &  15.43 &  15.25 &  14.65 & - & &  & &  \\
     252 &   4.09& -1.10 &  15.44 &  14.54 &  14.21 &  13.82 & &  & &  \\
     253 &  -1.76& -5.11 &  15.45 &  15.06 &  14.60 &  14.18 & &  & &  \\
     254 &  -3.80&  7.38 &  15.47 &  14.90 &  14.46 &  13.90 & &  & &  \\
     255 &  -7.33&-10.35 &  15.49 &  14.79 &  14.25 &  14.29 & &  & &  \\
     256 &   5.63&  4.43 &  15.50 &  15.00 &  14.68 &  13.91 & &  & &  \\
\hline
\end{longtable}
\scriptsize
\tablecomments{Photometry and counterparts of 256 $JHK_{S}(L')$-detected sources brighter than $m_{J} = 15.5$ ($\sim$95\% completeness) in the central region $r \leq 13''$ of the star-forming cluster in NGC 3003.}
\begin{list}{}{}
\item[$^{\mathrm{[a]}}$] Source location is shown in an offset from the center of the cluster (RA = 11h15m05.90s, Dec = $-61^{\circ}$15$'$28.4$''$ [J2000]). 
\item[$^{\mathrm{[b]}}$] Spectral type is adopted from a list in \citet[][Table~8 therein]{cro98}, which consists of spectral types in their study combined with those in \citet{mof83} and in \citet{dri95}.
\item[$^{\mathrm{[c]}}$] Designated IDs of potential counterparts in \textit{HST} optical detections by \citet{mof94}. A source with prefix H is found to have a X-ray counterpart by \citet{mof02}. IDs originated in previous works such as \citet{mel89} (in parentheses), and letters were adopted from tables in \citet{mof94}. The letters A-F originated in \citet{vdb28} (G in Walborn 1973), and the source A was resolved into A1-3 by \citet{hof86}.
\item[$^{\mathrm{[c1]}}$] Star 21 and 52 together would correspond to MDS-61 (G/MTT-10).
\item[$^{\mathrm{[c2]}}$] Star 73 and 74 together would correspond to MDS-67 (MTT-36).
\item[$^{\mathrm{[c3]}}$] Multiple detections. MDS-58 and 62 correspond to MTT-15. 
\item[$^{\mathrm{[c4]}}$] MDS-8 and 9 correspond to MTT-21. 
\item[$^{\mathrm{[c5]}}$] MDS-49, 50, 52, and 53 correspond to D. 
\item[$^{\mathrm{[c6]}}$] MDS-68 and 69 correspond to MTT-44. 
\item[$^{\mathrm{[c7]}}$] MDS-10 and 11 correspond to MTT-22. 
\item[$^{\mathrm{[c8]}}$] MDS-12 and 13 correspond to MTT-30. 
\item[$^{\mathrm{[c9]}}$] A1, A2, and A3 together consist of a X-ray source within the match of the smoothed PSF in \citet{mof02}.
\item[$^{\mathrm{[c10]}}$] MDS-49 and 50 together consist of a X-ray source.
\item[$^{\mathrm{[d]}}$] Designated IDs of potential counterparts in mid-infrared detections by \citet{ns03}. The 11.9 $\mu$m sources 6A-F are in the concentrated cluster core. 
\item[$^{\mathrm{[d1]}}$] Star 19, 33 and 51 together would correspond to 6G.
\item[$^{\mathrm{[d2]}}$] Star 18 and 43 together would correspond to 6K.
\item[$^{\mathrm{[d3]}}$] Star 17 and 22 together would correspond to 6N. 
\item[$^{\mathrm{[e]}}$] IDs in \citet[][Table~2 therein]{sun04}. Sources brighter than $V$ = 14 mag detected in their analysis of the \textit{HST} PC1 archival data set.
\item[$^{\mathrm{[e1]}}$] Star 21 and 52 together would correspond to 10207.
\end{list}
\label{tbl-photometry-list}


\clearpage


\begin{thebibliography}{}
\bibitem[Armitage et al.(2003)]{arm03} Armitage P. J., Clarke C. J., Palla F. 2003, \mnras, 342, 1139
\bibitem[Ascenso et al.(2007)]{asc07} Ascenso, J., Alves, J., Beletsky, Y., Lago, M. T. V. T. 2007, \aap, 466, 137
\bibitem[Balick et al.(1980)]{bal80} Balick, B., Boeshaar, G. O., \& Gull, T. R. 1980, \apj, 242, 584
\bibitem[Baraffe et al.(1998)]{bar98} Baraffe, I., Chabrier, G., Allard, F., \& Hauschildt, P. H.  1998, \aap, 337, 403
\bibitem[Baraffe et al.(2002)]{bar02} Baraffe, I., Chabrier, G., Allard, F., \& Hauschildt, P. H.  2002, \aap, 382, 563
\bibitem[Barlow(2003)]{bar03} Barlow, R. J. 2003, preprint (arXiv:physics/0306168)
\bibitem[Bertelli et al.(1994)]{ber94} Bertelli, G., Bressan, A., Chiosi, C., Fagotto, F., \& Nasi, E. 1994, \aaps, 106, 275
\bibitem[Bessell \& Brett(1988)]{bes88} Bessell, M. S., \& Brett, J. M. 1988, \pasp, 100, 1134
\bibitem[Binney \& Tremaine(1987)]{bin87} Binney, J., Tremaine, S. 1987, Galactic Dynamics, Princeton University Press, 514
\bibitem[Bonatto et al.(2004)]{bon04} Bonatto, Ch., Bica, E., \& Girardi, L. 2004, \aap, 415, 571
\bibitem[Bonnell \& Davies(1998)]{bon98}Bonnell, I. A., \& Davies, M. B. 1998, \mnras, 295, 691
\bibitem[Bouvier et al.(1998)]{bou98} Bouvier, J., Stauffer, J. R., Martin, E. L., Barrado y Navascues, D., Wallace, B., \& Bejar, V. J. S. 1998, \aap, 336, 490
\bibitem[Brandl et al.(1996)]{bra96} Brandl, B., Sams, B. J., Bertoldi, F., et al. 1996, \apj, 466, 254
\bibitem[Brandl et al.(1999)]{bra99} Brandl, B., Brandner, W., Eisenhauer, F., Moffat, A. F. J., Palla, F., \& Zinnecker, H. 1999, \aap, 352, L69 
\bibitem[Brandner et al.(1997)]{bra97} Brandner, W., Grebel, E. K., Chu, Y.-H., \& Weis, K. 1997, \apj, 475, L45
\bibitem[Brice\~{n}o et al.(2002)]{bri02} Brice\~{n}o, C., Luhman, K. L., Hartmann, L., Stauffer, J. R., \& Kirkpatrick, D. 2002, \apj, 580, 317
\bibitem[Brown(1998)]{bro98} Brown, A. G. A. 1998, ASP Conference Series, 142, ed. G. Gilmore \& D. Howell (San Francisco), 45
\bibitem[Clayton(1986)]{cla86} Clayton, C. A. 1986, \mnras, 219, 895
\bibitem[Crowther \& Dessart(1998)]{cro98} Crowther, P. A., \& Dessart, L. 1998, \mnras, 296, 622
\bibitem[Crowther et al.(2006)]{cro06} Crowther, P. A., Lennon, D. J., Walborn, N. R., \& Smartt, S. J. 2006, preprint (arXiv:astro-ph/0606717)
\bibitem[Chabrier(2003)]{cha03} Chabrier, G. 2003, \pasp, 115, 763
\bibitem[Demarque et al.(2004)]{dem04} Demarque, P., Woo, J.-H., Kim, Y.-C., \& Yi, S. K. 2004, \apjs, 155, 667
\bibitem[De Pree et al.(1999)]{dep99} De Pree, C. G., Nysewander, M. C., \& Goss, W. M. 1999, \aj, 117, 2902
\bibitem[Dib et al.(2007)]{dib07} Dib, S., Kim, J., \& Shadmehri, M. 2007, \mnras, 381, L40
\bibitem[Diolaiti et al.(2000)]{dio00} Diolaiti, E., Bendinelli, O., Bonaccini, D., Close, L., Currie, D., \& Parmeggiani, G.  2000, \aap, 147, 335
\bibitem[Djorgovski et al. (1995)]{djo95} Djorgovski, S., et al. 1995, \apj, 438, L13
\bibitem[Drissen et al.(1995)]{dri95} Drissen, L., Moffat, A. F. J., Walborn, N. R., \& Shara, M. M. 1995, \aj, 110, 2235
\bibitem[Duch\^{e}ne et al.(1999)]{duc99} Duch\^{e}ne, G., Bouvier, J., \& Simon, T. 1999, \aap, 343, 831
\bibitem[Duquennoy \& Mayor(1991)]{duq91} Duquennoy, A., \& Mayor, M. 1991, \aap, 248, 485
\bibitem[Eisenhauer et al.(1998)]{eis98} Eisenhauer, F., Quirrenbach, A., Zinnecker, H., \& Genzel, R.  1998, \apj, 498, 278
\bibitem[Elson, Fall, \& Freeman(1987)]{els87} Elson R. A. W., Fall M. S., \& Freeman K. C. 1987, \apj, 323, 54
\bibitem[Figer et al.(1999)]{fig99} Figer, D. F., Kim, S. S., Morris, M., Serabyn, E., Rich, R. M., \& McLean, I. S. 1999, \apj, 525, 750
\bibitem[Figuer\^{e}do et al.(2002)]{fig02} Figuer\^{e}do, E., Blum, R. D., Damineli, A., \& Conti, P. S. 2002, \aj, 124, 2739
\bibitem[Fischer \& Marcy(1992)]{fis92} Fischer, D. A., \& Marcy, G. W. 1992, \apj, 396, 178
\bibitem[Frogel et al.(1977)]{fro77} Frogel, J. A., Persson, S. E., \& Aaronson, M. 1977, \apj, 213, 723
\bibitem[Genzel et al.(2003)]{gen03} Genzel, R., et al. 2003, \apj, 594, 812
\bibitem[Girardi et al.(2000)]{gir00} Girardi, L., Bressan, A., Bertelli, G., \& Chiosi, C. 2000, A\&AS, 141, 371
\bibitem[Goss \& Radhakrishnan(1969)]{gos69} Goss, W. M., \& Radhakrishnan, V. 1969, Astrophys. Lett., 4, 199
\bibitem[Grabelsky et al.(1988)]{gra88} Grabelsky, D. A., Cohen, R. S., Bronfman, L., \& Thaddeus, P. 1988, \apj, 331, 181
\bibitem[Grebel(2004)]{gre04} Grebel, E. K. 2004, ASP Conference Series, 322, ed. Lamers et al. (San Francisco), 101
\bibitem[Hernandez et al.(2007)]{her07} Hernandez, J., et al. 2007, \apj, 662, 1067
\bibitem[Hillenbrand et al.(1998)]{hil98} Hillenbrand, L. A., Strom, S. E., Calvet, N., Merrill, K. M., Gatley, I., Makidon, R. B., Meyer, M. R., \& Skrutskie, M. F. 1998, \aj, 116, 1816
\bibitem[Hillenbrand \& Hartmann(1998)]{hil98b} Hillenbrand, L. A., \& Hartmann, L. W. 1998, \apj, 492, 540
\bibitem[Hillenbrand \& White(2004)]{hil04} Hillenbrand, L. A., \& White, R. J. 2004, \apj, 604, 741
\bibitem[Hofmann \& Weigelt(1986)]{hof86} Hofmann, K.-H., Weigelt, G. 1986, \aap, 167, L15
\bibitem[Hofmann et al.(1995)]{hof95} Hofmann, K.-H., Seggewiss, W., Weigelt, G. 1995, \aap, 300, 403
\bibitem[Hunter et al.(1995)]{hun95} Hunter, D. A., Shaya, E. J., Holtzman, J. A., Light, R. M., O'Neil, E. J., Jr., \& Lynds, R. 1995, \apj, 448, 179
\bibitem[Kennicutt(1984)]{ken84} Kennicutt, R. C., Jr. 1984, \apj, 287, 116
\bibitem[Kenyon \& Hartmann(1995)]{ken95} Kenyon, S. J., \& Hartmann, L. 1995, \apjs, 101, 117
\bibitem[Kenyon \& G\'{o}mez(2001)]{ken01} Kenyon, S. J., \& G\'{o}mez, M. 2001, \aj, 121, 2673 
\bibitem[Kerr \& Lynden-Bell(1986)]{ker86} Kerr, F. J. \& Lynden-Bell, D. 1986, \mnras, 221, 1023
\bibitem[Kim et al.(2006)]{kim06} Kim, S. S., Figer, D. F., Kudritzki, R. P., \& Najarro, F. 2006, \apj, 653, L113
\bibitem[King(1962)]{kin62} King, I. 1962, \aj, 67, 471
\bibitem[Kroupa(2001)]{kro01} Kroupa, P. 2001, \mnras, 322, 231
\bibitem[Kroupa(2002)]{kro02} Kroupa, P. 2002, Science, 295, 82
\bibitem[K\"{o}hler(2001)]{koe01} K\"{o}hler, R. 2001, \aj, 122, 3325
\bibitem[K\"{o}hler et al.(2006)]{koe06} K\"{o}hler, R., Petr-Gotzens, M. G., McCaughrean, M. J., et al. 2006, \aap, 458, 461
\bibitem[Kouwenhoven et al.(2005)]{kou05} Kouwenhoven, M. B. N., Brown, A. G. A., Zinnecker, H., Kaper, L., \& Portegies Zwart, S. F. 2005, \aap, 430, 137
\bibitem[Lada \& Adams(1992)]{lad92} Lada, C. J., \& Adams, F. 1992, \apj, 393, 278 
\bibitem[Lada et al.(2000)]{lad00} Lada, C. J., Muench, A. A., Haisch, K. E., et al. 2000, \aj, 120, 3162
\bibitem[Lejeune \& Schaerer(2001)]{lej01} Lejeune, T., \& Schaerer, D. 2001, \aap, 366, 538 
\bibitem[Luhman et al.(2003)]{luh03} Luhman, K. L., Stauffer, J. R., Muench, A. A., Rieke, G. H., Lada, E. A., Bouvier, J., \& Lada, C. J. 2003, \apj, 593, 1093
\bibitem[Mackey \& Gilmore(2003)]{mac03} Mackey, A. D., \& Gilmore, G. F. 2003, \mnras, 338, 85
\bibitem[Maercker \& Burton(2005)]{mae05} Maercker, M., \& Burton, M. G. 2005, \aap, 438, 663
\bibitem[Maercker et al.(2006)]{mae06} Maercker, M., Burton, M. G., \& Wright, C., M. 2006, \aap, 450, 253
\bibitem[Malkov \& Zinnecker(2001)]{mal01} Malkov, O., Zinnecker, H. 2001, \mnras, 321, 149
\bibitem[Massey \& Hunter(1998)]{mas98} Massey, P., \& Hunter, D. A. 1998, \apj, 493, 180
\bibitem[Melnick \& Grosb{\o}l(1982)]{mel82} Melnick, J., \& Grosb{\o}l, P. 1982, \aap, 107, 23
\bibitem[Melnick et al.(1989)]{mel89} Melnick, J., Tapia, M., \& Terlevich, R. 1989, \aap, 213, 89
\bibitem[Meynet \& Maeder(2003)]{mey03} Meynet, G., \& Maeder, A. 2003, \aap, 404, 975
\bibitem[Miller \& Scalo(1979)]{mil79} Miller G. E., Scalo J. M. 1979, ApJS, 41, 513
\bibitem[Moffat(1983)]{mof83} Moffat, A. F. J. 1983, \aap, 124, 273
\bibitem[Moffat et al.(1994)]{mof94} Moffat, A. F. J., Drissen, L., \& Shara, M. M. 1994, \apj, 436, 183
\bibitem[Moffat et al.(2002)]{mof02} Moffat, A. F. J., et al. 2002, \apj, 573, 191
\bibitem[Moffat et al.(2004)]{mof04} Moffat, A. F. J., Poitras, V., Marchenko, S. V., Shara, M. M., Zurek, D. R., Bergeron, E., Antokhina, E. A. 2004, \aj, 128, 2854
\bibitem[Muench et al.(2002)]{mue02} Muench, A. A., Lada, E. A., Lada, C. J., Alves J. 2002, \apj, 573, 366
\bibitem[N\"{u}rnberger \& Petr-Gotzens(2002)]{npg02} N\"{u}rnberger, D. E. A., \& Petr-Gotzens, M. G. 2002, \aap, 382, 537
\bibitem[N\"{u}rnberger et al.(2002)]{nue02} N\"{u}rnberger, D. E. A., Bronfman, L., Yorke, H. W., \& Zinnecker, H. 2002, \aap, 394, 253
\bibitem[N\"{u}rnberger \& Stanke(2003)]{ns03} N\"{u}rnberger, D. E. A., \& Stanke, T. 2003, \aap, 400, 223
\bibitem[Palla \& Stahler(1999)]{pal99} Palla, F., \& Stahler, S. W.  1999,
\apj, 525, 772
\bibitem[Pandey et al.(2000)]{pan00} Pandey, A. K., Ogura, K., \& Sekiguchi, K. 2000, PASJ, 52, 847
\bibitem[Patience et al.(1998)]{pat98} Patience, J., Ghez, A. M., Reid, I. N., Weinberger, A. J., Matthews, K. 1998, \aj, 115, 1972
\bibitem[Paumard et al.(2006)]{pau06} Paumard, T., et al. 2006, \apj, 643, 1011
\bibitem[Petr et al.(1998)]{pet98} Petr-Gotzens, M. G., Coud\'{e} du Foresto, V., Beckwith, S. V. W., Richichi, A., \& McCaughrean, M. J. 1998, \apj, 500, 825
\bibitem[Pfalzner et al.(2006)]{pfa06} Pfalzner, S., Olczak, C., Eckart, A. 2006, \aap, 454, 811
\bibitem[Phelps \& Janes(1993)]{phe93} Phelps, R. L., \& Janes, K. A. 1993, \aj, 106, 1870
\bibitem[Pinfield et al.(1998)]{pin98} Pinfield, D. J., Hodgkin, S. T., \& Jameson, R. F. 1998, \mnras, 299, 955
\bibitem[Portegies Zwart et al.(2002)]{por02} Portegies Zwart, S. F., Makino, J., McMillan, S. L. W., \& Hut, P. 2002, \apj, 565, 265
\bibitem[Preibisch et al.(1999)]{pre99} Preibisch, Th., Balega, Y., Hofmann, K.-H., Weigelt, G., Zinnecker, H. 1999, New Astronomy, 4, 531
\bibitem[Preibisch et al.(2003)]{pre03} Preibisch, T., Stanke, T., \& Zinnecker, H. 2003, \aap, 409, 147 
\bibitem[Prisinzano et al.(2003)]{pri03} Prisinzano, L., Micela, G., Sciortino, S., \& Favata, F. 2003, \aap, 404, 927
\bibitem[Prosser et al.(1994)]{pro94} Prosser, C. F., Stauffer, J. R., Hartmann, L., et al. 1994, \apj, 421, 517
\bibitem[Reid \& Gizis(1997)]{rei97} Reid, I. N., \& Gizis, J. E. 1997, \aj, 113, 2246
\bibitem[Rieke \& Lebofsky(1985)]{rie85} Rieke, G. H., \& Lebofsky, M. J. 1985, \apj, 288, 618
\bibitem[Rudolph et al.(2006)]{rud06} Rudolph, A. L., Fich, M., Bell, G. R., Norsen, T., Simpson, J., P., Haas, M. R. \& Erickson, E. F 2006, \apjs, 162, 346
\bibitem[Sagar et al.(2001)]{sag01} Sagar, R., Munari, U., \& de Boer, K. S. 2001, \mnras, 327,23
\bibitem[Salpeter(1955)]{sal55} Sapleter, E. E. 1955, \apj, 121, 161
\bibitem[Sanner et al.(2000)]{san00} Sanner, J., Altmann, M., Brunzendorf, J., \& Geffert, M. 2000, \aap, 357, 471
\bibitem[Scalo(1986)]{sca86} Scalo, J. M. 1986, Fundamentals of Cosmic Physics, 11, 1
\bibitem[Scalo(1998)]{sca98} Scalo, J. M. 1998, in ASP Conf. Ser. 142, The Stellar Initial Mass Function (38yh Herstmonceux Conference), ed. G. Gilmore \& D. Howell (San Francisco: Astron. Soc. Pac.), 201
\bibitem[Scalo(2005)]{sca05} Scalo, J. M. 2005, in IMF@50: The Initial Mass Function 50 Years Later, ed. E. Corbelli, F. Palla, \& H. Zinnecker (Dordrecht: Kluwer), 23 
\bibitem[Schmutz \& Drissen(1999)]{sch99} Schmutz, W., Drissen, L. 1999, Rev. Mexicana Astron. Astrofis. Ser. Conf., 8, 41
\bibitem[Shatsky \& Tokovinin(2002)]{sha02} Shatsky, N., \& Tokovinin, A. 2002, \aap, 382, 92
\bibitem[Sher(1965)]{she65} Sher, D. 1965, \mnras, 129, 237
\bibitem[Siess et al.(2000)]{sie00} Siess, L., Dufour, E., \& Forestini, M. 2000, \aap, 358, 593 
\bibitem[Sirianni et al.(2000)]{sir00} Sirianni, M., Nota, A., De Marchi, G., Leitherer, C., \& Clampin, M. 2000, \apj , 533, 203
\bibitem[Slesnick et al.(2002)]{sle02} Slesnick, C. L., Hillenbrand, L. A., \& Massey, P. 2002, \apj, 576, 880
\bibitem[Sollima et al.(2007)]{sol07} Sollima, A., Ferraro, F. R., \& Bellazzini, M. 2007, arXiv:astro-ph/0708072
\bibitem[Stolte et al.(2004)]{sto04} Stolte, A., Brandner, W., Brandl, B., Zinnecker, H., \& Grebel, E. K. 2004, \aj, 128, 765
\bibitem[Stolte et al.(2005)]{sto05} Stolte, A., Brandner, W., Grebel, E. K., Lenzen, R., \& Lagrange, A.-M. 2005, \apj, 628, L113
\bibitem[Stolte et al.(2006)]{sto06} Stolte, A., Brandner, W., Brandl, B., Zinnecker, H., \& Grebel, E. K. 2006, \aj, 132, 253
\bibitem[Sung \& Bessell(2004)]{sun04} Sung, H., \& Bessell, M. S. 2004, \aj, 127, 1014
\bibitem[Tapia et al.(2001)]{tap01} Tapia, M., Bohigas, J., P\'{e}rez, B., Roth, M., Ruiz, Ma. T. 2001, RMxAA, 37, 39
\bibitem[Tout et al.(1999)]{tou99} Tout, C., Livio, M., \& Bonnell, I. 1999, \mnras, 310, 360
\bibitem[van den Bergh(1978)]{vdb78} van den Bergh, S. 1978, \aap, 63, 275
\bibitem[van den Bos(1928)]{vdb28} van den Bos, W. H. 1928, Bull. Astron. Inst. Netherlands, 4, 261 
\bibitem[Walborn(1973)]{wal73} Walborn, N. R. 1973, \apj, 182, L21
\bibitem[Walborn et al.(2002)]{wal02} Walborn, N. R., Howarth, I. D., Lennon, D. J., et al. 2002, \aj, 123, 2754
\bibitem[Wilking et al.(2004)]{wil04} Wilking, B. A., Meyer, M. R., Greene, T. P., Mikhail, A., \& Carlson, G. 2004, \aj, 127, 1131
\bibitem[Yan et al.(1998)]{yan98} Yan, L., McCarthy, P. J., Storrie-Lombardi, L. J., \& Weymann, R. J. 1998, \apj, 503, L19
\bibitem[Yi et al.(2001)]{yi01} Yi, S., Demarque, P., Kim, Y. -C., Lee, Y.-W., Ree, C. H., Lejeune, T., \& Barnes, S. 2001, \apjs, 136, 417 
\end{thebibliography}
\end{document}